\documentclass[prd,amssymb,amsmath,amsfonts,nofootinbib,reprint,showpacs,longbibliography]{revtex4-1}

\usepackage{graphicx}
\usepackage{lmodern}
\usepackage{amsmath,amssymb}
\usepackage{mathrsfs}
\usepackage{amsfonts}
\usepackage[utf8]{inputenc}
\usepackage{url}
\usepackage[colorlinks]{hyperref}
\usepackage[table]{xcolor}
\usepackage{multirow}
\usepackage[normalem]{ulem}
\usepackage{lipsum}
\normalsize
\usepackage{booktabs}
\usepackage{tabularx}
\usepackage{paralist}

\usepackage{marvosym}
\usepackage{enumerate}
\usepackage{color,soul}
\usepackage{acronym}

\usepackage{bm}
\usepackage{color}
\usepackage{commath}
\allowdisplaybreaks
\usepackage{multirow}

\begin{document}
\title{Parametrized model for gravitational waves from eccentric, precessing binary black holes: 
       theory-agnostic tests of General Relativity with pTEOBResumS}


\newcommand{\RG}[1]{{\textcolor{dodgerblue}{{RG: #1}} }}
\newcommand{\JL}[1]{{\textcolor{purple}{{JL: #1}} }}
\newcommand{\DC}[1]{{\textcolor{orange}{{DC: #1}} }}
\newcommand{\NC}[1]{{\textcolor{magenta}{{NC: #1}} }}
\newcommand{\kc}[1]{{\textcolor{purple}{{KC: #1}} }}

\newcommand{\todo}[1]{\textcolor{red}{\texttt{TODO: #1}}} 
\newcommand{\red}[1]{\textcolor{red}{#1}} 
\newcommand{\bl}[1]{\textcolor{blue}{#1}} 
\newcommand{\cor}[2]{\sout{#1}\textcolor{red}{#2}} 
\newcommand{\old}[1]{\textcolor{gray}{\sout{#1}}}
\newcommand{\new}[1]{\red{#1}}

\newcommand{\eo}[0]{\hat{E}_0}
\newcommand{\lo}[0]{\hat{L}_0}
\newcommand{\bphi}[0]{\bar{\varphi}}
\newcommand{\dbphi}[0]{\dot{\bar{\varphi}}}
\newcommand{\dphi}[0]{\dot{\varphi}}
\newcommand{\amrg}[1]{{A_{#1}^{\rm mrg}}}
\newcommand{\anqc}[1]{{A_{#1}^{\rm NQC}}}
\newcommand{\omgmrg}[1]{{\omega_{#1}^{\rm mrg}}}
\newcommand{\apeak}[1]{{A_{#1}^{\rm peak}}}
\newcommand{\omgpeak}[1]{{\omega_{#1}^{\rm peak}}}

\newcommand{\avg}[1]{{\langle {#1} \rangle}}
\newcommand{\scri}{{\mathrsfs{I}}}
\newcommand{\stf}[1]{{\langle {#1} \rangle}}
\newcommand{\ord}{\mathcal{O}}
\newcommand{\f}{\frac}
\newcommand{\gsf}{\text{1GSF}}
\newcommand{\bk}{\text{0GSF}}
\newcommand{\el}{\ell}
\newcommand{\nn}{\nonumber}
\newcommand{\mbf}[1]{\mathbf{#1}}
\newcommand{\LOhat}{\!\hat{\,\mathbf{L}}_0}
\newcommand{\Lhat}{\!\hat{\,\mathbf{L}}_\text{N}}
\newcommand{\Jhat}{\hat{\mathbf{J}}_\text{N}}
\newcommand{\Sa}{\mbf{S}_1}
\newcommand{\Sb}{\mbf{S}_2}
\newcommand{\hs}{\hat{s}}
\newcommand{\hS}{\hat{S}}
\newcommand{\Lhatdot}{\dot{\hat{\,\mathbf{L}}}_\text{N}}
\newcommand{\LhatdotNew}{\dot{\hat{\,\mathbf{L}}}_{\text{N},\perp}}
\newcommand{\Sadot}{\dot{\mbf{S}}_1}
\newcommand{\Sbdot}{\dot{\mbf{S}}_2}
\newcommand{\Jdot}{\dot{\mbf{J}}}
\newcommand{\rdot}{\dot{\mbf{r}}}
\newcommand{\vdot}{\dot{\mbf{v}}}
\newcommand{\ud}{\mathrm{d}}
\newcommand{\Ldot}{\dot{\mbf{L}}_{\text N}}
\newcommand{\LN}{\mbf{L}_{\text N}}
\newcommand{\nnm}{\nonumber}
\newcommand{\mce}{\mathcal{E}}
\newcommand{\mcb}{\mathcal{B}}
\newcommand{\bayes}[2]{\mathcal{B}^{#1}_{#2}}

\newcommand{\dali}[0]{\texttt{TEOBResumS-Dalí}}
\newcommand{\gwecc}[0]{\texttt{gw\_eccentricity}}
\newcommand{\bilby}[0]{\texttt{bilby}}
\newcommand{\pseob}[0]{\texttt{pSEOBNRv5PHM}}
\newcommand{\dcssim}[1]{\texttt{dCS_0p#1}}
\def\TEOBResumS{\texttt{TEOBResumS}}
\def\TEOBResumSGIOTTO{\texttt{TEOBResumS-GIOTTO}}
\def\bajes{\texttt{bajes}}
\def\Boldtheta{\boldsymbol{\theta}}
\def\Boldd{\textbf{d}}
\def\TEOBResumSDali{\texttt{TEOBResumS-Dalí}}
\def\TEOBB{\texttt{pTEOBResumS}}
\def\cthree{c_{\rm{N}^3\rm{LO}}}
\def\mbhf{M_{\rm BH}^f}
\def\abhf{a_{\rm BH}^f}
\def\alphalm0{\alpha_{\ell m 0}}
\def\taulm0{\tau_{\ell m 0}}
\def\omegalm0{\omega_{\ell m 0}}
\def\lm{{\ell m}}

\def\Nevents{11}

\definecolor{cyan}{rgb}{0,0.9,0.9}
\definecolor{orange}{rgb}{0.9,0.5,0}
\definecolor{magenta}{rgb}{1,0,1}
\definecolor{purple}{rgb}{0.8,0.4,0.8}
\definecolor{gray}{rgb}{0.8242,0.8242,0.8242}
\definecolor{dodgerblue}{rgb}{0.12, 0.56, 1.0}
\newacro{adm}[ADM]{Arnowitt-Deser-Misner}
\newacro{bbh}[BBH]{binary black hole}
\newacroplural{bbh}[BBHs]{binary black holes}
\newacro{bh}[BH]{black hole}
\newacroplural{bh}[BHs]{black holes}
\newacro{bhns}[BHNS]{black hole-neutron star}
\newacro{bhpt}[BHPT]{black hole perturbation theory}
\newacro{bns}[BNS]{binary neutron star}
\newacro{bf}[BF]{Bayes factor}
\newacro{cbc}[CBC]{compact binary coalescence}
\newacro{ce}[CE]{Cosmic Explorer}
\newacro{da}[DA]{data analysis}
\newacro{et}[ET]{Einstein Telescope}
\newacro{eob}[EOB]{Effective-One-Body}
\newacro{eom}[EOM]{equations of motion}
\newacro{fd}[FD]{frequency domain}
\newacro{fft}[FFT]{Fast Fourier transform}
\newacro{gw}[GW]{gravitational wave}
\newacroplural{gw}[GWs]{gravitational waves}
\newacro{gr}[GR]{general relativity}
\newacro{grb}[GRB]{gamma-ray burst}
\newacro{grhd}[GRHD]{general-relativistic hydrodynamics}
\newacro{gwosc}[GWOSC]{Gravitational Wave Open Science Center}
\newacro{gwtc1}[GWTC-1]{the first gravitational-wave transients catalog}
\newacro{gsf}[GSF]{Gravitational Self Force}
\newacro{hm}[HM]{Higher mode}
\newacroplural{hm}[HMs]{Higher modes}
\newacro{ifo}[IFO]{interferometer}
\newacro{imr}[IMR]{inspiral-merger-ringdown}
\newacro{im}[IMR]{inspiral-to-merger}
\newacro{kagra}[KAGRA]{Kamioka Gravitational Wave Detector}
\newacro{ligo}[LIGO]{Laser Interferometer Gravitational-Wave Observatory}
\newacro{lisa}[LISA]{Laser Interferometer Space Antenna}
\newacro{lr}[LR]{Light Ring}
\newacro{lso}[LSO]{Last Stable Orbit}
\newacro{lvc}[LVC]{LIGO-Virgo Collaboration}
\newacro{lvk}[LVK]{LIGO-Virgo-KAGRA}
\newacro{lo}[LO]{leading order}
\newacro{ns}[NS]{neutron star}
\newacroplural{ns}[NSs]{neutron stars}
\newacro{nr}[NR]{numerical relativity}
\newacro{nqc}[NQC]{next-to-quasicircular}
\newacro{nlo}[NLO]{next-to-leading order}
\newacro{nnlo}[NNLO]{next-to-next-to-leading order}
\newacro{n3lo}[N3LO]{next-to-next-to-next-to-leading order}
\newacro{n4lo}[N3LO]{next-to-next-to-next-to-next-to-leading order}
\newacro{ode}[ODE]{Ordinary Differential Equation}
\newacroplural{ode}[ODEs]{Ordinary Differential Equations}
\newacro{pn}[PN]{post-Newtonian}
\newacro{pm}[PM]{post-Minkowskian}
\newacro{pe}[PE]{parameter estimation}
\newacro{psd}[PSD]{power spectral density}
\newacroplural{psd}[PSD]{power spectral densities}
\newacro{pa}[PA]{post-adiabatic}
\newacro{qnm}[QNM]{quasi-normal mode}
\newacro{qc}[QC]{quasi-circular}
\newacro{snr}[SNR]{signal-to-noise ratio}
\newacro{spa}[SPA]{stationary-phase approximation}
\newacro{sxs}[SXS]{Simulating eXtreme Spacetimes}
\newacro{td}[TD]{time domain}
\newacro{ng}[NG]{Nect Generation}
\newacro{dcs}[dCS]{dynamical Chern-Simons}
\newacro{bgr}[beyond-GR]{beyond-\ac{gr}}
\newacro{spec}[SpEC]{Spectral Einstein Code}
\newacro{edgb}[EdGB]{Einstein-dilaton-Gauss-Bonnet}
\newacro{emri}[EMRI]{extreme mass-ratio inspiral}
\newacro{gwtc}[GWTC]{Gravitational-Wave Transient Catalog}
\newacro{far}[FAR]{false alarm rate}
\newacro{ci}[CI]{credible interval}
\newacro{bs}[BS]{boson star}

\author{Danilo \surname{Chiaramello}${}^{1,2}$}
\author{Nicol\'o \surname{Cibrario}${}^{1,2}$}
\author{Jacob \surname{Lange}${}^{2}$}
\author{Koustav \surname{Chandra}${}^{3,5}$}
\author{Rossella \surname{Gamba}${}^{3,4}$}
\author{Raffaella \surname{Bonino}${}^{1,2}$}
\author{Alessandro \surname{Nagar}${}^{2}$}

\affiliation{${}^{1}$ Dipartimento di Fisica, Universit\`a di Torino, Via P. Giuria 1, 10125 Torino, Italy}
\affiliation{${}^{2}$ INFN Sezione di Torino, Torino, 10125, Italy}
\affiliation{${}^{3}$ Institute for Gravitation \& the Cosmos, Department of Physics \& Department of Astronomy and Astrophysics, The Pennsylvania State University, University Park, Pennsylvania 16802, USA}
\affiliation{${}^{4}$ Department of Physics, University of California, Berkeley, CA 94720, USA}
\affiliation{${}^{5}$ Max Planck Institute for Gravitational Physics (Albert Einstein Institute), Am M{\"u}hlenberg 1, Potsdam 14476, Germany}

\begin{abstract}
    Gravitational waves from binary black hole mergers allow us to test General Relativity in the strong-field, 
    high-curvature regime.
    However, existing gravitational waves-based tests for this have so far assumed non-eccentric signal sources, limiting their range of applicability.
    In this work, we present \TEOBB, a new parametrized inspiral-merger-ringdown model for null tests of General Relativity that incorporates both orbital eccentricity and spin precession, 
    building on the effective-one-body model \dali.
    We introduce parametrized deviations from General Relativity both in the inspiral and the merger-ringdown regimes.
    We validate the model through parameter estimation of synthetic signals generated with the model itself or from numerical simulations, including of systems featuring physics beyond General Relativity.
    We thus establish the model's consistency, demonstrate its capability to identify effects beyond General Relativity, and gauge the impact of eccentricity on the recovery of deviation parameters.
    We then analyze a set of binary black hole events from the first three LIGO-Virgo-KAGRA observing runs, under either the eccentric, spin-aligned or precessing, quasi-circular hypotheses, searching for deviations from General Relativity in the remnant black hole's fundamental quasi-normal mode.
    We find no statistically significant evidence for such deviations in any of the events.
    We combine single event results either by multiplying likelihoods, or via a hierarchical procedure, assuming an underlying normal distribution for the deviations.
    Consistent with previous works, we infer a mild preference for longer remnant quasi-normal mode
    damping times than expected, although the limited event sample and potential systematics reduce the significance of this finding.
    We find no support for orbital eccentricity among the analyzed events, except for GW200129.
    An eccentric description is strongly favored for this event, although as previous works have noted this result could be influenced by data quality issues.
\end{abstract}

\date{\today}
\maketitle

\acresetall

\section{Introduction}

Since the landmark direct detection of \acp{gw} 
from the merger of two \acp{bh} in 2015~\cite{LIGOScientific:2016aoc}, \ac{gw} astronomy has advanced at an accelerating pace.
Successive observing runs by the \ac{lvk} collaboration, enabled by steadily improving detector sensitivity~\cite{LIGOScientific:2025slb, LIGOScientific:2025yae, LIGOScientific:2025snk},
have reported an ever-growing catalog that now includes signals from \ac{bns}~\cite{LIGOScientific:2017vwq, LIGOScientific:2020aai} and mixed
\ac{bh}-\ac{ns} systems~\cite{LIGOScientific:2021qlt, LIGOScientific:2024elc} on top of the dominant \ac{bbh} population.
Meanwhile, advances in waveform modelling have produced increasingly physically complete \ac{cbc} models, including phenomenological
approaches~\cite{Pratten:2020ceb, Thompson:2023ase, Estelles:2021gvs, Colleoni:2024knd, Hamilton:2025xru, Planas:2025feq}, \ac{nr} surrogates~\cite{Blackman:2015pia, Varma:2019csw}, and \ac{eob}
models~\cite{Gamba:2024cvy,Nagar:2024oyk,Albanesi:2025txj,Pompili:2023tna,vandeMeent:2023ols,Khalil:2023kep,
Ramos-Buades:2023ehm,Gamboa:2024hli,Estelles:2025zah}. Together with improvements in detection~\citep{Allen:2005fk, Usman:2015kfa, Messick:2016aqy, Sachdev:2019vvd, Aubin:2020goo, Drago:2020kic, Chandra:2021wbw, Klimenko:2022nji, Kumar:2024bfe, Mishra:2024zzs, Joshi:2025nty}
and \ac{pe} algorithms~\cite{Veitch:2014wba, Ashton:2018jfp, Pankow:2015cra, Lange:2018pyp, Dax:2021tsq}, these developments
have enabled not only confident identification of compact binary signals, but also inference of their properties
and astrophysical origin.

Additionally, \ac{bbh} mergers are currently the only experimentally available laboratory for testing \ac{gr} in the strong-field regime. 
Such tests can be broadly split into \textit{theory-specific} and \textit{theory-independent}. 
In the former, observational data are directly compared against predictions from a particular alternative theory of
gravity~\cite{Will:2014kxa, Berti:2015itd, Yunes:2016jcc, Maselli:2023khq,Julie:2022qux,Julie:2024fwy}, including models
with extra fields and/or higher-order curvature corrections to the \ac{gr} action (e.g. \ac{edgb} and \ac{dcs} gravity), or ones postulating variation of Newton's
constant, a massive graviton, or the existence of extra dimensions (see e.g. Ref~\cite{Yunes:2024lzm} for a review). 
Direct tests of these kinds are limited due to a lack of waveform models
with accuracy and completeness comparable to state-of-the art \ac{gr}-ones, despite recent progress. 

Theory-independent tests adopt a more agnostic framework. They search for deviations from \ac{gr} predictions and encompass a broad suite of analyses,
including consistency tests and parametrized tests (See~\citep{Krishnendu:2021fga} for an overview).
The former check for agreement between the observed data and the \ac{gr}-predicted \ac{imr} signal.
Put simply, they search for any statistically significant ``trace'' in the data that is unlikely to be explained as
either part of a \ac{gr} signal or instrument noise. Currently, the \ac{lvk} collaboration performs two
kinds of consistency tests: the residual test, which searches for an anomalous signature left in the data after
subtracting the best-fit \ac{gr} signal~\cite{LIGOScientific:2016lio}, and the inspiral-merger-ringdown consistency test,
which compares independent remnant property estimates from the inspiral and post-inspiral signal~\cite{Hughes:2004vw, Ghosh:2016qgn, Ghosh:2017gfp}.
A related check tests the black hole area theorem by verifying whether the final black hole's area exceeds the combined
area of the progenitors~\citep{Cabero:2017avf, Isi:2020tac, KAGRA:2025oiz}. Although not part of the \ac{lvk}'s suite of analyses, the \texttt{SCoRe} framework also measures consistency by looking for unmodelled signatures in \ac{gw} data by analyzing the cross-correlation of residual strains in pairs of detectors~\citep{Dideron:2022tap, Dideron:2024xwm}.

Parametrized tests, on the other hand, introduce model-agnostic phenomenological deviations in the waveform and use observed data
to infer statistical bounds on these parameters. The rationale is that, for fixed intrinsic parameters, the inspiral-merger-ringdown
dynamics are uniquely determined within \ac{gr} and are well understood via \ac{pn} theory, numerical relativity, and \ac{bh} perturbation theory.
Any departure from \ac{gr} modifies the dynamics and thus the waveform. Applications include constraints on semi-analytically modelled inspiral coefficients
through pipelines such as \texttt{TIGER}~\cite{Agathos:2013upa, Meidam:2017dgf, Roy:2025gzv} and \texttt{FTI}~\cite{Mehta:2022pcn, Sanger:2024axs}, often combined with
principal component analysis to reduce parameter correlations~\cite{Pai:2012mv, Saleem:2021nsb, Shoom:2021mdj}; tests of the spin-induced quadrupole moment
to probe potential departures from the Kerr black-hole expectation~\cite{Krishnendu:2017shb, Saleem:2021vph}; merger-ringdown analyses such as
\texttt{pSEOBNR} and \texttt{KerrPostmerger}~\cite{Brito:2018rfr, Ghosh:2021mrv, Maggio:2022hre, Maggio:2023vch, Gennari:2023gmx, Toubiana:2023cwr, Pompili:2025cdc},
which test the compatibility of the remnant's \ac{qnm} spectrum with that of a Kerr black hole.
In particular, \texttt{pSEOBNR} employs a parametrized \ac{eob} model~\cite{Pompili:2023tna, vandeMeent:2023ols, Khalil:2023kep, Ramos-Buades:2023ehm, Gamboa:2024hli, Estelles:2025zah}
to perform a null test on the frequency $\omega_{220}$ and damping time $\tau_{220}$ of the fundamental \ac{qnm} or
other higher-order modes, using the full \ac{imr} signal.
This pipeline tests whether the observed post-inspiral signal is consistent with a \ac{gr}-driven binary inspiral.
Its latest variant, \pseob~\cite{Pompili:2025cdc}, incorporates spin precession as it is based on multipolar precessing \ac{bbh}
waveform model \texttt{SEOBNRv5PHM}~\cite{Pompili:2023tna, vandeMeent:2023ols,Khalil:2023kep, Ramos-Buades:2023ehm}.
An alternative to this strategy is proposed in \citet{Chandra:2025ipu} where the authors introduce a generic prescription
and use the post-inspiral portion of \ac{nr} surrogate models as a test case.

As the number of detected events increases --- the fourth \ac{gwtc} now lists over 200 events~\cite{LIGOScientific:2025slb} --- and detector sensitivity improves,
constraints from parametrized tests are becoming increasingly stringent, with no evidence so far for physics beyond \ac{gr}.
Next-generation detectors, including the ground-based \ac{et}~\cite{Punturo:2010zz} and \ac{ce}~\cite{Reitze:2019iox},
and the space-based \ac{lisa}~\cite{2017arXiv170200786A}, will achieve unprecedented sensitivity across the frequency spectrum,
enabling observations of many thousands of high-\ac{snr}(\(\gtrsim 100\)) events. At such precision, statistical uncertainties
will approach the level of current modeling systematics, making accurate waveform modeling essential to avoid misinterpreting
mismodeling as signs of beyond-\ac{gr} physics~\cite{Gupta:2024gun}.
To meet this challenge, \ac{gr}-based waveform modelling has advanced rapidly. Spin precession is now incorporated in \ac{nr} surrogates~\cite{Varma:2019csw},
phenomenological~\cite{Pratten:2020ceb,Thompson:2023ase,Estelles:2021gvs,Colleoni:2024knd,Hamilton:2025xru}, and \ac{eob} 
models~\cite{Gamba:2024cvy,Ramos-Buades:2023ehm}. Parallel efforts have targeted accurate modelling of signals from eccentric
sources~\cite{Yunes:2009yz,Huerta:2014eca,Tanay:2016zog,Hinderer:2017jcs,Hinder:2017sxy,Cao:2017ndf,Loutrel:2017fgu,Huerta:2016rwp,Huerta:2017kez,Moore:2018kvz,Moore:2019xkm,Tanay:2019knc,Tiwari:2019jtz,Tiwari:2020hsu,Setyawati:2021gom,Ramos-Buades:2021adz,Islam:2021mha,Islam:2024zqo,Planas:2025feq,Nee:2025nmh,Maurya:2025shc,Morras:2025nlp,Gamboa:2024hli,Gamboa:2024imd,Paul:2024ujx,Islam:2024rhm,Islam:2024bza,Islam:2025llx,Islam:2025rjl,Islam:2025bhf}.
This has lead to to the first \ac{lvk} constraints on eccentricity in an
event (GW250114)~\cite{KAGRA:2025oiz} via \texttt{SEOBNRv5EHM}~\cite{Gamboa:2024imd,Gamboa:2024hli} and \dali~\cite{Nagar:2024dzj}. 
Given that neglecting eccentricity can bias parameter inference~\cite{Ramos-Buades:2019uvh,OShea:2021faf,Divyajyoti:2023rht,Divyajyoti:2025cwq}, 
especially when combined with spin precession~\cite{CalderonBustillo:2020xms,Romero-Shaw:2022fbf}, it can also affect tests of \ac{gr},
be it consistency or parametrized tests ~\cite{Saini:2022igm,Saini:2023rto,Narayan:2023vhm,Shaikh:2024wyn,Bhat:2022amc,Bhat:2024hyb}.

In this work, we present \TEOBB, a parametrized extension of the \dali\ waveform model for compact binaries with precessing spins in generic
orbits~\cite{Chiaramello:2020ehz, Nagar:2020xsk, Nagar:2021gss, Nagar:2021xnh, Nagar:2024oyk, Gamba:2024cvy, Albanesi:2025txj}.
This model introduces controlled deviations from the \ac{nr}-informed values of key physical quantities that determine the inspiral dynamics,
the corresponding waveform, and merger-ringdown signal, thereby enabling null tests of \ac{gr} in the strong-field regime.
Following and extending the approach of the \texttt{pSEOBNR} family, which is restricted to non-eccentric \ac{bbh} mergers, 
we incorporate parametrized corrections to two high-order \ac{pn} coefficients in the gravitational potentials and spin-orbit coupling.
At the waveform level, we allow each mode to deviate in its peak amplitude, peak frequency, and fundamental \ac{qnm} modes.
Additionally, we can introduce independent deviations in the \ac{nr}-fitted mass and spin of the remnant black hole, as an additional
probe of the inspiral-post-inspiral connection.

A crucial feature of \TEOBB\ is its ability to probe non-\ac{gr} effects in eccentric inspirals, owing to \dali's treatment of
eccentric \ac{bbh} dynamics in the inspiral-plunge phase.
However, as \dali\ currently assumes that eccentric binaries circularize before merger, the merger-ringdown is modeled using a non-eccentric prescription.
In future work, we aim to relax this assumption.

We validate our model through simulation studies employing \ac{nr} waveforms,
and apply it to a selection of \ac{bbh} events from the first three \ac{lvk} observing runs, imposing constraints on possible deviations from \ac{gr}.
A companion work leverages the breadth of physics included in \TEOBB~to
perform precision tests of \ac{gr} with GW250114, the highest-\ac{snr} event detected to date~\cite{Chandra:2025jfc}.

The rest of this paper is structured as follows. Sec.~\ref{sec:dali_base} summarises the main features of our baseline \dali~model;
Sec.~\ref{sec:pteob} describes the deviation parameters we introduce in it, exploring their phenomenological
effect; Sec.~\ref{sec:pe} summarizes the Bayesian framework and setup we use for \ac{pe} studies, presented
in Sec.~\ref{sec:results}. The latter includes most of our results, beginning with model and \ac{nr} simulations designed to
test the new parametrized model, and including our re-analyses of \ac{bbh} events.
Finally, in Sec.~\ref{sec:conclusions} we discuss our findings and outline future directions.

\paragraph*{Conventions}
Throughout this paper, we use geometric units in which $G=c=1$. For our spinning two-body systems, we define the component masses $m_1 \geq m_2$ and spin vectors $\mbf{S}_{1,2} = m_{1,2} \bm{a}_{1,2} =
m_{1,2}^2 \bm{\chi}_{1,2}$. We denote by $M=m_1 + m_2$ the total mass of the system, the mass ratio
by $q = m_1/m_2 \geq 1$, and the symmetric mass ratio by $\nu = m_1m_2/M^2 = \mu/M$, where $\mu$ is the reduced
mass. The chirp mass is defined as $\mathcal{M}_c = (m_1 m_2)^{3/5}/M^{1/5} = M \nu^{3/5}$.
We split the spin vectors into components parallel, $\chi_{1,2}^{\parallel}$, and perpendicular,
$\chi_{1,2}^{\perp}$, to the orbital angular momentum $\bm{L}$, and define the effective aligned \(\chi_\mathrm{{eff}}\) and precessing spin \(\chi_\mathrm{p}\)
parameters by~\cite{Racine:2008qv, Ajith:2009bn, Schmidt:2014iyl}:
\begin{subequations}
\begin{align}
\chi_{\rm eff} &= \frac{m_1 \chi_1^{\parallel} + m_2 \chi_2^{\parallel}}{M} \, , \\
\chi_{\rm p}   &= \max \biggl\{ \chi_1^\perp, \dfrac{4+3q}{4q^2+3q} \chi_2^\perp \biggr\}.
\end{align}
\end{subequations}
Spin effects in the orbital dynamics model are parametrized by the combinations $\tilde{a}_0 = \tilde{a}_1 +
\tilde{a}_2 = (m_1\chi_1^\parallel + m_2 \chi_2^\parallel)/M$ and $\tilde{a}_{12} = \tilde{a}_1 - \tilde{a}_2$,
as well as the total spin $\hat{\bm{S}} = (\bm{S}_1 + \bm{S}_2)/M^2$ and the vector $\hat{\bm{S}}_* = 
\left[(m_2/m_1) \bm{S}_1 + (m_1/m_2) \bm{S}_2\right]/M^2$.
We decompose the \ac{gw} strain $h$ into a sum of spin-weighted spherical harmonics of weight $-2$:
\begin{subequations}
    \begin{align}
        h          &= h_+ - i h_\times = \dfrac{1}{D_L} \sum_{\ell, m} h_{\ell m} {}_{-2}Y_{\ell m} (\iota, \phi)\ ,
    \end{align}
\end{subequations}
where $D_L$ is the system's luminosity distance from the observer, and $\iota, \phi$ determine its position
in the source's sky.
Finally, we denote the eccentricity of the orbit by \(e\) and the mean anomaly by $\zeta$.
Given that eccentricity, anomaly and spin components evolve with time for a generic precessing, eccentric binary,
we specify them at an initial orbit-averaged reference frequency \(f_{\rm ref}\).

\section{Parametrized effective-one-body model for eccentric, precessing binary black holes}

We build upon the \TEOBResumSDali~\ac{bbh} model, which describes both non-circular
orbits and spin precession~\cite{Nagar:2024oyk, Gamba:2024cvy, Albanesi:2025txj}. 
Before outlining the new features of \TEOBB, we summarize the relevant
aspects of the baseline \dali~model below (see~\cite{Nagar:2024oyk} and references therein for more details).

\subsection{The baseline \dali~model}
\label{sec:dali_base}

\Ac{eob}~\cite{Buonanno:1998gg, Buonanno:2000ef, Damour:2000we, Damour:2001tu} models are built from three main
ingredients: a conservative Hamiltonian governing the orbital dynamics, a waveform model, and radiation reaction
forces that describe dissipative effects in the former due to \ac{gw} emission and horizon absorption.
\dali~also covers spin precession by complementing these with a model for the evolution of the spin
and orbital angular momentum vectors.

The \dali~Hamiltonian for \ac{bbh} systems can be written as:
\begin{subequations}
\begin{align}
H_{\rm EOB}       &= M \sqrt{1 + 2 \nu \bigl(\hat{H}_{\rm eff} - 1\bigr)} \\
\hat{H}_{\rm eff} &= \dfrac{H_{\rm eff}}{\mu} = \hat{H}_{\rm eff}^{\rm orb} + \hat{H}_{\rm eff}^{\rm SO} \\
\hat{H}_{\rm eff}^{\rm orb} &= \sqrt{p_{r_*}^2 + A(r) \left(1 + p_\varphi^2 u_c^2 + Q(r, p_{r_*})\right)} \\
\hat{H}_{\rm eff}^{\rm SO}  &= \left(G_S \hat{\bm{S}} + G_{S_*} \hat{\bm{S}}_*\right) \cdot \hat{\bm{L}} \ \,
\end{align}
\end{subequations}
where we are using dimensionless phase space variables, $r = R/M$, $p_{r_*} = P_{r_*}/\mu$, $p_\varphi =
P_\varphi/(\mu M)$ and $t = T/M$. The radial momentum conjugate to the tortoise coordinate $r_*$ is
$p_{r_*} = p_r \sqrt{A(r)/B(r)}$~\cite{Damour:2014sva}, while $u_c = r_c^{-1}$ is the inverse centrifugal radius
~\cite{Damour:2014sva, Nagar:2024dzj}. $A(r), B(r) = D(r)/A(r)$ are the gravitational potentials,
given by \ac{pn} expansions resummed according to Ref.~\cite{Nagar:2024oyk}, while $Q(r, p_{r_*})$
is taken in Taylor-expanded form, neglecting non-local-in-time terms~\cite{Nagar:2021xnh, Nagar:2024dzj}.
$G_S$ and $G_{S_*}$
are the gyrogravitomagnetic functions that encode the spin-orbit coupling~\cite{Damour:2014sva}. Notably,
two coefficients in the Hamiltonian are not fixed by \ac{pn} theory, but are kept as free parameters
and calibrated by time-domain comparisons with representative samples of \ac{nr} waveforms, as described
in~\cite{Nagar:2024oyk}. These are the effective 5\ac{pn} coefficient $a_6^c$ entering the potential $A(r)$,
fitted as a function of $\nu$, and the next-to-next-to-next-to-leading-order coefficient $c_{\rm{N}^3\rm{LO}}$
found in both $G_S$ and $G_{S_*}$, represented as a function of $\nu, \tilde{a}_0, \tilde{a}_{12}$.

Initial conditions are defined by a starting orbit-averaged $(2,2)$-mode frequency $f_{\rm start}$, and the initial eccentricity $e$ and mean anomaly $\zeta$.
The \dali~definition of eccentricity ties it to the turning points of the radial motion, and $e$ is used along with $f_{\rm start} \simeq \bar{\Omega}/\pi$, where $\bar{\Omega}$ is the average orbital frequency, to compute the initial energy and angular momentum of the system; $\zeta$ then identifies the initial separation and radial momentum.
The dynamical evolution of the system is governed by a modified version of Hamilton's equations,
incorporating the radiation reaction forces:
\begin{subequations}
\begin{align}
    \dot{r} &= \sqrt{\dfrac{A}{B}} \dfrac{\partial H_{\rm EOB}}{\partial p_{r_*}} \qquad 
    \dot{\varphi} = \Omega = \dfrac{\partial H_{\rm EOB}}{\partial p_\varphi} \\
    \dot{p}_{r_*} &= -\sqrt{\dfrac{A}{B}} \biggl(\dfrac{\partial H_{\rm EOB}}{\partial r} + \mathcal{\hat{F}}_r\biggr)\qquad 
    \dot{p}_\varphi = \mathcal{\hat{F}}_\varphi
\end{align}
\end{subequations}
$\mathcal{\hat{F}}_\varphi, \mathcal{\hat{F}}_r$ are dissipative terms that cause the system's
energy and angular momentum to decrease over time. They are computed by invoking the balance
between these dynamical losses by the system and the fluxes of energy and angular momentum
carried by the emitted \ac{gw} signal at infinity and through each \ac{bh}'s horizon~\cite{Iyer:1993xi,
Gopakumar:1997ng,Iyer:1995rn,Alvi:2001mx}.
In \dali, they are given by:
\begin{subequations}
\begin{align}
    \mathcal{\hat{F}}_\varphi &= -\dfrac{32}{5} \nu r_\Omega^4 \Omega^5 \hat{f}(x) + \mathcal{\hat{F}}^{\rm H}_\varphi \\
    \mathcal{\hat{F}}_r       &= -\dfrac{5p_{r_*}}{3p_\varphi} \mathcal{\hat{F}}_\varphi \hat{f}_{p_{r_*}}\, ,
\end{align}
\end{subequations}
where $r_\Omega$ is an effective radius satisfying Kepler's third law on a circular orbit~\cite{Damour:2014sva},
$\mathcal{\hat{F}}_{\rm H}$ is the horizon absorption contribution~\cite{Damour:2014sva}, and $\hat{f}_{p_{r_*}} (r)$ is 
a resummed 2\ac{pn} polynomial relating the radiation reaction components in the quasi-circular limit~\cite{Albertini:2022rfe}.
The factor $\hat{f} (x)$, with $x = (r_\Omega \Omega)^2 \sim \Omega^{2/3}$, inherits the highly
effective factorization and resummation of the waveform multipoles to describe the main dissipative
effect of \acp{gw} on the orbit:
\begin{equation}
    \hat{f} (x) = \sum_{\ell = 2}^8 \sum_{m = -\ell}^{\ell} \dfrac{F_{\ell m}^{\rm Newt}}{F_{22}^{\rm Newt}}
    |\hat{h}_{\ell m}|^2 \hat{f}_{\ell m}^{\rm noncircular}\, .
\end{equation}
Each multipolar contribution is rescaled by the leading $(2,2)$ mode flux prefactor, with $\hat{h}_{\ell m}$
being the \ac{pn} waveform factor as described below. Noncircular effects in the radiation reaction
are included through the leading, Newtonian factor $\hat{f}_{22}^{\rm noncircular}$, as prescribed
in Refs.~\cite{Chiaramello:2020ehz,Nagar:2021xnh} after testing in the test-mass limit. This term
involves high-order time derivatives of $r$ and the orbital frequency $\Omega$, which are computed
directly from the full, resummed \ac{eob} equations of motion, rather than being re-expanded via
their \ac{pn} form; this proves key in attaining good agreement with numerical fluxes~\cite{Albanesi:2021rby,
Albanesi:2022ywx}.

The waveform model implements for each spherical harmonic mode of the \ac{gw} strain the effective
factorization and resummation of \ac{pn} results introduced in~\cite{Damour:2008gu}:
\begin{subequations}
    \begin{align}
        h_{\ell m} &= h_{\ell m}^{\rm insp}\theta(t_{\ell m}^{\rm match} - t) + 
                      h_{\ell m}^{\rm rng} \theta(t - t_{\ell m}^{\rm match})\\
        h_{\ell m}^{\rm insp} &= h_{\ell m}^N \hat{h}_{\ell m} h_{\ell m}^{\rm NQC} 
    \end{align}
\end{subequations}
The signal is split
into an inspiral-plunge-merger part $h_{\ell m}^{\rm insp}$, completed after a suitable time $t_{\ell m}^{\rm match}$ by the post-merger
ringdown portion $h_{\ell m}^{\rm rng}$. The inspiral waveform is computed on the orbital dynamics and
factorized into a leading Newtonian term $h_{\ell m}^N$, a resummed \ac{pn} factor
$\hat{h}_{\ell m}$, and the \ac{nqc} corrections $h_{\ell m}^{\rm NQC}$.
The Newtonian factor $h_{\ell m}^N$, given in its circular-orbit form in~\cite{Damour:2008gu}, also
incorporates the dominant noncircular effects similarly to the angular radiation reaction~\cite{Chiaramello:2020ehz,
Nagar:2021xnh,Albanesi:2021rby}, by employing the full \ac{eob} dynamics to compute the \ac{lo} noncircular
contribution to each multipole.

The \ac{nqc} corrections~\cite{Damour:2007xr,Damour:2009kr} are designed to help smoothly transition from
the late inspiral to the ringdown. The factor reads:
\begin{equation}
    \hat{h}_{\ell m}^{\rm NQC} = (1 + a_1^{\ell m} n_1^{\ell m} + a_2^{\ell m} n_2^{\ell m}) e^{i (b_1^{\ell m} n_3^{\ell m} + b_2^{\ell m} n_4^{\ell m})}\, ,
\end{equation}
where the functions $n_k^{\ell m}$ depend on the radial velocity and acceleration~\cite{Nagar:2019wds,Nagar:2021gss,
Nagar:2024dzj,Nagar:2024oyk}, and the coefficients $a_k^{\ell m}, b_k^{\ell m}$ are determined by imposing a $C^1$
match between the inspiral-plunge-merger and ringdown
waveforms' amplitude $A_{\ell m}$ and frequency $\omega_{\ell m}$ at the matching time $t_{\ell m}^{\rm match}$:
\begin{equation}
    u_{\ell m}^{\rm insp} (t_{\ell m}^{\rm match}) \equiv u_{\ell m}^{\rm rng} (t_{\ell m}^{\rm match})\, ,
\end{equation}
where $u \in \{A, \dot{A}, \omega, \dot{\omega}\}$.

The ringdown signal in \dali~is modeled via the multiplicative decomposition introduced in~\cite{Damour:2014yha}.
The fundamental \ac{qnm} contribution is factored out, and the complex remainder is represented by an analytical
model fitted to \ac{nr} data:
\begin{subequations}
\begin{align}
    h_{\ell m}^{\rm rng}(t) &= e^{-\sigma_{\ell m 0} \bar{t} - i \phi_{\ell m}^0} \bar{h}_{\ell m} (\bar{t}) \\
    \bar{h}_{\ell m} (\bar{t}) &= A_{\bar{h}} (\bar{t}) e^{i \phi_{\bar{h}}(\bar{t})}\, .
\end{align}
\end{subequations}
Here, $\sigma_{\ell m n} = \alpha_{\ell m n} + i \omega_{\ell m n}$ are the \ac{qnm} complex frequencies,
labeled by mode and overtone indices, $\alpha_{\ell m n} = 1/\tau_{\ell m n}$ being the inverse damping time.
$\phi_{\ell m}^0$ is the phase of the waveform mode at its matching time, and $\bar{t} =
(t - t_{\ell m}^{\rm peak})/\mbhf$ is rescaled by the remnant \ac{bh} mass; the latter and the
remnant's spin $\abhf$
\footnote{By $\abhf$ we denote the dimensionless, mass-rescaled spin of the remnant;
explicitly, if $S_{\rm BH}^f$ is the \ac{bh}'s angular momentum, $\abhf = S_{\rm BH}^f/(\mbhf)^2 \in [-1,1]$.}
are determined from accurate fits of \ac{nr} data~\cite{Jimenez-Forteza:2016oae}.
The ansätze for the \ac{qnm}-factorized waveform are the following~\cite{Damour:2014yha}:
\begin{subequations}
\begin{align}
    A_{\bar{h}} (\bar{t}) &= c_1^A \tanh (c_2^A \bar{t} + c_3^A) + c_4^A \\
    \phi_{\bar{h}} (\bar{t}) &= -c_1^\phi \ln \biggl(\dfrac{1 + c_3^\phi e^{-c_2^\phi \bar{t}} + c_4^\phi e^{-2c_2^\phi \bar{t}}}{1 + c_3^\phi + c_4^\phi}\biggr)\, ,
\end{align}
\end{subequations}
where 3 of the 8 coefficients are fitted to \ac{nr} waveforms~\cite{Nagar:2019wds,Nagar:2020pcj}, while the remaining
5 are fixed by as many physically motivated constraints:
\begin{subequations}
    \begin{align}
        A_{\bar{h}}(0) &\equiv A_{\ell m}^{\rm peak} \label{eq:apeak}\\
        \dfrac{dA_{\bar{h}}}{d \bar{t}} \biggl\vert_{\bar{t} = 0} &\equiv 0 \\
        2c_2^A = c_2^\phi &\equiv \alpha_{\ell m 1} - \alpha_{\ell m 0} \label{eq:alpha21}\\
        \dfrac{d \phi_{\bar{h}}}{d \bar{t}} \biggl\vert_{\bar{t} = 0} &\equiv \omega_{\ell m 0} - \mbhf \omega_{\ell m}^{\rm peak}\, . \label{eq:omgpeak}
    \end{align}
\end{subequations}
Eqs.~\eqref{eq:apeak} and~\eqref{eq:omgpeak} above particularly fix the model waveform's amplitude and
frequency at their peak to their \ac{nr}-fitted values $\apeak{\ell m}, \omgpeak{\ell m}$.

Both the location of the \ac{nqc} point $t_{\ell m}^{\rm match}$ and the determination
of the relevant waveform quantities at that time in \TEOBResumSDali~differ between modes.
We have three different default behaviors among the modes for which a post-merger model is implemented,
each choice motivated by comparisons with \ac{nr} data:
\begin{itemize}
\item[(i)] For the $(2,2), (3,2), (4,2)$ and $(4,3)$ modes, the \ac{nqc} point is tied to the peak time of the $(2,2)$
mode amplitude of the EOB waveform, $t_{\ell m}^{\rm NQC} = t_{A_{22}^{\rm peak}}^{\rm EOB} + 2 + \Delta t_{\ell m}$~\cite{Damour:2014sva,Nagar:2019wds}, 
where $t_{A_{22}^{\rm peak}}^{\rm EOB} = t_{\Omega_{\rm orb}^{\rm peak}} - 2 - \Delta t_{\rm NQC}$, $t_{\Omega_{\rm orb}^{\rm peak}}$
is the time when the \textit{pure} orbital frequency\footnote{The pure orbital frequency is defined by the
equation of motion for $\dot{\varphi}$, but omitting the contribution of $\hat{H}_{\rm eff}^{\rm SO}$ to the Hamiltonian~\cite{Damour:2014sva}.}
peaks, $\Delta t_{\rm NQC} = 1$~\cite{Nagar:2023zxh},
and $\Delta t_{\ell m}$ is an \ac{nr}-fitted parameter encoding the delay between the peak of a generic mode
$(\ell, m)$ with respect to the $(2,2)$ mode~\cite{Nagar:2019wds,Nagar:2020pcj}.
The NQC extraction point thus falls in the post-merger part of the waveform, where the ringdown
template is valid; so, the amplitude, frequency and their derivatives are computed by evaluating the
template at the appropriate time.
\item[(ii)] For the $(5,5)$ mode, the \ac{nqc} point is located as defined above, but the \ac{nqc} quantities
are computed through direct \ac{nr} fits~\cite{Nagar:2020pcj}.
\item[(iii)] For the $(2,1), (3,3)$ and $(4,4)$ modes, the NQC point coincides
with the peak of the $(2,2)$ mode~\cite{Nagar:2023zxh}, $t_{\ell m}^{\rm NQC} = t_{A_{22}^{\rm peak}}^{\rm EOB}$;
direct \ac{nr} fits~\cite{Pompili:2023tna} are used in these cases as well.
\end{itemize}

We note that the merger-ringdown model used in \dali~is informed exclusively by \ac{qc} \ac{nr} simulations, under the
assumption that an eccentric binary system has sufficiently circularized via \ac{gw} emission by the time of merger.
\dali~has been validated by comparison with \ac{nr} waveforms up to moderate eccentricities ($e \lesssim 0.4$)~\cite{Nagar:2024oyk,Albanesi:2025txj}.
Additionally, its prescription for the \ac{gw} modes on noncircular orbits has been compared with a few \ac{nr}
simulations of highly eccentric or initially unbound systems in~\cite{Andrade:2023trh}, and, along with
the radiation reaction force, has been extensively studied in the test-mass limit~
\cite{Albanesi:2021rby,Placidi:2021rkh,Albanesi:2022ywx,Albanesi:2023bgi}.
Refs.~\cite{Gamboa:2024hli,Gamboa:2026jht} show comparisons with additional eccentric \ac{nr} simulations from the \ac{sxs} collaboration
that became public after work on this project started.\footnote{Note that~\cite{Gamboa:2024hli,Gamboa:2026jht} use an earlier version of~\dali
~\cite{Nagar:2024dzj} that predates the improvements shown in App.~C of~\cite{Nagar:2024oyk}.}

\dali~employs the twist method~\cite{Schmidt:2010it,OShaughnessy:2011pmr,Schmidt:2012rh,Pekowsky:2013ska}
to also describe systems undergoing spin precession, following the approach of~\cite{Gamba:2024cvy} to incorporate
the main effects of orbital eccentricity in the evolution of the precessing angular momentum vectors. In addition, as mentioned
in~\cite{Gamba:2025qfg}, we model residual post-merger precession effects in the waveform according to~\cite{OShaughnessy:2012iol,Ossokine:2020kjp,Gamba:2021ydi,Hamilton:2023znn}.

\subsection{The \TEOBB~model}
\label{sec:pteob}

\begin{table*}[t]
    \newcolumntype{C}{>{\centering\arraybackslash}X}
    \newcolumntype{L}{>{\raggedright\arraybackslash}X}
    \renewcommand{\arraystretch}{1.25}
    \centering
    \caption{Summary of the deviation parameters implemented in \TEOBB, including for each
    limits imposed by physical considerations (merger-ringdown) or sane model performance (inspiral). The bounds for the inspiral parameters
    are highly dependent on the system's mass ratio and spins; the ones reported here are approximate, conservative estimates.
    Visual inspection of the waveforms is advised to evaluate appropriate ranges for any application.
    \label{tab:dev_pars}}
    \begin{tabularx}{0.85\textwidth}{l c C c}
        \toprule
         & \textbf{Parameter} & \centering{\textbf{Description}} & \textbf{Limits} \\
        \midrule
        \multirow{2}{2.5cm}{\centering \textbf{Inspiral}} 
        & $\delta a_6^c$   & Effective 5PN coefficient in $A$ potential        & $-20 \lesssim \delta a_6^c \lesssim 5$ \\
        & $\delta \cthree$ & $\rm{N}^3\rm{LO}$ spin-orbit coupling coefficient & $\cthree \gtrsim -10$ \\
        \midrule
        \multirow{7}{2.5cm}{\centering \textbf{Merger-Ringdown}} 
        & $\delta \mbhf$            & Remnant mass                                        & $\mbhf (1 + \delta \mbhf) > 0$ \\
        & $\delta \abhf$            & Remnant spin                                        & $|a_{\rm BH}^f (1 + \delta \abhf)| \leq 1$ \\
        & $\delta \taulm0$          & Damping time of fundamental $(\ell, m)$ QNM         & $\delta \taulm0 > -1$    \\
        & $\delta \alphalm0$        & Inverse damping time of fundamental $(\ell, m)$ QNM & $\delta \alphalm0 > -1$  \\
        & $\delta \omegalm0$        & Frequency of fundamental $(\ell, m)$ QNM            & $\delta \omegalm0 > -1$  \\
        & $\delta \apeak{\ell m}$   & Amplitude at peak of mode $(\ell, m)$               & $\delta \apeak{\ell m} > -1$  \\
        & $\delta \omgpeak{\ell m}$ & Frequency at peak of mode $(\ell, m)$               & $\delta \omgpeak{\ell m} > -1$  \\
        \bottomrule
    \end{tabularx}
\end{table*}

\begin{figure}
    \includegraphics[width=0.47\textwidth]{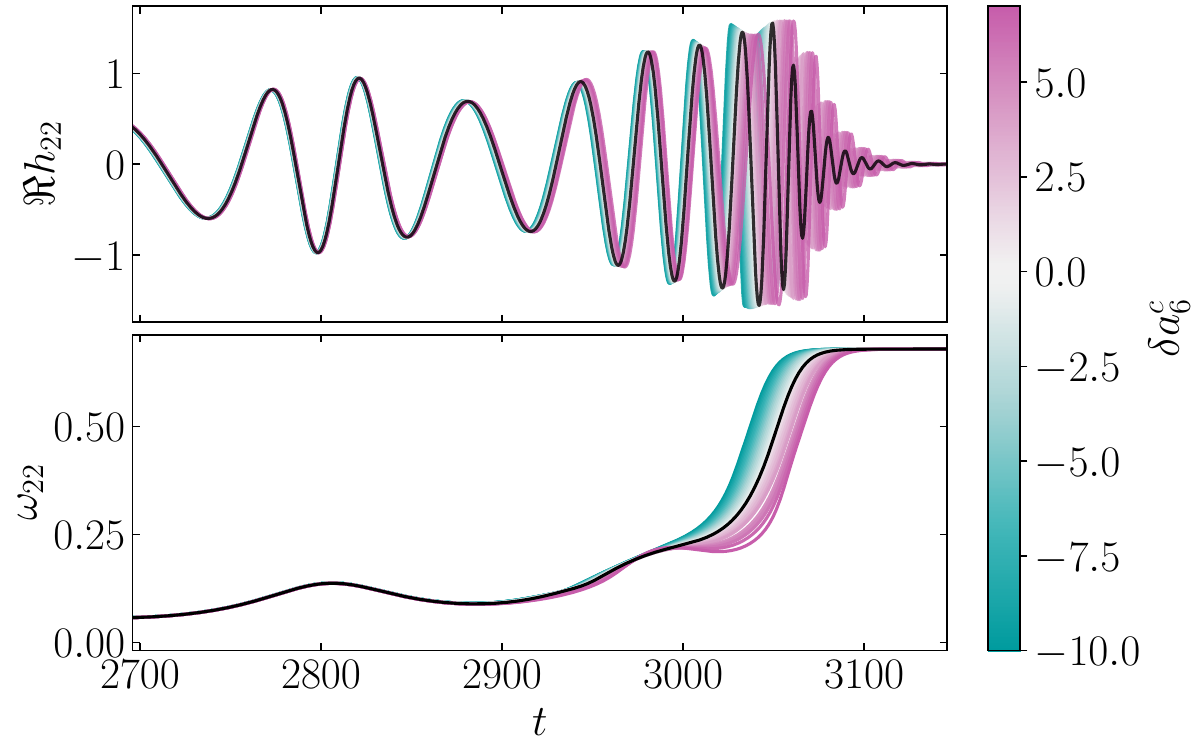}
    \caption{Effect of the variation of $a_6^{\rm c}$ on the real part (top) and frequency (bottom) of $h_{22}$
    for a binary with $q = 1, \chi_1 = \chi_2 = 0.6$, and initial eccentricity $e_0 = 0.5$ at a reference frequency
    of $20$ Hz for a total mass of $40 M_\odot$. Waveforms here are aligned to begin at $t = 0$ to
    highlight the cumulative effect of the deformed dynamics. $\delta a_6^c$ causes a
    delayed ($\delta a_6^c > 0$) or accelerated ($\delta a_6^c < 0$) late inspiral and plunge. The model is
    sensitive to the increase of $a_6^c$, delivering unphysical results for deviations $\gtrsim 8$ in this case.
    \label{fig:a6_ecc}}
\end{figure}

\begin{figure*}
    \includegraphics[width=0.47\textwidth]{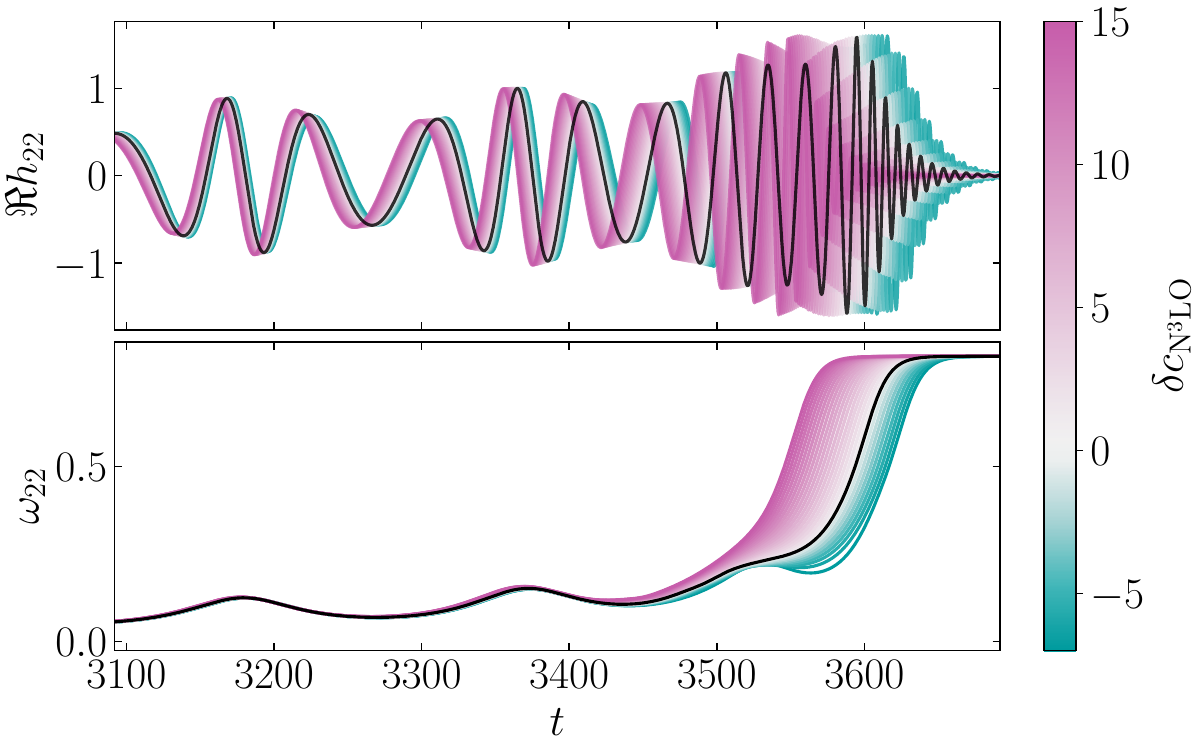}
    \includegraphics[width=0.47\textwidth]{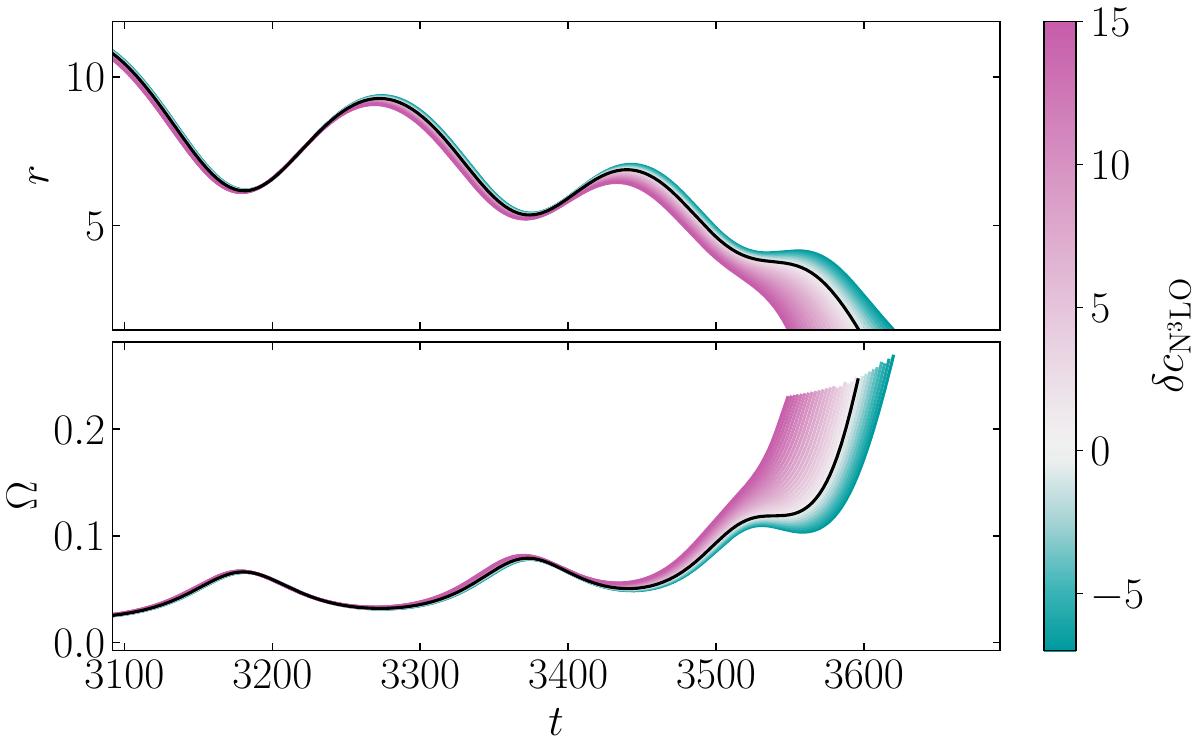}
    \caption{Changing $\delta \cthree$ for a moderately eccentric ($e_0 = 0.5$ at an orbit-averaged frequency of 20 Hz with
    total mass $M = 40 M_\odot$) binary with $q = 1$ and $\chi_1 = \chi_2 = 0.95$. \textit{Left:} real part and
    instantaneous frequency of the $(2,2)$ mode; \textit{right:} dynamics (orbital separation $r$ and
    frequency $\Omega$). Waveforms are aligned to start at $t=0$. 
    The high spins enhance the effect of the deviation in $\cthree$. A negative value of $\delta \cthree$ delays
    the merger for positive spins; the opposite would happen if they were anti-aligned with the orbital angular
    momentum.
    Within the displayed range the effective potential warped by the
    changing $\cthree$ can induce an additional, smaller orbit after what would have been the onset of the plunge.
    \label{fig:c3_ecc}}
\end{figure*}
\begin{figure}
    \includegraphics[width=0.47\textwidth]{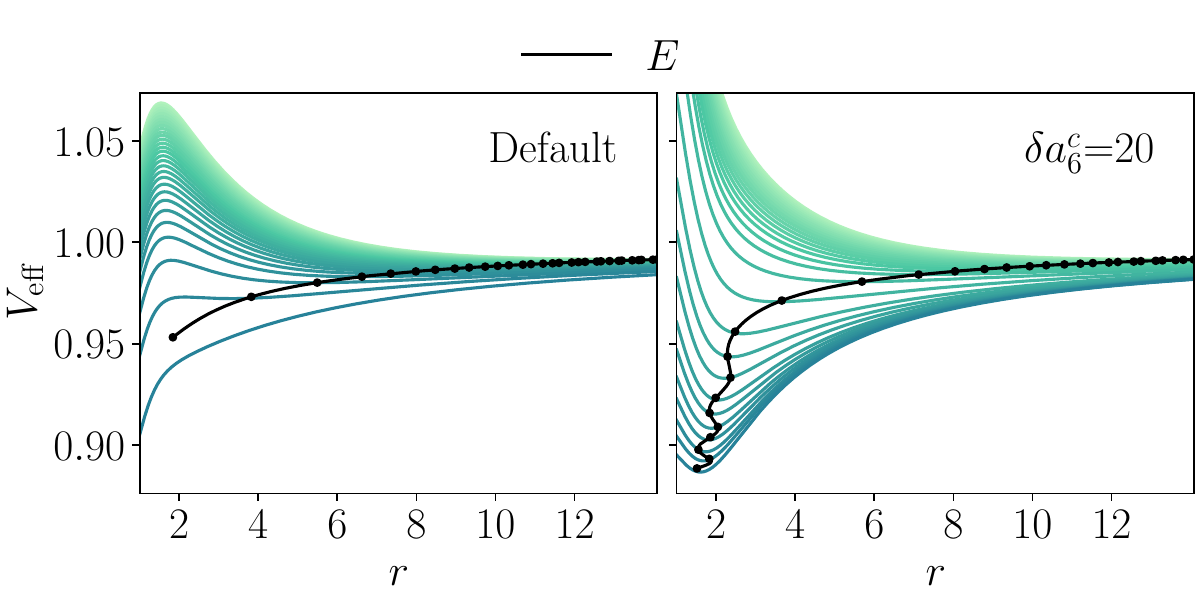}
    \caption{\label{fig:a6_eff_pot}Effective potential $V_{\rm eff} = H_{\rm EOB} (r, p_\varphi, p_{r_*} = 0)/M$ throughout the inspiral of an
        equal-mass \ac{bbh} system with $\chi_1 = \chi_2 = 0.6$; these are the same parameters used in Fig.~\ref{fig:a6_ecc}, save
        for the eccentricity, set here to 0 for easier visualization. The black line is the evolving energy
        of the system, $E = H_{\rm EOB}(t)/M$, the dots corresponding to the times of the potential plots. \textit{Left:} default \dali~model; 
        \textit{Right:} \TEOBB~with a large, positive deviation to the fitted $a_6^c$.}
\end{figure}

The parametrized model is constructed by allowing deviations from a few of the \ac{nr}-informed quantities
that enter the dynamics and waveform template. Below, we discuss in detail their impact
on the waveform morphology. A summary is provided in Table~\ref{tab:dev_pars}.

Concerning the dynamical sector and the inspiral, we consider deviations from the
\ac{nr}-calibrated parameters $a_6^c$ and $c_{\rm{N}^3\rm{LO}}$.
This parallels the approach introduced in~\citet{Pompili:2025cdc} for the equivalent \pseob~coefficients.
We use in this case additive deviations from the fitted values:
\begin{subequations}
\begin{align}
    a_6^c &\rightarrow a_6^c + \delta a_6^c \\
    \cthree &\rightarrow \cthree + \delta \cthree \, .
\end{align}
\end{subequations}
Figs.~\ref{fig:a6_ecc} and~\ref{fig:c3_ecc} display the effect of these deviations on the waveform for sample equal-mass,
moderately eccentric spinning systems. Most of the changes accumulate during the late stages of the inspiral, especially in the
case of $\delta a_6^c$, leading to an accelerated or delayed plunge. The impact of these parameters is also modulated
by the system's intrinsic properties: both $a_6^c$ and $\cthree$ appear in their respective functions weighted by a $\nu$
factor, so their role diminishes moving to higher mass ratios. In addition, as $\cthree$ enters the spin-orbit coupling
term in the Hamiltonian, its influence on the dynamics is enhanced for fast spinning systems, such as the one shown
in Fig.~\ref{fig:c3_ecc}. Highly eccentric orbits can also showcase more pronounced differences accumulating throughout
the inspiral due to the repeated close periastron passages, particularly with high spins.

Fine tuning of the inspiral deviation parameters can sometimes markedly alter the waveform morphology around plunge
in the eccentric case, as shown in Figs.~\ref{fig:a6_ecc} and~\ref{fig:c3_ecc}. This happens when a 
periastron passage transforms into a plunge or, vice versa, a missed periastron becomes a turning point. 
This effect adds or removes a peak in the instantaneous frequency of the waveform just before merger, a feature that could
increase sensitivity to these deviations in \ac{pe}.

The model can be quite sensitive to deviations from the fitted values, being particularly prone to delivering unphysical
waveforms when $a_6^c$ is increased.
This is a consequence of the form of the $A$ potential implemented in \dali, and specifically its resummation scheme.
Details in the resulting function's strong field behavior can markedly affect the character of the late dynamics, on which the attachment of the
post-merger waveform and its smooth connection to the inspiral are reliant.
The same can be said for the gyrogravitomagnetic functions $G_S, G_{S_*}$ in the case of $\cthree$.
The threshold in $\delta a_6^c$ for the onset of these issues depends on the mass ratio and spins.
The equal-mass case is generally the most stable, with the approximate upper limit varying between $\sim 6$ for high, positive (aligned) spins,
$\sim 9$ for $\chi_{\rm eff} = 0$, and $\sim 15$ for high, negative spins.
Near the extreme spins ($|\chi_{\rm eff}| \geq 0.8$), these values change only slightly with $q$.
Conversely, moderate mass ratios ($3 \lesssim q \lesssim 10$) and effective spins ($|\chi_{\rm eff}| \lesssim 0.7$) prove to be the most
problematic region of the parameter space: the upper bound decreases to $\sim 4-5$ for $q = 3$, and even further to $\lesssim 1$ for $5 < q < 8$,
slightly growing again for more asymmetric systems.
The upper limit of the range used in Fig.~\ref{fig:a6_ecc} was chosen as high as possible while avoiding problematic behavior.
We find that a lower bound of $\delta a_6^c \geq -20$ is safe across the parameter space.
As for $\delta \cthree$, large, positive deviations are supported for all spins and mass ratios ($\delta \cthree \lesssim 50$).
Negative values can be problematic in the case of positive $\chi_{\rm eff}$, where we advise as a conservative lower bound $\delta \cthree \gtrsim -10$.

Wider ranges for the inspiral parameters would in general be supported by the dynamical model, but as the deviations warp the effective potential,
that eventually breaks down as well and unphysical behavior occurs in the strong field.
Fig.~\ref{fig:a6_eff_pot} exemplifies this by contrasting the form of the evolving effective potential $V_{\rm eff} = H_{\rm EOB} (r, p_\varphi, p_{r_*} = 0)$
during the quasi-circular inspiral of an equal-mass system with $\chi_1 = \chi_2 = 0.6$ according to the either the baseline \dali~model or \TEOBB~with a large,
positive value of $\delta a_6^c$.
While in the former case the system proceeds to plunge as the potential minimum vanishes, in the latter it is
trapped in a series of radial oscillations at small separation.

\begin{figure}
    \includegraphics[width=0.47\textwidth]{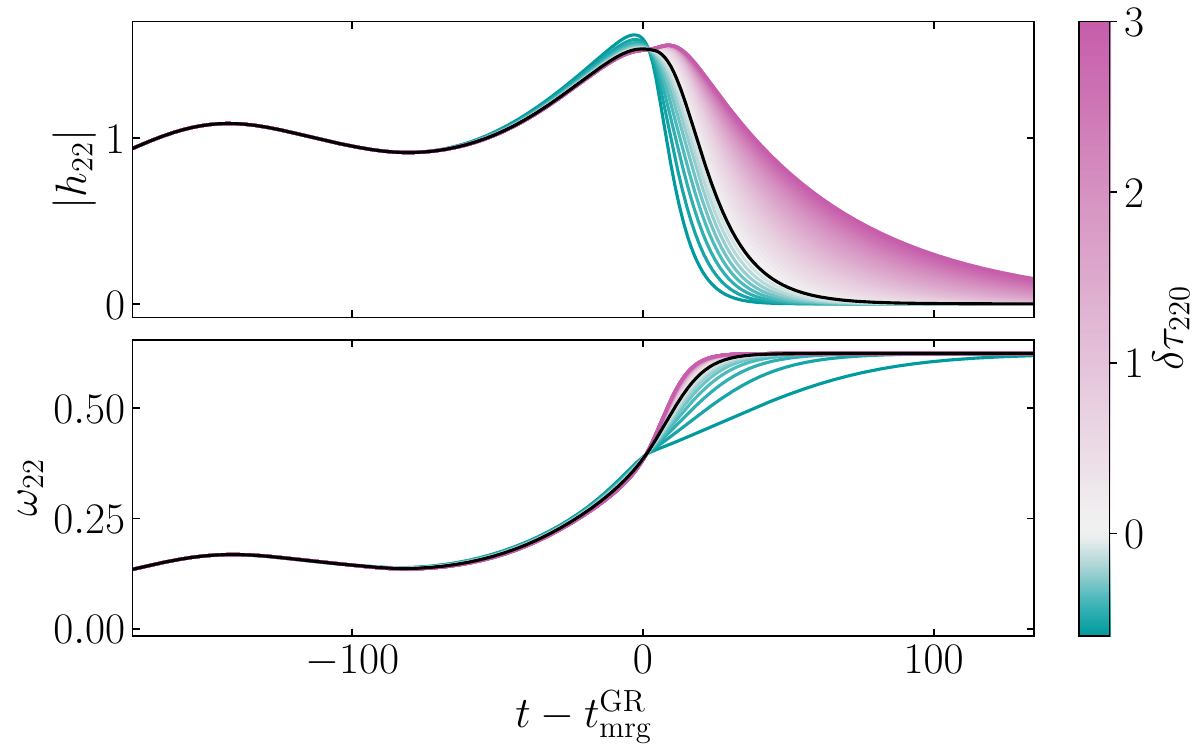}
    \caption{Effect of the variation of $\tau_{220}$ on the $(2,2)$ mode amplitude and frequency
    for a binary with $q = 1, \chi_1 = \chi_2 = 0.6$ and initial eccentricity $e_0 = 0.5$ at a reference frequency of
    $20$ Hz for a total mass of $40 M_\odot$. $t = 0$ corresponds to the peak of $|h_{22}|$ for the zero-deviation waveform.
    A decrease in the damping time leads to a faster
    decaying ringdown, and vice-versa; this also affects the frequency evolution due to the structure of the
    ringdown model.
    \label{fig:qnmdev_tau}}
\end{figure}

\begin{figure}
\includegraphics[width=0.47\textwidth]{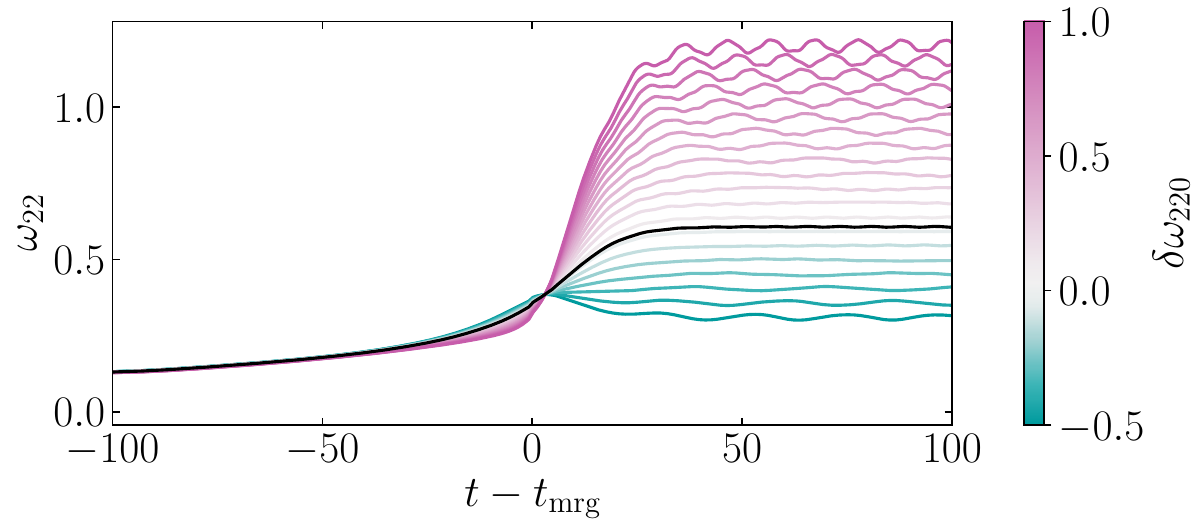}
\caption{Effect of the variation of $\omega_{220}$ on the $(2,2)$ mode frequency
    for an eccentric, precessing binary system with $q = 3, \bm{\chi_1} = (0.3, -0.2, 0.45), \bm{\chi_2} = (-0.5, -0.1, 0.2)$,
    and initial eccentricity $e_0 = 0.2$, measured at a reference frequency of $20$ Hz for a total mass of $40 M_\odot$. 
    $t = 0$ corresponds to the peak of $|h_{22}|$ for the zero-deviation waveform.
    The changing \ac{qnm} frequency can also be appreciated in the oscillations of $\omega_{22}$ during the ringdown, which are due to the
    mixing of the co-precessing $\ell = 2$ modes in the inertial frame.
    \label{fig:qnmdev_omg}}
\end{figure}

\begin{figure*}
    \includegraphics[width=0.47\textwidth]{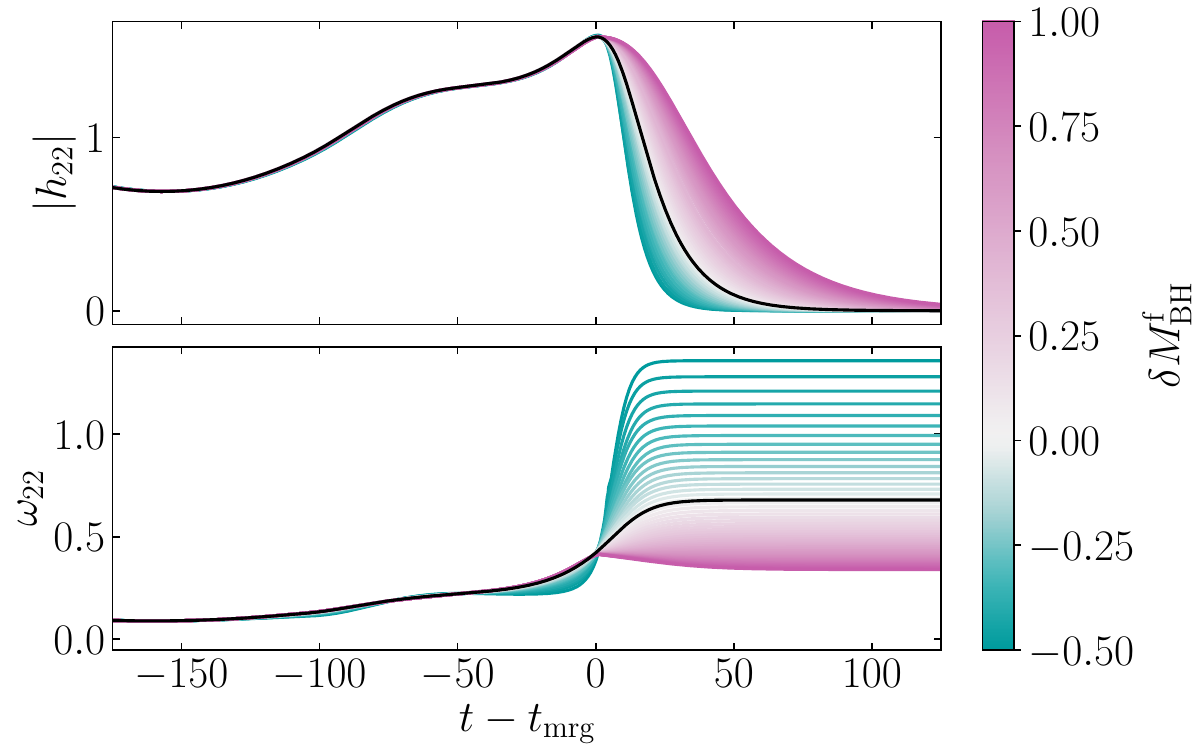}
    \includegraphics[width=0.47\textwidth]{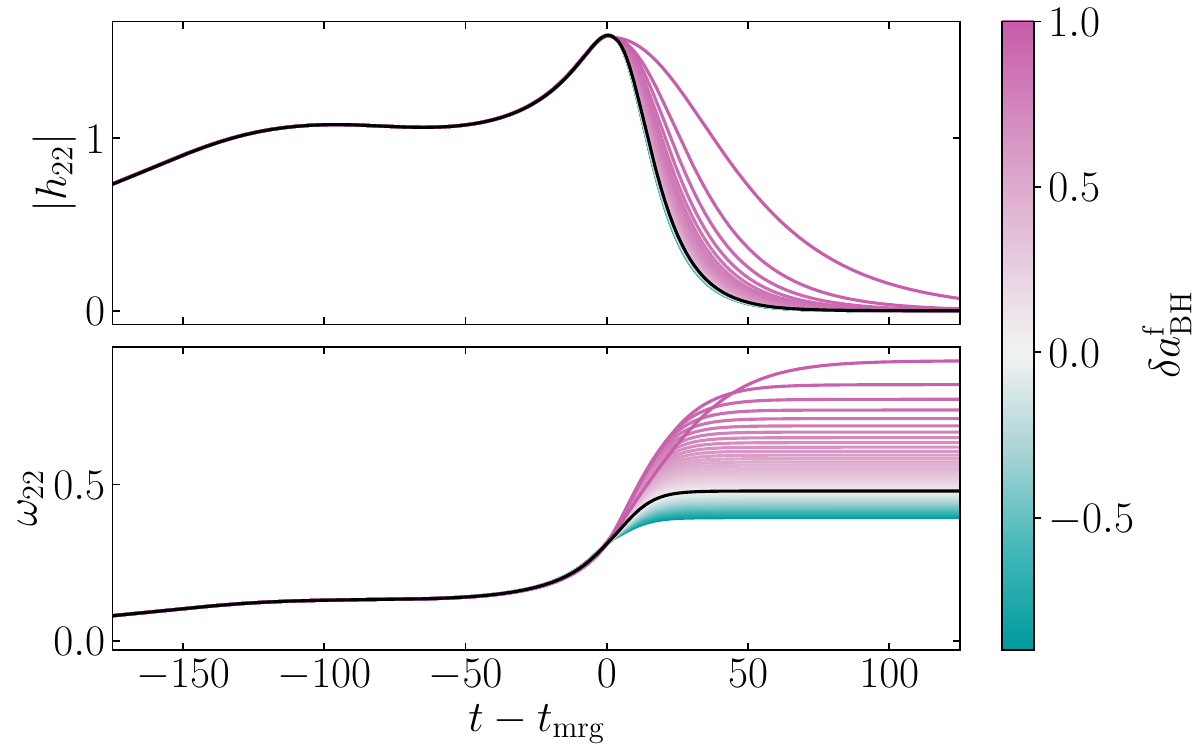}
    \caption{Effect of the variation of $\mbhf$ (\textit{left}) and $\abhf$ (\textit{right}) on the $(2,2)$ mode amplitude and frequency
    for a binary with $q = 1, \chi_1 = \chi_2 = 0.6$, and initial eccentricity $e_0 = 0.5$
    at a reference frequency of $20$ Hz for a total mass of $40 M_\odot$.
    Here $t = 0$ corresponds to the peak of $|h_{22}|$ for the zero-deviation waveform.
    By propagating to the \acp{qnm}, changes to the remnant properties can be degenerate with $\delta \omega_{\ell m 0}$ and $\delta \tau_{\ell m 0}$.
    \label{fig:remnantdev}}
\end{figure*}

\begin{figure}
    \includegraphics[width=0.47\textwidth]{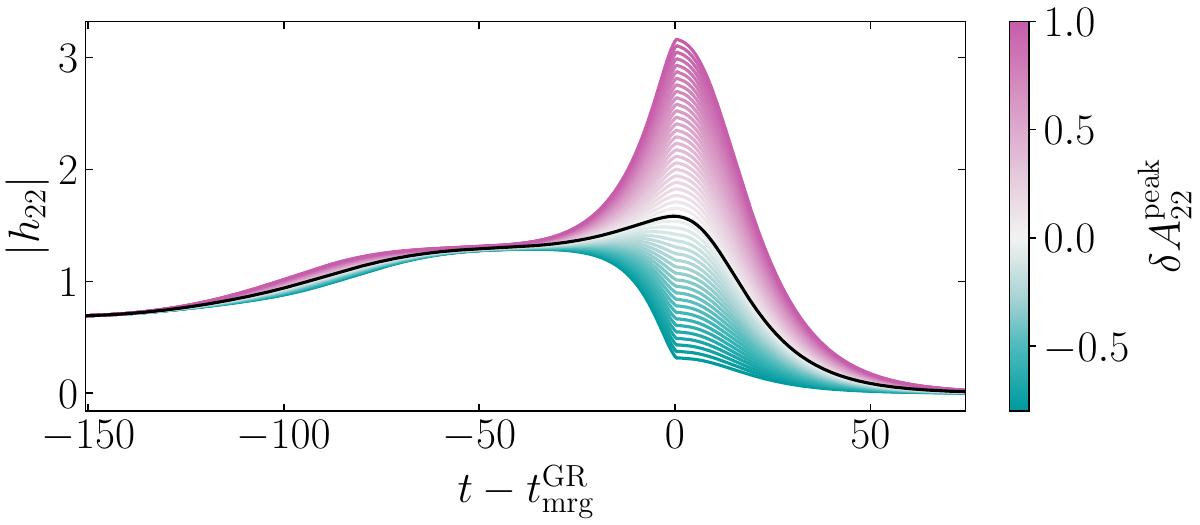}
    \includegraphics[width=0.47\textwidth]{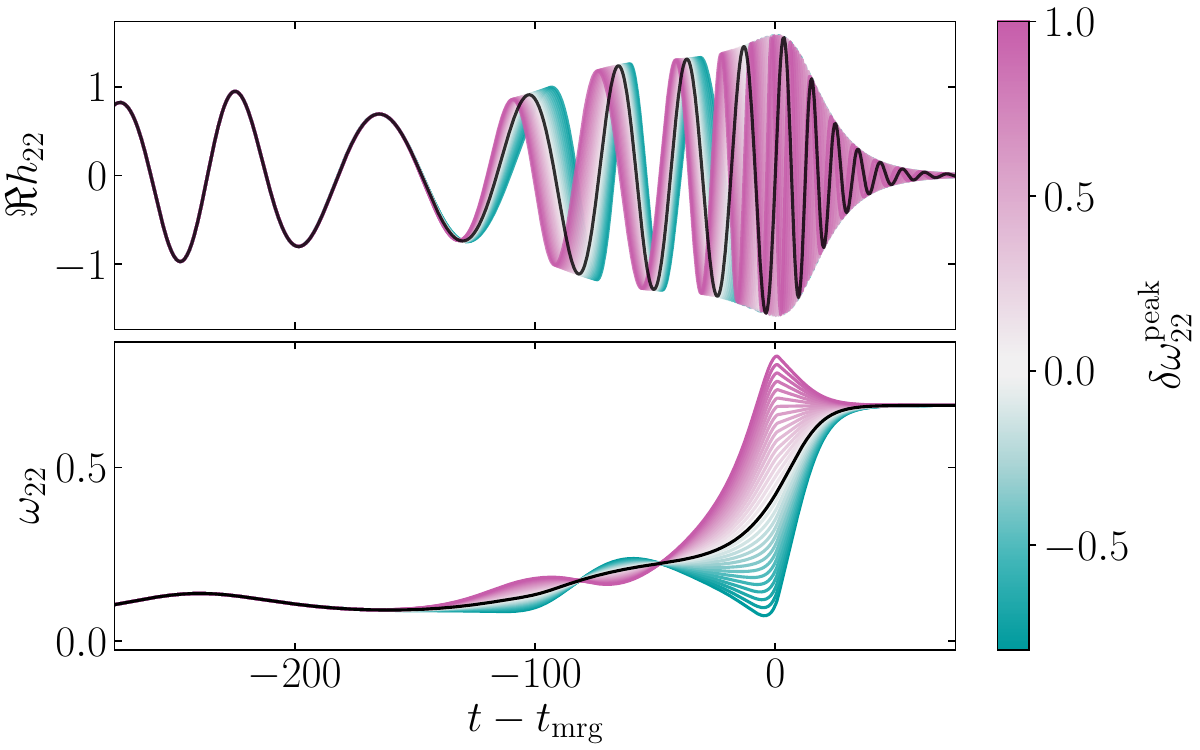}
    \caption{Variation of $A_{22}^{\rm peak}$ (\textit{top}) and $\omega_{22}^{\rm peak}$ (\textit{bottom}) for a binary with $q = 1, \chi_1 = \chi_2 = 0.6$, and initial eccentricity $e_0 = 0.5$ at a reference frequency
    of $20$ Hz for a total mass of $40 M_\odot$. Here $t = 0$ corresponds to the peak of $|h_{22}|$ for the zero-deviation waveform. The peak deviations propagate backward to the plunge and late inspiral via the \ac{nqc} corrections, and can greatly deform the morphology of the signal around merger.
    \label{fig:peakdev_amp}}
\end{figure}

Moving on to the waveform model, the deviations we allow are the following:
\begin{itemize}
    \item[(a)]Fractional deviations from the \ac{nr}-fitted mass and spin of the final remnant \ac{bh}:
    \begin{subequations}
    \begin{align}
        \mbhf &\rightarrow \mbhf (1 + \delta \mbhf) \\
        \abhf &\rightarrow \abhf (1 + \delta \abhf) \, .
    \end{align}
    \end{subequations}
    The modified parameter $\abhf$ is constrained to satisfy
    $|\abhf| < 1$, to avoid extremal and over-extremal \acp{bh}.
    Fig.~\ref{fig:remnantdev} shows the effect of altering $\mbhf, \abhf$.
    \item[(b)]Deviations from the fundamental QNM parameters $\sigma_{\ell m 0}$ of each
    mode. The default values of the $\sigma_{\ell m 0}$ are fits of numerically computed \acp{qnm},
    described in~\cite{Nagar:2019wds}. 
    For any mode, we separately allow fractional deformations of $\omegalm0$ and either $\tau_{\ell m 0}$
    or its inverse $\alpha_{\ell m 0}$:
    \begin{subequations}
    \begin{align}
    \alphalm0 &\rightarrow \alphalm0 (1 + \delta \alphalm0) \\
    \taulm0   &\rightarrow \taulm0   (1 + \delta \taulm0) \\
    \omegalm0 &\rightarrow \omegalm0 (1 + \delta \omegalm0)
    \end{align}
    \end{subequations}
    We avoid exponentially growing post-merger amplitudes by requiring that the (inverse) damping time
    deviations satisfy $\delta \taulm0 > -1\  (\delta \alphalm0 > -1)$. Any modification of the damping time
    of the fundamental QNM is consistently carried over to the difference $\alpha_{\ell m 1} - \alphalm0$ 
    that enters the phenomenological ringdown model~\cite{Damour:2014sva, Nagar:2019wds, Nagar:2020pcj}
    (see Eq.~\eqref{eq:alpha21}). Figs.~\ref{fig:qnmdev_tau} and~\ref{fig:qnmdev_omg} showcase the effect
    these deviations have on the waveform, respectively for an eccentric, spin-aligned and an eccentric,
    precessing system.
    This is identical to the approach currently used by \ac{lvk} to analyze non-eccentric \ac{bbh} signals within the \texttt{pSEOBNR} framework~\citep{Brito:2018rfr, Ghosh:2021mrv, Maggio:2022hre, Pompili:2023tna}.
    \item[(c)]For each mode, we consider deviations from \ac{nr}-fitted values of the amplitude and frequency of the
    waveform at its peak. If a waveform mode is decomposed
    as $h_{\ell m} = A_{\ell m} e^{-i \phi_{\ell m}}$ and $\omega_{\ell m} = d\phi_{\ell m}/dt$,
    \begin{subequations}
    \begin{align}
    A_{\ell m}^{\rm peak} &\rightarrow A_{\ell m}^{\rm peak} (1 + \delta A_{\ell m}^{\rm peak}) \ , \\
    \omega_{\ell m}^{\rm peak} &\rightarrow \omega_{\ell m}^{\rm peak} (1 + \delta \omega_{\ell m}^{\rm peak}) \ .
    \end{align}
    \end{subequations}
    See Fig.~\ref{fig:peakdev_amp} for the outcome of these deviations on a moderately
    eccentric system. Increasing $A_{\ell m}^{\rm peak}$ leaves the location of the maximum
    itself largely unchanged. A decrease instead tends to delay it, in addition to inducing a significantly different
    morphology in the signal, such as a much wider peak, or plateaus in the subsequent amplitude decay.
    Similarly drastic changes occur in the evolution of the frequency when using non-zero $\delta \omgpeak{\ell m}$.
\end{itemize}

The deviations in the \ac{qnm} spectrum and the remnant properties are degenerate, as the
latter will also affect the fundamental mode frequencies used for the ringdown waveform, for every multipole. 
Conceptually, however, they represent different kinds of probes into the physics of \ac{bbh} mergers.
By only varying the \ac{qnm} frequencies, one assumes the coalescence to result in a non-Kerr remnant.
The variation of the remnant properties themselves, instead, posits that the final object is a Kerr \ac{bh},
but that the emission of energy and momentum during the merger process deviates from the \ac{gr} expectation,
informed by \ac{nr} simulations, resulting in a remnant with different characteristics.

Following the discussion at the end of Sec.~\ref{sec:dali_base}, utilizing the peak amplitude and frequency
deviations requires special care to avoid undesireable features in the waveforms. For all modes listed under point (i) above,
applying either $\delta \apeak{\ell m}$ or $\delta \omgpeak{\ell m}$ will shift the post-merger model, and correspondingly induce
changes in the \ac{nqc} time quantities, such that the \ac{nqc} corrections will always attempt to enforce a smooth
connection between the late inspiral and the ringdown (this effect can be appreciated in Fig.~\ref{fig:peakdev_amp}).
This is not true for the multipoles under points (ii) and (iii). 
In these cases, the plunge-merger waveform is blind to the ringdown's deformation, and discontinuities 
will appear in the complete signals under the model's default behavior. 
When using peak amplitude and/or merger deviations, the affected modes
should be added to the list of those for which the \ac{nqc} quantities are evaluated
using the post-merger template, which is an optional user input in the model.\footnote{This can be done by changing the \texttt{kpostpeak} parameter passed to the \dali~Python wrapper.}

\section{Analysis Setup}
\label{sec:pe}

To assess \TEOBB's ability to identify beyond-\ac{gr} signals and to search for potential deviations from \ac{gr} in observed \ac{gw} events, we perform Bayesian inference using the \bilby\ library~\cite{Ashton:2018jfp}. Bayesian parameter estimation aims to estimate the posterior distribution
\begin{equation}
    p(\bm{\theta} | d, H) = \frac{\pi(\bm{\theta})\, \mathcal{L}(d | \bm{\theta}, H)}{\mathcal{Z}_H} \, ,
\end{equation}
for model parameters $\bm{\theta}$ given data $d$ under model hypothesis $H$, where $\pi(\bm{\theta})$ is the prior, $\mathcal{L}$ the likelihood, and $\mathcal{Z}_H=\int d\bm{\theta} \pi(\bm{\theta})\mathcal{L}(d | \bm{\theta}, H)$ the evidence. The latter can be considered as the average of the likelihood function over the prior space.

Assuming wide-sense stationary, Gaussian detector noise with a one-sided power spectral density $S_d(f)$, the log of the likelihood takes the standard form
\begin{equation}
    \ln \mathcal{L}(d | \bm{\theta}, H) \propto -\frac{1}{2} \sum_{k} \frac{2\,|\tilde{d}_k - \tilde{h}_k(\bm{\theta} \mid H)|^2}{T\, S_d(f_k)} \, ,
\end{equation}
where $\tilde{h}_k(\bm{\theta} \mid H)$ is the fourier domain model waveform in frequency bin $f_k$, $T$ is the duration of the analyzed data segment, and $\tilde{d}_k$ is the Fourier transform of the detector strain.

Model selection between two competing hypotheses $H_1$ and $H_2$ is performed using the \ac{bf},
\begin{equation}
    \log_{10} \mathcal{B}^1_2 = \log_{10} \mathcal{Z}_1 - \log_{10} \mathcal{Z}_2 \, ,
\end{equation}
where $\mathcal{Z}_{1,2}$ are the corresponding evidence values, with $\log_{10} \mathcal{B}^1_2 > 1$ often interpreted as strong preference for $H_1$~\cite{Kass01061995}. 

When sampling explicitly on deviation parameters $\bm{\delta}$ in \TEOBB, the \ac{bf} in favor of the \ac{gr} (null) hypothesis can be estimated via the Savage--Dickey ratio~\cite{Dickey:1970}:
\begin{equation}
    \mathcal{B}_{\mathrm{GR}} = \frac{p(\bm{\delta} = 0 | d, \texttt{pTEOB})}{\pi(\bm{\delta} = 0 | \texttt{pTEOB})} \, , \label{eq:sdr_bf}
\end{equation}
where the numerator and denominator are the marginalized posterior and prior probabilities, respectively, of $\bm{\delta}$ evaluated at the \ac{gr} value $\bm{\delta} = 0$. The Savage–Dickey ratio applies here because the \ac{gr} hypothesis corresponds to a fixed point \(\bm{\delta}=0\) within the parameter space of the more general \TEOBB~model, allowing the null \ac{bf} to be obtained directly from the ratio of posterior to prior density at that point.

For our simulation studies, we use a three-detector network consisting of Advanced LIGO and Virgo operating at design sensitivity. 
When injecting a synthetic signal, i.e., projecting it onto each detector with the appropriate response function and time delay, we do not add Gaussian noise to the simulated strain, so that the analysis is performed on zero-noise data.
Under the assumption of wide-sense stationary Gaussian detector noise, the resulting posterior distributions correspond to those obtained by averaging over many independent noise realizations. Further details on the simulated signals are summarized in Sec.~\ref{sec:validate}.

As for the \ac{bbh} events, we analyze, depending on the specific case, either 4 or 8 s of publicly available data, down-sampled to \(1024\) Hz, using the associated power spectrum and calibration envelopes.
We employ the nested sampling algorithm \texttt{dynesty} with the \texttt{rwalk} proposal, a minimum of $100$ and maximum of $5000$ MCMC steps, and $\texttt{nact}=50$~\citep{Speagle:2019ivv}.
The number of live points is set to $1000$, $1500$, or $2048$ for simulations, and to $1500$ for the \ac{bbh} event analyses.
The minimum frequency for the likelihood integration is always set to $f_{\rm low} = 20\, {\rm Hz}$.
The starting (orbit-averaged) $(2,2)$ mode frequency for waveform generation is $f_{\rm start} = 20\, {\rm Hz}$ for synthetic signals only including the quadrupolar $(2,\pm 2)$ modes, and $f_{\rm start} = 13.33\, {\rm Hz}$ otherwise; exceptions are explicitly stated when presenting results.

We adopt standard prior distributions for the binary parameters, as listed below.
Following~\citet{Romero-Shaw:2020owr}, we sample in chirp mass $\mathcal{M}_c$ and mass ratio $q$ space, enforcing $1/q \in [0.05, 1]$ and ensuring a prior uniform in component masses. 
For the dimensionless spin vectors $\bm{\chi}_{i}$, precessing analyses use priors uniform in magnitude and isotropic in orientation. For spin-aligned runs, we instead use uniform priors on the aligned components $|\bm{\chi}_{i} \cdot \bm{L}| \in [0, 0.99]$.  
Eccentricity parameters are sampled uniformly with $e \in [0, 0.5]$ and mean anomaly $\zeta \in [0, 2\pi]$, both defined at the waveform’s starting frequency, as are the spin angles.
The chosen range in $e$ mostly reflects \dali's range of validity, as defined by comparisons with \ac{nr} waveforms, only extending to higher $e$ to fully
encompass the broader posterior distributions of the less loud signals; the model delivers fully sensible waveforms in the entire range.
For the luminosity distance we use a uniform prior in the source frame, with \(\Lambda\)CDM cosmology with parameters \(H_0 = 67.9 \, \mathrm{km} \, \mathrm{s}^{-1} \, \mathrm{Mpc}^{-1}\) and \(\Omega_m = 0.3065\), as in \citet{KAGRA:2021vkt}, and $D_L$ constrained between 100 Mpc and 10 Gpc. 
We use isotropic priors for the sky location angles and binary orientation parameters. 
For the \ac{qnm} deviation parameters $\delta\omega_{220}$ and $\delta\tau_{220}$, we adopt uniform priors in $[-0.8, 2.0]$, expanding the interval when necessary to avoid posterior railing.
Priors for other deviation parameters are specified when used in analyses.

We can combine results from multiple events to obtain stronger constraints on the deviation parameters;
following~\cite{LIGOScientific:2020tif,LIGOScientific:2021sio,Pompili:2025cdc,LIGOScientific:2026qni}, we adopt two methods to do so.
(i) Assuming that the deviations' true values are common to all the combined events, we derive a joint posterior distribution
by multiplying the individual likelihoods; using uniform priors, this is equivalent to computing the normalized product of
the single event posteriors:
\begin{equation}
    p (\bm{\delta} | \{d_k\}, H) \propto \prod_{k = 1}^{N_{\rm ev.}} p (\bm{\delta} | d_k, H)\, .
\end{equation}
While suited to the hypothesis of \ac{gr} being correct, the underlying assumption of constant deviations is a strong one,
incompatible with most alternative theories of gravity, where differences from \ac{gr} depend on the \acp{bh}' properties.
(ii) We can relax this assumption by employing a hierarchical procedure~\cite{Zimmerman:2019wzo,Isi:2019asy}; we adopt
the approach of~\cite{Zhong:2024pwb}, which was first applied by the \ac{lvk} in~\cite{LIGOScientific:2026qni}.
We posit that the deviation parameters $\delta_{i}$ are drawn from an underlying population;
we approximate their distribution via a multivariate Gaussian, characterized by a set of \emph{hyperparameters}: the means $\mu_i$, variances $\sigma_i$, and correlations $\rho_{ij}$.
If \ac{gr} were correct, a perfect measurement would yield a posterior with zero means and zero width, $\mu_i = \sigma_i = 0$;
a departure from this in the hyperparameter posteriors would signal evidence for a potential deviation in the data.
After estimating the hyperparameters from our individual event results, we can marginalize over them to generate samples for the
deviations themselves.
These can give us a clearer idea of their range of variability as inferred from the available data, although care should be
taken when interpreting them as they can be misleading:
even when the mean and variance of the underlying Gaussian point to a deviation from \ac{gr}, the population-marginalized 
distributions can still include the null hypothesis at higher credibility.

\section{Application to data analysis}
\label{sec:results}

We begin by validating \TEOBB\ against simulated \ac{bbh} signals in zero noise generated with the model itself or via \ac{nr} simulations, 
both consistent with and deviating from \ac{gr}.
This allows us to benchmark the model's self-consistency and ability to identify \ac{gr} and beyond-\ac{gr} signals.
Following this step, we apply \TEOBB\ to \Nevents\ \ac{bbh} signals from the \ac{gwtc}-3 catalog~\citep{KAGRA:2021vkt, LIGOScientific:2021sio}, performing parametrized post-inspiral
tests to search for potential deviations from \ac{gr}.

\subsection{Cross-validation of \TEOBB}
\label{sec:validate}

\begin{figure}[t]
    \includegraphics[width=0.47\textwidth]{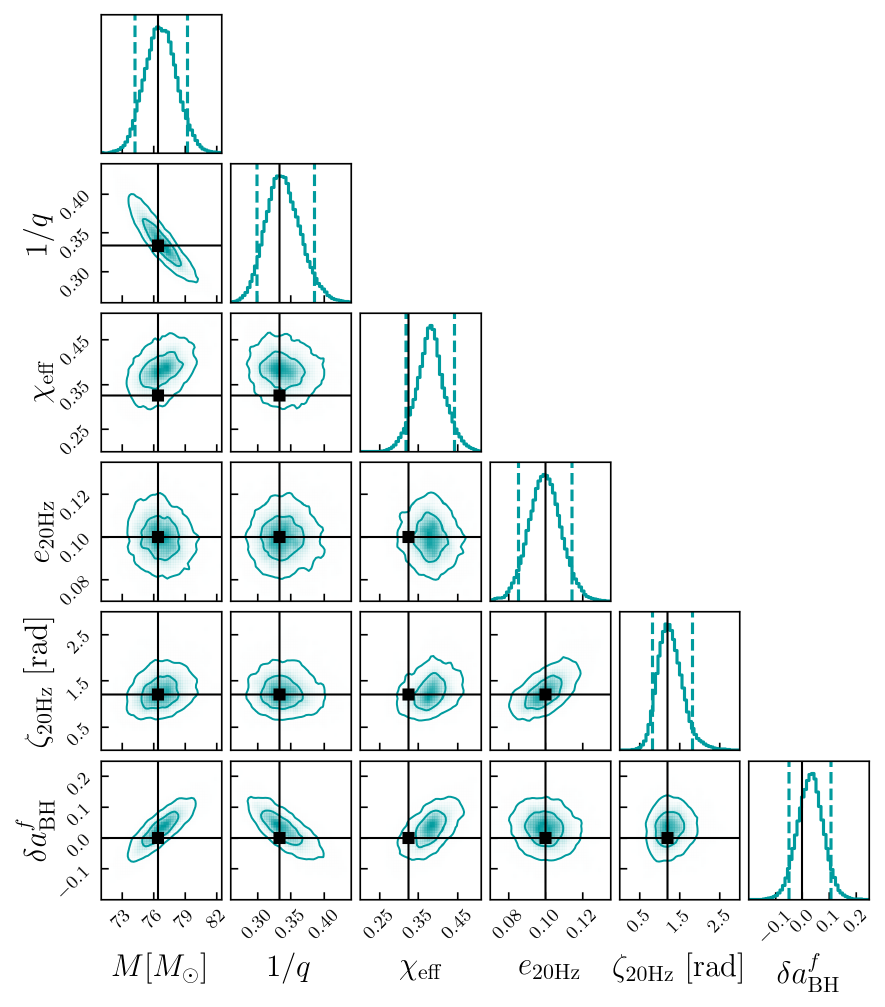}
    \caption{\ac{pe} results for the injection and recovery of the $(2,\pm 2)$ modes of a \ac{gr}-based eccentric \ac{bbh} signal, with $q=3, \mathcal{M}
    = 28 M_\odot, \chi_1 = 0.6,\chi_2 = -0.5$, and $e=0.1$ at 20 Hz, using \TEOBB~and sampling on the $\delta \abhf$ deviation.
    We show here the one- and two-dimensional posteriors for the total mass, mass ratio, effective spin, eccentricity, mean anomaly and deviation parameter.
    The black markers and straight lines indicate the injected values; the vertical dashed blue lines mark the 90\% \acp{ci}; the solid contours correspond to 50 and 90\% credibility.
    \label{fig:gr_ecc_abhf}}
\end{figure}
    
\begin{figure}[t]
    \centering
    \includegraphics[width=0.47\textwidth]{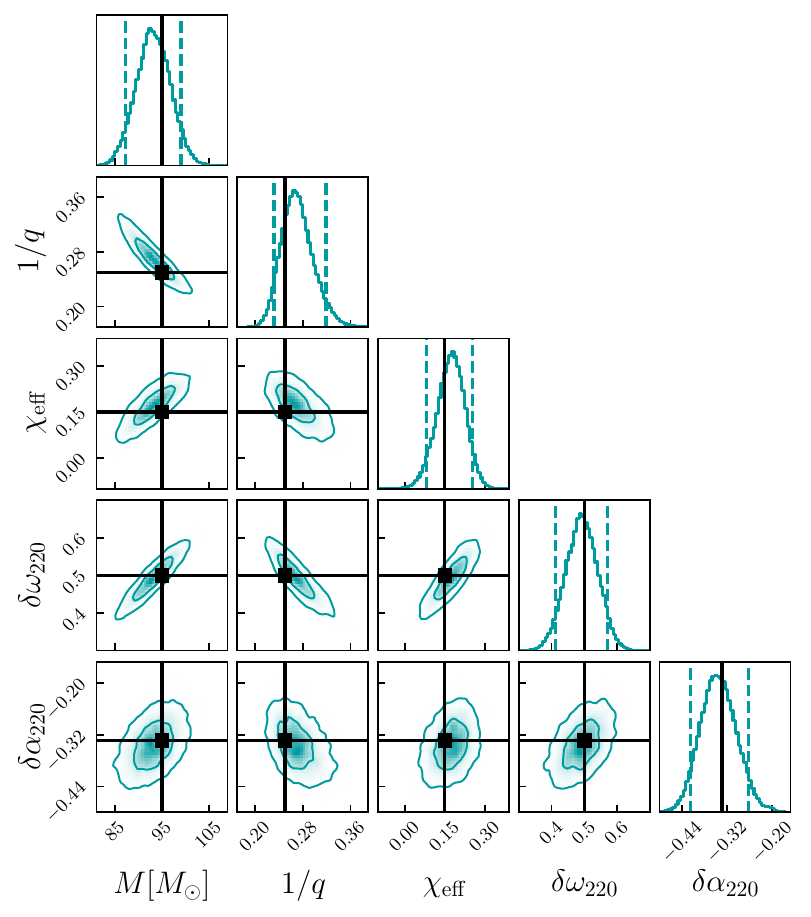}
    \caption{Results of an injection-recovery study with \TEOBB. The injected signal comprises the $(2,\pm 2)$ modes of the waveform produced
    by a spin-aligned binary with total mass $95 M_\odot$, mass ratio $q = 4$, and spins $\chi_1 = 0.3, \chi_2 = -0.45$. 
    The signal uses deviations from the fundamental QNM's frequency and damping time, $\delta \omega_{220} = \delta \tau_{220}
    = 0.5$; we sample on the former and the inverse damping time deviation, $\delta \alpha_{220}$. Dashed vertical lines bound the 90\% \acp{ci}, while solid contours enclose the 50 and 90\% credible regions; injected values are in black.
    \label{fig:bgr_results}}
\end{figure}

\begin{figure}[t]
    \centering
    \includegraphics[width=0.47\textwidth]{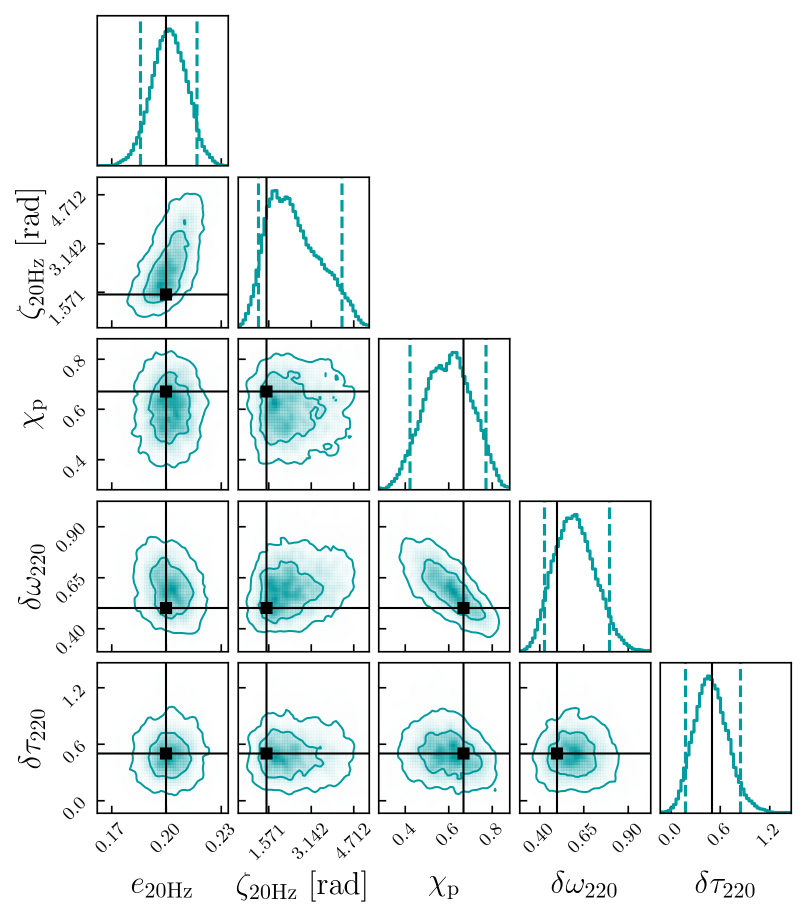}
    \caption{Posterior distributions from the analysis of an eccentric, precessing \ac{bbh} signal incorporating deviations from the fundamental $(2,2,0)$ \ac{qnm}. We show one-dimensional histograms and two-dimensional posteriors for the eccentricity, mean anomaly, effective precessing spin, and the \ac{qnm} deviations. We mark 90\% \acp{ci} with dashed vertical lines, 50 and 90\% credible regions with solid contours, and the expected values with black lines and markers.
    \label{fig:eccprec_inj}}
\end{figure}
    
\begin{figure}[t]
    \centering
    \includegraphics[width=0.47\textwidth]{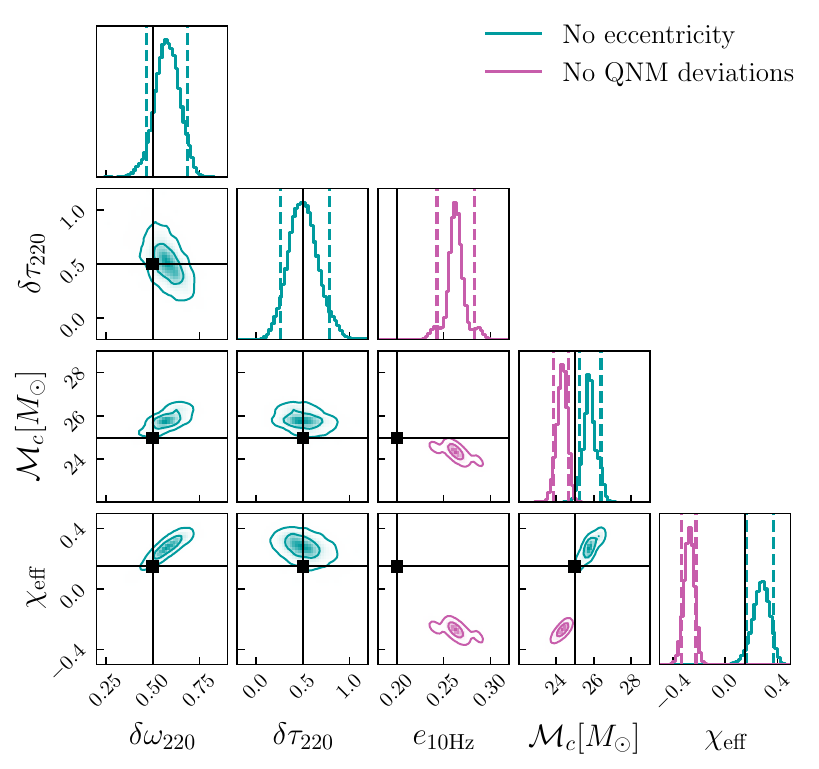}
    \caption{Results of two different \ac{pe} runs for a simulated signal corresponding to a system with $M = 75 M_\odot$, $q=4$, $\chi_1 = 0.3, \chi_2 = -0.45$, and eccentricity at 10 Hz $e_{\rm 10 Hz} = 0.2$. Deviations from the \ac{qnm} spectrum are also added, with $\delta \omega_{220} = \delta \tau_{220} = 0.5$.
    We sample on all the baseline \ac{bbh} parameters, alongside either the \ac{qnm} deviations (green) or the eccentricity and mean anomaly (pink). 90\% credible intervals are indicated by vertical dashed lines, while the injected values are represented by black lines and markers.
    \label{fig:bgr_qnm_ecc_defective}}
\end{figure}

We assess \TEOBB's self-consistency through zero-noise analyses of three synthetic quadrupole-only \ac{bbh} signals, each generated and recovered with the quadrupole-only model itself to probe distinct aspects of its performance with increasing complexity. For all simulations, we fix the binary’s sky location to right ascension and declination of $0.1,\mathrm{rad}$, the inclination and reference azimuth to $0.1,\mathrm{rad}$, and the polarization angle to $\psi = 1.2,\mathrm{rad}$.

The first is a \ac{gr}-consistent, mildly eccentric binary with $\mathcal{M}_c = 28 M\odot$, $q = 3$, $\chi_1 = 0.6$, $\chi_2 = -0.5$, $e = 0.1$, $\zeta = 1.2$ at 20 Hz with a network \ac{snr} of 59.
Sampling on the deviation parameter \(\delta \abhf\) (uniform in $[-0.5,0.5]$) and all other intrinsic and extrinsic parameters, we find that the recovered posteriors (Fig.~\ref{fig:gr_ecc_abhf}) faithfully reproduce the simulated values, including $\delta \abhf = 0$ at $\sim 47\%$ credibility,
thereby affirming  \TEOBB's capability of recovering a \ac{gr}-faithful signal. 

Next, we test the model against a simulated beyond-GR signal with a network SNR of 52, produced by a binary with $M = 95\,M_\odot$, $q = 4$, $(\chi_1, \chi_2) = (0.3, -0.45)$, and fractional deviations $\delta\omega_{220} = \delta\tau_{220} = 0.5$.
The inferred posteriors, shown in Fig.~\ref{fig:bgr_results}, accurately recover the source parameters within statistical uncertainties and exhibit clear correlations reflecting intrinsic waveform degeneracies.
For instance, increasing $M$ or $q$ lowers $\omega_{220}$, whereas increasing the effective spin $\chi_{\rm eff}$ has the opposite effect. 
This behavior follows from the remnant mass and spin's dependence on the binary configuration --- higher total mass yields a heavier remnant; 
higher $\chi_{\rm eff}$ results in a faster spinning final \ac{bh}, whose mass is also decreased due to the longer inspiral radiating more energy.
Similar things happen varying the mass ratio.
Consequently, larger $\delta\omega_{220}$ and $\chi_{\rm eff}$ correlate with higher $M$ and more asymmetric binaries, compensating the associated shift in the quasi-normal-mode frequency.
The comparison of Figs.~\ref{fig:qnmdev_tau} and~\ref{fig:remnantdev} further shows that negative deviations in the damping time ($\delta\alpha_{220} < 0$) can mimic amplitude variations driven by total mass changes.
Hence, lower $\delta\alpha_{220}$ values are favored at the lower end of the mass posterior, typically coupled with smaller $q$ and $\delta\omega_{220}$ to offset the frequency shift due to $M$.

Finally, we analyze a signal that simultaneously features \ac{qnm} deviations, orbital eccentricity, and spin-induced precession.
The signal with a network \ac{snr} of 37 has \(M=80 M_\odot\), \(q=2.5\), \(e=0.2\), \(\zeta=1.5\), effective spin parameters of $\chi_{\rm eff} \simeq -0.33$ and $\chi_{\rm p} = 0.67$ and \ac{qnm} deviations of $\delta\omega_{220} = \delta\tau_{220} = 0.5$.
Fig.~\ref{fig:eccprec_inj} shows the recovered posterior distributions for $e$, $\zeta$, $\chi_{\rm p}$, $\delta\omega_{220}$, and $\delta\tau_{220}$, all of which are consistent with the simulated values within the 90\% \acp{ci}.
These results demonstrate that the model can faithfully recover the binary parameters, even in a more challenging scenario where eccentricity, spin precession, and \ac{qnm} deviations jointly influence the waveform morphology.
Moreover, Fig.~\ref{fig:eccprec_inj} demonstrates the presence of a clear correlation between $\chi_p$ and $\delta\omega_{220}$.
This can be partly explained by the fact that non-zero in-plane spins contribute to the final spin estimate, which in turn affects the baseline \ac{qnm} frequency and -- to a lesser degree -- damping time.

\subsubsection*{Eccentricity, deviations and systematics}
To better understand the interplay between different parameters and potential systematics, we analyze a synthetic signal including both orbital eccentricity and \ac{qnm} deviations under two different assumptions:
(i) varying only the \ac{qnm} parameters, and 
(ii) varying only $e$ and $\zeta$.
This comparison enables us to assess whether neglecting either eccentricity or \ac{qnm} deviations in the signal reconstruction introduces systematic biases in the estimated parameters.

The synthetic signal is generated with the $(2,\pm 2)$, $(2, \pm 1)$, $(3, \pm 3)$, and $(4, \pm 4)$ waveform modes, starting at a frequency $f_{\rm start} = 10\,{\rm Hz}$, with redshifted total mass $M = 75 M_\odot$, eccentricity $e_{\rm 10 Hz} = 0.2$, and mean anomaly $\zeta_{\rm 10 Hz} = 1.5$. The network \ac{snr} is 44. 
Results are shown in Fig.~\ref{fig:bgr_qnm_ecc_defective}, with (i) in green and (ii) in pink.

When eccentricity is neglected in the analysis [case~(i)], the damping time deviation $\delta\tau_{220}$ is recovered consistently with the simulated value, and $\delta\omega_{220}$ remains within the 90\% credible interval, though outside the 67\%, indicating a mild preference for higher values.
This trend arises because eccentricity influences the waveform predominantly during the inspiral phase, with only indirect effects on the post-merger morphology where the \ac{qnm} deviations enter, owing to \TEOBB' explicit assumption that the binary has circularized by the time of merger.
Degeneracy between eccentricity and the chirp mass~\cite{Favata:2021vhw, OShea:2021faf, Hegde:2023yoz} implies that neglecting $e$ leads to an overestimation of $\mathcal{M}_c$ in \ac{pe}, biasing $\chi_{\rm eff}$ upwards while leaving $q$ well recovered.
This shift in $\mathcal{M}_c$ slightly lowers the predicted post-merger frequency, which the $\delta\omega_{220}$ deviation then compensates.

At the \ac{snr} considered here, this interplay does not suffice to push the \ac{gr} value of $\delta\omega_{220}$ outside the 90\% credible interval. 
However, at higher \ac{snr} ($\gtrsim 100$), the same bias would exclude the \ac{gr} value at high confidence.
This analysis, therefore, illustrates that neglecting orbital eccentricity in strong-field \ac{gr} tests can lead to biased inferences of \ac{qnm} deviations for loud signals.
This situation will become increasingly common with next-generation detectors.

On the other hand, when \ac{qnm} deviations are neglected instead [case~(ii)], we find significant biases in the recovered binary parameters (left panel of Fig.~\ref{fig:bgr_qnm_ecc_defective}), including an underestimation of the chirp mass, overestimation of the eccentricity (exhibiting a trimodal distribution), and an almost inverted effective spin $\chi_{\rm eff}$.
This is expected because, at this total mass, the post-inspiral signal has a decisive impact on parameter inference, and neglecting \ac{qnm} deviations forces the model to absorb disagreement into inspiral parameters.

\subsection{\ac{nr} waveforms}
\label{sec:nrwf}

The previous section validated \TEOBB\ against the baseline \dali\ model in a controlled environment.
We now extend this analysis to \ac{nr} simulations of
an eccentric \ac{gr} binary waveform from the SXS catalog~\cite{Hinder:2017sxy, Scheel:2025jct},
a binary boson star merger waveform~\cite{Evstafyeva:2024qvp},
and a set of binary waveforms modified according to dynamical Chern-Simons gravity~\cite{Okounkova:2019zjf}.

\subsubsection{SXS:BBH:1363}

\begin{figure}[t]
    \centering
    \includegraphics[width=0.47\textwidth]{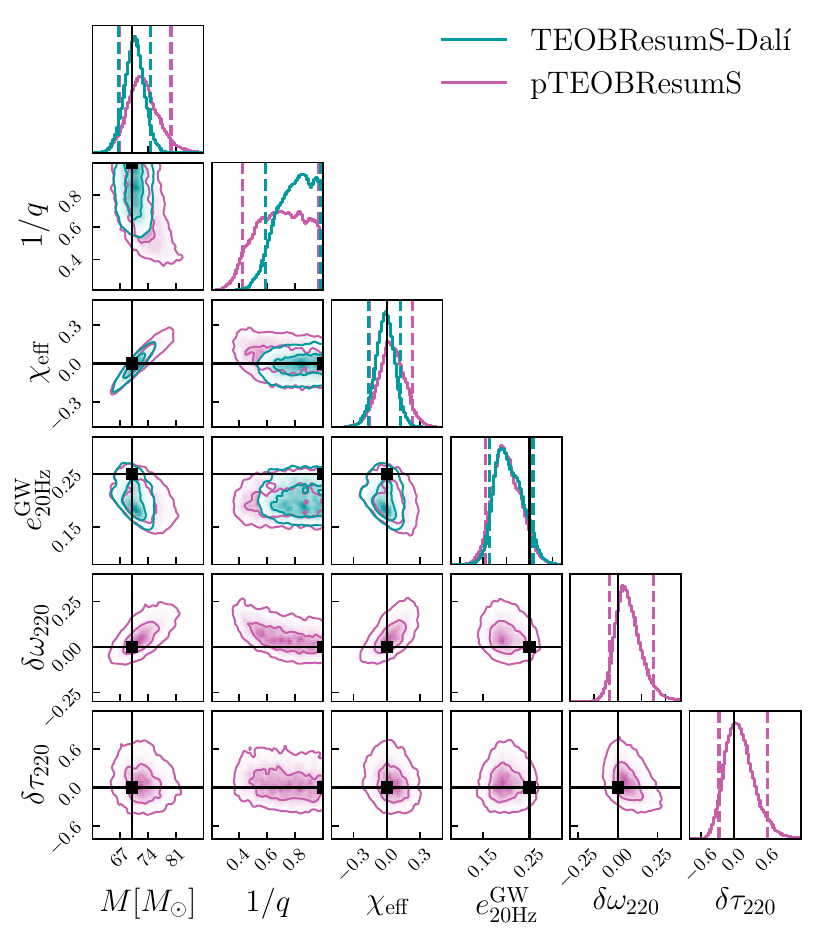}
    \caption{\ac{pe} for a synthetic signal generated from the \ac{nr} simulation SXS:BBH:1363, an equal-mass,
    nonspinning, moderately eccentric \ac{bbh} system. We show here the one- and two-dimensional poterior distributions
    for the total mass, mass ratio, effective spin, eccentricity at 20 Hz and the \ac{qnm} frequency and damping time deviation
    parameters obtained employing \dali~(green) or the \TEOBB~model (pink). The black marker
    and straight lines correspond to the expected values. The vertical dashed lines indicate the bounds of the 90\% \acp{ci},
    while the contours correspond to the 50 and 90\% credible regions.
    \label{fig:sxs1363_corner}}
\end{figure}

We analyze the non-spinning, moderately eccentric numerically simulated \ac{bbh} signal, SXS:BBH:1363~\cite{Hinder:2017sxy, Scheel:2025jct}.
This simulation has \(q=1\), \(e=0.35\) and \(\zeta=2.31\).
Since $e$ and $\zeta$ are gauge-dependent parameters, we cannot directly compare these values with \dali's internal definitions of them.
We instead use the \gwecc~\cite{Shaikh:2023ypz, Shaikh:2025tae} package, which provides estimates of the eccentricity and mean anomaly of a \ac{gw} signal measured directly from the $(2,2)$ mode waveform.
Locating the turning points via the oscillations in the amplitude or frequency, the package interpolates in time the evolving periastron, apastron and orbit-averaged \ac{gw} frequencies; the estimates $e^{\rm GW}, \zeta^{\rm GW}$ are computed from the former two, evaluated at a chosen reference value for the latter.
We consider here a reference orbit-averaged $(2,2)$ mode frequency of 20 Hz, obtaining $e_{\rm 20 Hz}^{\rm GW} \simeq 0.25$ and $\zeta_{\rm 20 Hz}^{\rm GW} \simeq 4.27$.

We again analyze this signal in zero noise after scaling the total mass to $M = 70 M_\odot$ and placing the source at $D_L \simeq 2307 \ \rm{Mpc}$, a right ascension of 1.375 rad, and a declination of $-1.211$ rad.
We set the inclination angle with respect to the line of sight to $\iota = 0$, and $\psi$ and $\phi_{\rm ref}$ both to 0 rad.
The signal's optimal network \ac{snr} is 19.
We carry out two analyses, one with the standard \dali~model and one using the parametrized version, sampling on $\delta \omega_{220}, \delta \tau_{220}$; we perform the recovery using only the leading $(2,\pm 2)$ modes in both cases. 
For each posterior sample produced by the \ac{pe} runs, we estimate $e_{\rm 20 Hz}^{\rm GW}$ and $\zeta_{\rm 20 Hz}^{\rm GW}$ with \texttt{gw\_eccentricity}, after evolving the waveform backward in time to have sufficient data around the reference frequency to do so.

A selection of the results are reported in Fig.~\ref{fig:sxs1363_corner}.
All intrinsic parameters are recovered within the 90\% \ac{ci} in both runs. Only the expected eccentricity and mean anomaly
fall outside the 50\% credible region, with minimal differences in their recovery between the models.
The \TEOBB~analysis finds no evidence for deviations from \ac{gr} in the \ac{qnm} parameters, with both $\delta \omega_{220}$ and $\delta \tau_{220}$
compatible with zero. The introduction of the beyond-\ac{gr} parameters widens the posteriors for many of the
intrinsic ones, to compensate for the effect of the deviations.
For instance, the higher $M$, $q$ and $\chi_{\rm eff}$ reached by the \TEOBB~posteriors are associated with the tail towards positive values in $\delta \omega_{220}$, an interaction we already pointed out in the previous section.
The increased flexibility of \TEOBB~does not however lead to a better fit to the data: the maximum likelihoods for the two analyses are close, $\log_{10}\mathcal{L}_{\rm{GR}}^{\rm max}/\mathcal{L}_{\rm BGR}^{\rm max} = 0.13$.
The \ac{bf} between the two hypotheses thus points to strong preference for the \ac{gr} description, with $\log_{10} \bayes{\rm GR}{} = 1.66$, mostly due to the larger prior volume of the \TEOBB~model.

\subsubsection{SXS:BBH:3717}
\begin{figure}[t]
    \centering
    \includegraphics[width=0.47\textwidth]{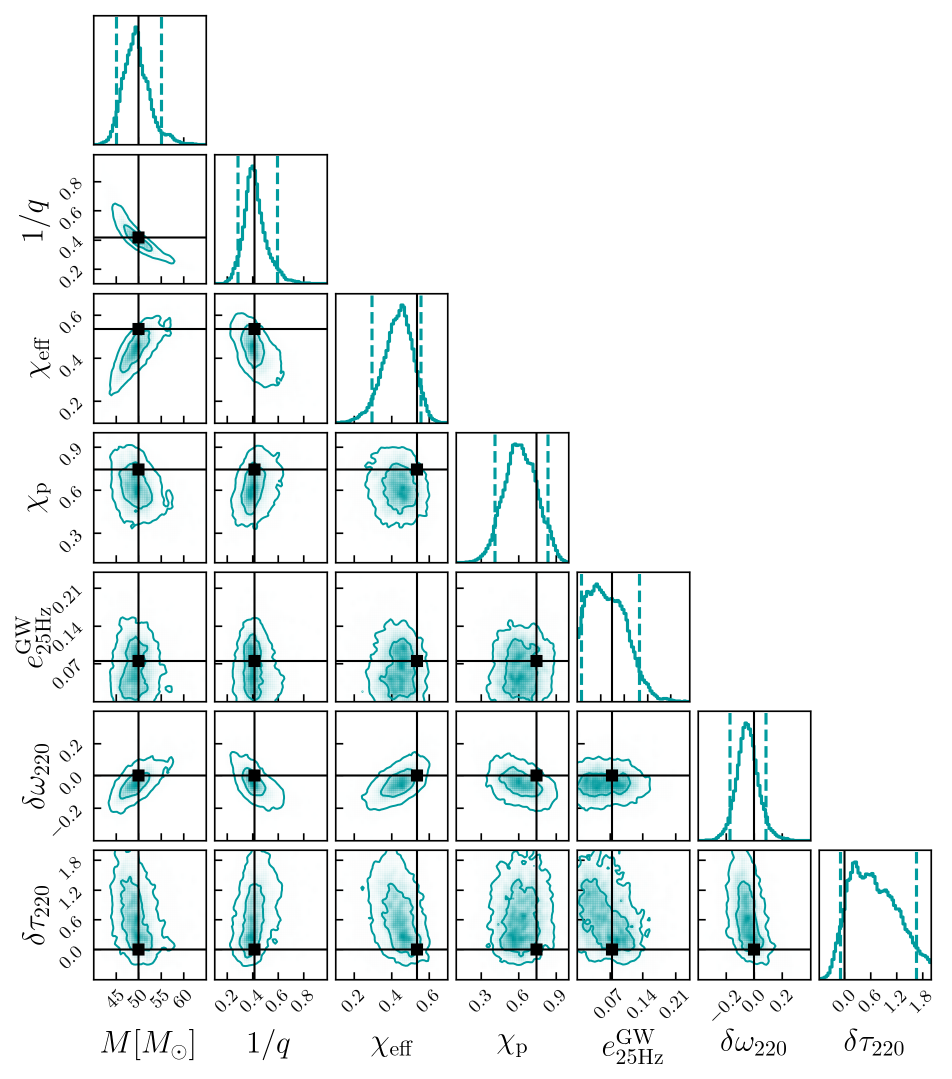}
    \caption{Corner plot showing the results of \ac{pe} on a synthetic signal generated from the eccentric, precessing \ac{nr} simulation SXS:BBH:3717.
    Contours mark the 90\% and 50\% credible levels, vertical dashed lines the 90\% \acp{ci}, and the injected values are in black.
    With an \ac{snr} of 13, all intrinsic parameters are recovered at 90\% credibility, although the effective spin and $\delta \tau_{220}$ sit at the edge of their respective intervals.
    The latter's posterior is also railing against the upper limit of the prior range.
    \label{fig:sxs3717_corner}}
\end{figure}
We analyze the eccentric, precessing \ac{nr} simulation SXS:BBH:3717 with \TEOBB, again sampling on the \ac{qnm} deviations.
The simulation is of a \ac{bbh} with mass ratio $q = 2.4$, $\chi_{\rm eff} = 0.54$, and $\chi_{\rm p} = 0.74$.
The initial eccentricity reported in the \ac{nr} metadata is $e = 0.096$.
We inject the signal in zero noise, setting the total mass to $M=50M_\odot$, $D_L$ to 1000 Mpc, the inclination to $\iota = 1.2$, $\phi_{\rm ref}$ and $\psi$ both to 0 rad.
We place the system at a right ascension of 0.33 rad, and a declination of 0.6 rad.
Generating the signal using all $\ell \leq 4, m \neq 0$ modes, the resulting \ac{snr} is 13.
Once the physical scale is set, we estimate the eccentricity of the waveform at 25 Hz with \texttt{gw\_eccentricity}, finding $e_{25 \rm{Hz}}^{\rm GW} = 0.075$.
We set the waveform starting frequency to $f_{\rm start} = 10\, {\rm Hz}$, and generate \TEOBB~waveforms featuring the (coprecessing) $(2, \pm 1), (2, \pm 2), (3, \pm 3)$, and $(4, \pm 4)$ modes.

Fig.~\ref{fig:sxs3717_corner} shows the results for the main intrinsic parameters.
While the total mass, mass ratio, and eccentricity show little to no bias, the effective spin parameters are both underestimated, although the respective expected values still lie within the 90\% \acp{ci}.
$\delta \omega_{220}$ peaks at a small negative value, but the posterior is widely compatible with the \ac{gr} expectation of 0.
As for the damping time, we find a marked preference for positive values, with the distribution railing against the upper limit of the prior range.\footnote{
    We have not rerun this analysis with a wider prior on $\delta \tau_{220}$ because of its prohibitively long computational time; the original \ac{pe} took several months to complete.
}
While 0 is not excluded at 90\% credibility, given this signal's low \ac{snr}, this result could sharpen into the detection of a false violation of \ac{gr} in the case of a louder event.
This test thus showcases \TEOBB's capability to model a challenging system, while also highlighting the need for further upgrades to the underlying \dali~in this physically complex regime.

\subsubsection{Boson star binary}
\begin{figure}
    \includegraphics[width=0.47\textwidth]{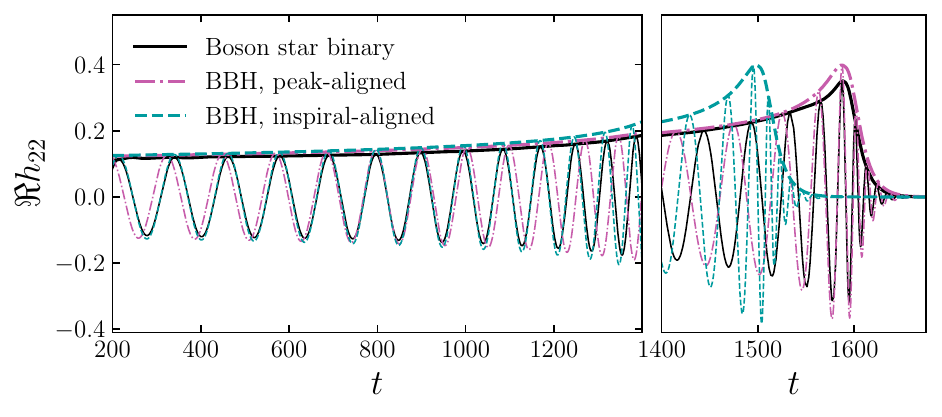}
    \caption{Time-domain comparison of the $(2,2)$ mode of the binary boson star signal (black) with a \ac{bbh} waveform generated with \dali\ with the nominal parameters of the boson star system ($q = 1$, no spin). We show the \ac{bbh} waveform twice: once aligned with the boson stars' so that their amplitude peaks coincide (pink); once time- and phase-shifted to minimize the cumulative squared phase difference with respect to the \ac{nr} signal over the inspiral (green). We draw for each waveform the amplitude and the real part.
    \label{fig:bsnr_wf}}
\end{figure}
\begin{figure}
\includegraphics[width=0.47\textwidth]{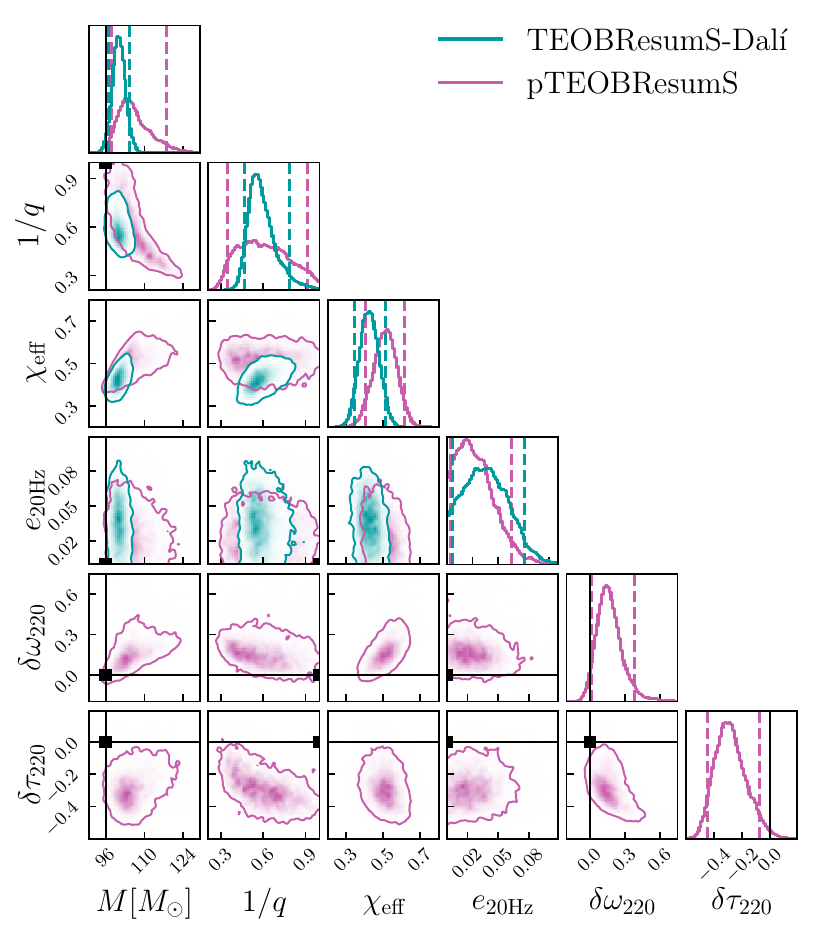}
\caption{Results of \ac{pe} performed on the zero-noise injection of a mock signal from a binary boson star merger resulting
    in a remnant \ac{bh}. We show one- and two-dimensional posterior distributions for the total mass, mass ratio, effective spin, eccentricity,
    and the \ac{qnm} deviations. Black markers and lines correspond to the expected values; vertical dashed lines bound the 90\% \acp{ci},
    while the contours correspond 90\% credibility.
    The model recognizes the non-\ac{bh} nature of the signal, finding a best match with non-zero
    deviations from the Kerr values of the \ac{qnm} parameters. These are coupled with biases in the remaining parameters that are due to the \ac{bbh}-based
    models attempting to match the intrinsically different properties of the \ac{bs} binary signal.
    \label{fig:bsnr_corner}}
\end{figure}

We analyze the numerically simulated binary boson star waveform, \texttt{A17-d17-p180}, from \citet{Evstafyeva:2024qvp} using both our parametrized and non-parametrized waveform models under the assumption that the signal source is eccentric but non-precessing. This waveform corresponds to the \ac{imr} of two equal-mass boson stars with central scalar density $\sqrt{G}A_{\mathrm{ctr}}(0)=0.17$ and phase offset $\delta\phi=\pi$, merging to form a Kerr black hole with spin $\sim0.7$. 
As shown in Fig.~\ref{fig:bsnr_wf}, the waveform resembles that of a non-spinning quasi-circular \ac{bbh} merger but with a delayed coalescence, primarily owing to the interaction of the out-of-phase scalar fields.
Following~\citet{Pompili:2025cdc}, we fix the detector frame total mass to \(M=96 M_\odot\), $D_L=1200\,\mathrm{Mpc}$, $(\alpha,\delta)=(0.33,\,-0.6)\,\mathrm{rad}$, $\iota=\pi/3$, $\psi=0.7\,\mathrm{rad}$, and $\phi_\mathrm{ref}=1.2\,\mathrm{rad}$, such that the signal's network \ac{snr} is 34. 
We restrict our analysis to the $\ell = 2$ modes, which are the only ones computed in the \ac{nr} simulation.

As shown in Fig~\ref{fig:bsnr_corner}, we find that \TEOBB\ favours non-zero \ac{qnm} deviations, with the \ac{gr} predictions $(\delta\omega_{220},\,\delta\tau_{220})=(0,0)$ lying outside the 90\% credible region.
In particular, we find that the waveform reconstructions favour stronger damping and slightly higher frequency than expected from a Kerr remnant produced by an equivalent \ac{bbh} merger.
Additionally, we find that both the \dali\ and \TEOBB\ runs show notable biases in the recovered binary parameters: positive effective spin, with the exclusion of $\chi_{\mathrm{eff}}=0$ at the 90\% level, inaccurate mass ratio (with the \ac{gr} analysis particularly preferring asymmetric configurations), and overestimated $D_L$.
Interestingly, \TEOBB\ yields a less accurate chirp-mass recovery than \dali.
These results are expected, and they agree with the systematic trends reported in \cite{Evstafyeva:2024qvp,Pompili:2025cdc}.
The biases arise from the intrinsic differences between the binary boson-star signal and the \ac{bbh}-based templates used in the recovery.
The modified phasing in the \ac{bs} binary is the probable source of the biases in the spin and mass ratio, since asymmetric configurations and positive $\chi_{\rm eff}$ both cause a slower inspiral, delaying the plunge in an attempt to better fit the injected signal.
The less accurate chirp-mass recovery with \TEOBB\ likely reflects the interplay between the added \ac{qnm} degrees of freedom and inspiral mismatches, which absorb part of the deviation without improving mass consistency.

\subsubsection{Dynamical Chern-Simons gravity}
\begin{figure}
    \includegraphics[width=0.47\textwidth]{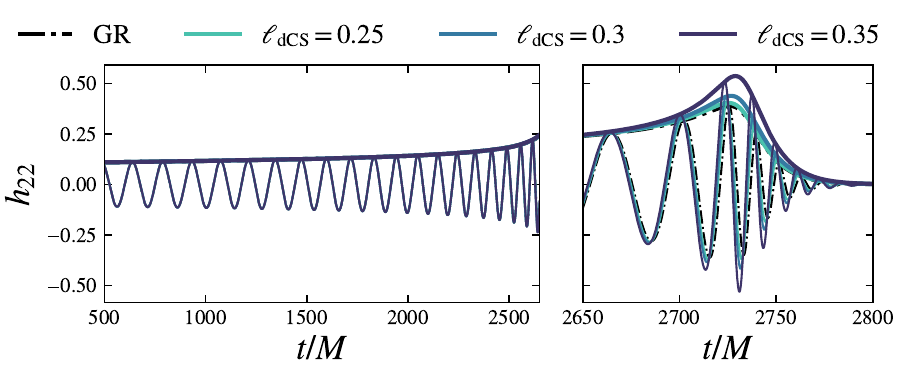}
    \caption{$(2,2)$ mode of the waveform for the base, GW150914-like \ac{nr} simulation and three \ac{dcs}-corrected versions with increasing coupling length $\ell_{\rm dCS}$. We draw for each the amplitude and the real part. Deviations from the \ac{gr}-consistent signal quickly accumulate during plunge, with noticeable peak amplitude differences that scale roughly as $\ell_{\rm dCS}^4$.
    \label{fig:dcs_waveform}}
\end{figure}
\begin{figure}
\includegraphics[width=0.47\textwidth]{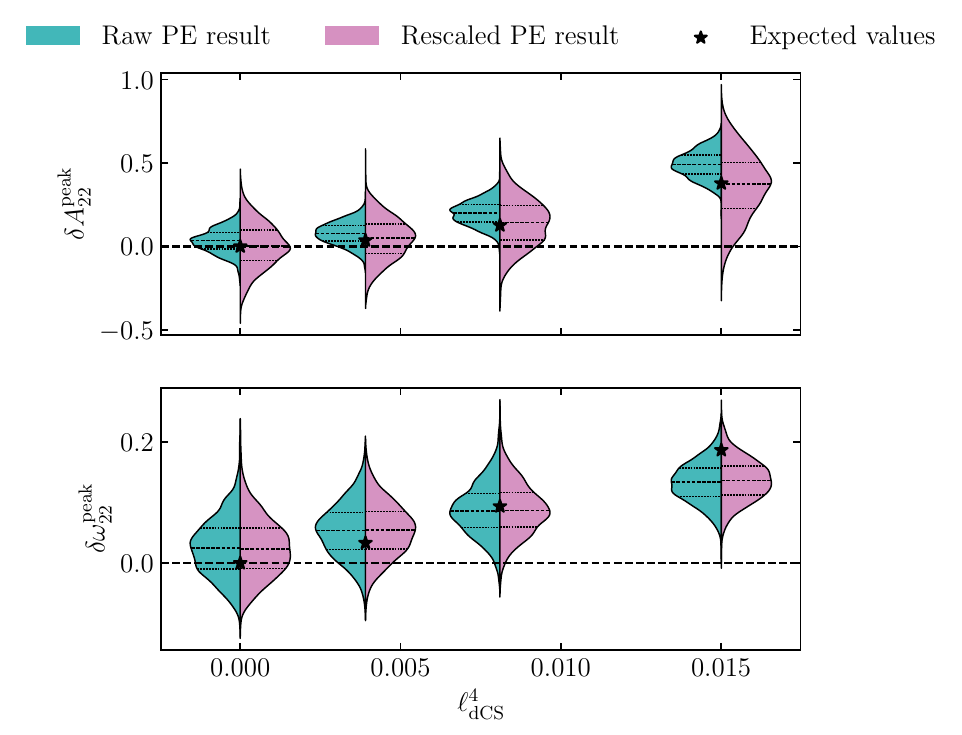}
\caption{\ac{pe} results for the peak time deviations for a GW150914-like \ac{nr} simulation and three \ac{dcs}-corrected versions.
    We show in green the raw posteriors from the analysis. The pink distributions are rescaled to account for the concurrent biased recovery of the total mass, which acts simply as a scale parameter for both $A_{22}^{\rm peak}$ and $\omega_{22}^{\rm peak}$, and luminosity distance, which similarly affects the amplitude. Expected deviations estimated from the numerical waveforms are accurately reproduced in the case of the amplitude, less so for the frequency.
    Horizontal dashed lines in the violins mark the median, the dotted ones the quartiles of each distribution.
    \label{fig:dcs_delta_posts}}
\end{figure}
We finally analyze a series of numerically simulated signals that include corrections to \ac{gr} according to \acf{dcs} gravity~\citep{Jackiw:2003pm} with \TEOBB. 
This theory adds higher-order curvature corrections to the Einstein-Hilbert action, coupled to a dynamical scalar field.
It introduces a new parameter, the length scale $\ell_{\rm dCS}$, below which deviations from \ac{gr} become significant; $\ell_{\rm dCS}=0$ recovers pure \ac{gr}.
We employ waveforms from~\citet{Okounkova:2019zjf, Okounkova:2022grv}, produced using an order-reduction scheme in which a \ac{gr} background is evolved alongside a leading-order scalar field, thereby modifying the \ac{gw} strain.
In these simulations, \ac{dcs} corrections are activated only late in the inspiral, primarily impacting the plunge, merger, and ringdown stages, making them an ideal test bed for our analysis; Fig.~\ref{fig:dcs_waveform} displays the resulting $(2,2)$ mode waveforms.
Deviations from the \ac{gr} base scale as $\ell_{\rm dCS}^4$.

The signals correspond to a GW150914-like binary with $q = 1.22$, $\chi_1 = 0.33$, $\chi_2 = -0.44$, $M = 68\,M_\odot$, $(\alpha, \delta) = (1.74, -1.22)$, $\iota = 2.72$, $\psi = 1.53$, and $\phi = 0$.
From this baseline, we generate \ac{dcs}-corrected waveforms with $\ell_{\rm dCS}/M \in \{0.25, 0.3, 0.35\}$.
We include all angular modes up to $(\ell, |m|) \leq (5,5)$ except those with $m=0$, and adjust $D_L$ to produce in each case a network \ac{snr} of 50.
We analyze each signal three times, always working in the quasicircular, spin-aligned limit with standard priors:
(i) under a \ac{gr}-consistent hypothesis;
(ii) sampling on the remnant deviations $\delta \mbhf, \delta \abhf$, employing uniform priors in $[-0.9, 1]$;
(iii) sampling on the peak time deviations in the amplitude and frequency of the dominant $(2,2)$ mode, $\delta A^{\rm peak}_{22}, \delta \omega^{\rm peak}_{22}$, both with a uniform prior in the interval $[-0.8, 4]$.
Our choice of parameters to take into account is driven by the nature of the signals, where \ac{dcs} corrections accumulate in the final wave cycles during plunge, mostly manifesting as an increased amplitude around and after merger.
We generate \TEOBB~waveforms including the $(2, \pm 1), (2, \pm 2), (3, \pm 3)$, and $(4, \pm 4)$ modes.

\ac{pe} results are shown in detail in Figs.~\ref{fig:dcs_gr_corner},~\ref{fig:dcs_rmn_corner},~\ref{fig:dcs_mrg_corner} in Appendix~\ref{app:dcs}.
The \ac{gr} analyses return growing biases as the coupling strength increases, particularly in the effective spin, luminosity distance, and inclination, the latter two likely allowing at the same time a comparatively better recovery of the total mass.
Sampling on the remnant properties, the model successfully offloads the \ac{gr} corrections in the signals onto the deviation parameters, delivering mostly accurate results for the standard binary parameters.
Posteriors for $\delta \mbhf, \delta \abhf$ widely exclude 0 for all non-zero values of $\ell_{\rm dCS}$, while no strong evidence for deviations is found in the base case.
Using the peak time deviations, we find increasingly more biased results than in the previous case for some of the intrinsic parameters (total mass, effective spin) as $\ell_{\rm dCS}$ grows, while also obtaining the most accurate recovery of $D_L$ for the \ac{dcs}-corrected signals.
Results for the deviation parameters display the expected trend, clearly moving away from 0 as the \ac{dcs} coupling is increased.
When rescaling for the bias in the $M$ and $D_L$ posteriors, the results for $\delta A^{\rm peak}_{22}$ closely align with the values that can be estimated from the numerical $(2,2)$ mode waveform, as displayed in Fig.~\ref{fig:dcs_delta_posts}.
The same cannot be said however for the peak time frequency deviation, possibly due to its dependence on the also biased effective spin, which is harder to similarly undo.

The \ac{gr}-consistent description is favored by the data for the $\ell_{\rm dCS} = 0$ signal, with $\log \mathcal{B}^{\rm GR} \sim 2-3$ over the analyses incorporating deviations from \ac{gr}.
This is still the case for $\ell_{\rm dCS} = 0.25$, but for larger coupling lengths the \ac{gr} description is strongly disfavored.
For all of the $\ell_{\rm dCS} > 0$ cases, we find a preference for the peak time deviations scenario over the remnant deviations, dramatically so for the two largest values of $\ell_{\rm dCS}$ ($\log \mathcal{B}^{\rm peak}_{\rm remnant} = 4.25$ and $22$ for $\ell_{\rm dCS} = 0.3$ and $0.35$, respectively).
This is despite the remnant analyses displaying lesser biases in most parameters, but it is consistent with the effect of the \ac{dcs} corrections on the waveform, as seen in Fig.~\ref{fig:dcs_waveform}.

To summarize, in this test \TEOBB~correctly recognizes a \ac{gr}-consistent signal (the $\ell_{dCS} = 0$ case) while sampling on two alternative sets of deviation parameters.
When corrections to \ac{gr} are present in the data, the model favors non-zero values for the deviations, in one case accurately reproducing their expected values as estimated directly from the waveforms, while at times also returning biased estimates of the standard binary parameters.

\subsection{Analysis of \ac{bbh} events}
\label{sec:events}
\newcolumntype{C}{>{\centering\arraybackslash}X}
\def\minus{\hphantom{-}}
\newcommand{\bmath}[1]{{\boldmath #1 \unboldmath}}
\begin{table*}[t]
	\centering 
	\caption{Main results of \ac{pe} with \TEOBB~for a collection of \ac{gw} events, under either a precessing or eccentric hypothesis. We report here the recovered
		median values of the \ac{qnm} deviation parameters, $\delta \omega_{220}$ and $\delta \tau_{220}$, with errors corresponding to their 90\% credible
		intervals. For each run, we also report the \ac{bf}, $\log_{10}\bayes{\rm{GR}}{}$, between the \ac{gr} and \ac{bgr} scenarios, as defined by the Savage-Dickey
		ratio in Eq.~\eqref{eq:sdr_bf}; positive values indicate a preference for the null hypothesis. The last rows share the combined constraints on
		the deviation parameters, obtained by multiplying the individual posteriors of each event.
		\label{tab:real_results}}
	\begin{tabularx}{0.9\linewidth}{l *{5}{C} *{5}{C} *{5}{C} *{5}{C} *{5}{C} *{5}{C}}
		\toprule
		\multirow{2}{2.5cm}{\centering \textbf{Event}} & \multicolumn{3}{c}{\textbf{Precessing}} & \multicolumn{3}{c}{\textbf{Eccentric}} \\ 
		& $\bm{\delta \omega_{220}}$ & $\bm{\delta \tau_{220}}$ & \bmath{$\log_{10}\bayes{\rm{GR}}{}$} & $\bm{\delta \omega_{220}}$ & $\bm{\delta \tau_{220}}$ & \bmath{$\log_{10}\bayes{\rm{GR}}{}$} \\
        \midrule
        GW150914          & $\minus0.04^{+0.12}_{-0.07}$   & $\minus0.16^{+0.45}_{-0.34}$   & $1.52$ & $\minus0.05^{+0.11}_{-0.07}$    & $\minus0.15^{+0.38}_{-0.31}$ & $1.52$ \\
		\midrule
        GW170104          & $-0.04^{+0.19}_{-0.12}$        & $\minus0.78^{+1.64}_{-1}$      & $1.28$ & $-0.02^{+0.23}_{-0.12}$         & $\minus0.55^{+1.34}_{-0.83}$ & $1.46$ \\
        \midrule
		GW190519          & $-0.01^{+0.21}_{-0.20}$        & $\minus0.07^{+0.72}_{-0.34}$   & $1.56$ & $-0.06^{+0.16}_{-0.17}$         & $\minus0.09^{+0.47}_{-0.32}$ & $1.73$ \\
		\midrule
		GW190521\_074359  & $\minus0.06^{+0.12}_{-0.1}$    & $\minus0.03^{+0.42}_{-0.32}$   & $1.78$ & $\minus0.09^{+0.12}_{-0.09}$    & $-0.03^{+0.30}_{-0.24}$      & $1.64$ \\
		\midrule
		GW190630          & $-0.04^{+0.42}_{-0.18}$        & $-0.05^{+0.98}_{-0.52}$        & $1.40$ & $-0.02^{+0.4}_{-0.18}$          & $-0.08^{+0.87}_{-0.48}$      & $1.45$ \\
		\midrule
		GW190828\_063405  & $\minus0.08^{+0.24}_{-0.13}$   & $\minus0.16^{+0.72}_{-0.54}$   & $1.26$ & $\minus0.12^{+0.22}_{-0.13}$    & $\minus0.13^{+0.69}_{-0.53}$ & $1.04$ \\
		\midrule
		GW190910          & $\minus0.02^{+0.11}_{-0.09}$   & $\minus0.96^{+0.8}_{-0.72}$    & $0.52$ & $\minus0.02^{+0.12}_{-0.09}$    & $\minus0.93^{+0.75}_{-0.65}$ & $0.27$ \\
		\midrule
        GW200129          & $-0.027^{+0.066}_{-0.067}$     &  $\minus0.23^{+0.71}_{-0.35}$  & $1.96$ & $-0.023^{+0.064}_{-0.06}$       & $\minus0.04^{+0.28}_{-0.23}$ & $2.21$ \\
		\midrule
		GW200208\_130117  & $\minus0.14^{+1.03}_{-0.28}$   & $-0.1^{+0.91}_{-0.44}$         & $1.16$ & $\minus0.13^{+0.78}_{-0.23}$    & $-0.06^{+0.8}_{-0.45}$       & $1.20$ \\
		\midrule
		GW200224          & $\minus0.01^{+0.15}_{-0.12}$   & $\minus0.31^{+0.69}_{-0.39}$   & $1.47$ & $\minus0.01^{+0.15}_{-0.09}$    & $\minus0.30^{+0.48}_{-0.34}$ & $1.49$ \\
		\midrule
		GW200311          & $\minus0.01^{+0.17}_{-0.09}$   & $\minus0.64^{+1.63}_{-0.86}$   & $1.49$ & $\minus0.06^{+0.18}_{-0.12}$    & $\minus0.07^{+0.57}_{-0.38}$ & $1.85$ \\
		\midrule
		\multicolumn{7}{l}{\textbf{Combined}}  \\
		\midrule
		Joint posterior   & $\minus0.012^{+0.032}_{-0.032}$& $\minus0.17^{+0.16}_{-0.14}$   &        & $\minus0.020^{+0.032}_{-0.027}$ & $\minus0.13^{+0.14}_{-0.13}$ &        \\
		\midrule
		Hierarchical      & $\minus0.011^{+0.067}_{-0.058}$& $\minus0.18^{+0.32}_{-0.26}$   &        & $\minus0.022^{+0.068}_{-0.065}$ & $\minus0.12^{+0.32}_{-0.25}$ &        \\
		\bottomrule
	\end{tabularx}
\end{table*}
\begin{table}[t]
	\centering 
    \caption{Results of the hierarchical combination of the posteriors for the \ac{qnm} deviation parameters obtained in the precessing and eccentric analyses of our sample of events.
    We report the median and 90\% \acp{ci} for the means $\mu$ and the correlation $\rho$, and the 90\% lowest credible interval for the variances $\sigma$ of the bivariate Gaussian.
    \label{tab:hierarchical}}
	\begin{tabularx}{0.8\linewidth}{*{1}{C} c *{1}{C} *{1}{C}}
		\toprule
		 & & \textbf{Precessing} & \textbf{Eccentric} \\ 
        \midrule \\[-0.95em]
        \multirow{2}{*}{\bmath{$\delta \omega_{220}$}} & \bmath{$\mu$}  & $0.012^{+0.038}_{-0.037}$ & $0.022^{+0.035}_{-0.034}$ \\[0.5em]
        
         & \bmath{$\sigma$}  & $<0.054$ & $<0.061$ \\[0.2em]
        \midrule \\[-0.95em]
        
        \multirow{2}{*}{\bmath{$\delta \tau_{220}$}} & \bmath{$\mu$}  & $0.19^{+0.2}_{-0.15}$ & $0.13^{+0.17}_{-0.14}$ \\[0.5em]
        
         & \bmath{$\sigma$}  & $<0.27$ & $<0.24$ \\[0.2em]
        \midrule \\[-0.95em]

        \multicolumn{2}{c}{\bmath{$\rho$}}  & $-0.093^{+0.96}_{-0.82}$ & $-0.035^{+0.94}_{-0.87}$ \\[0.2em]
		\bottomrule
	\end{tabularx}
\end{table}

\begin{figure*}[t]
	\centering
	\includegraphics[width=0.85\textwidth]{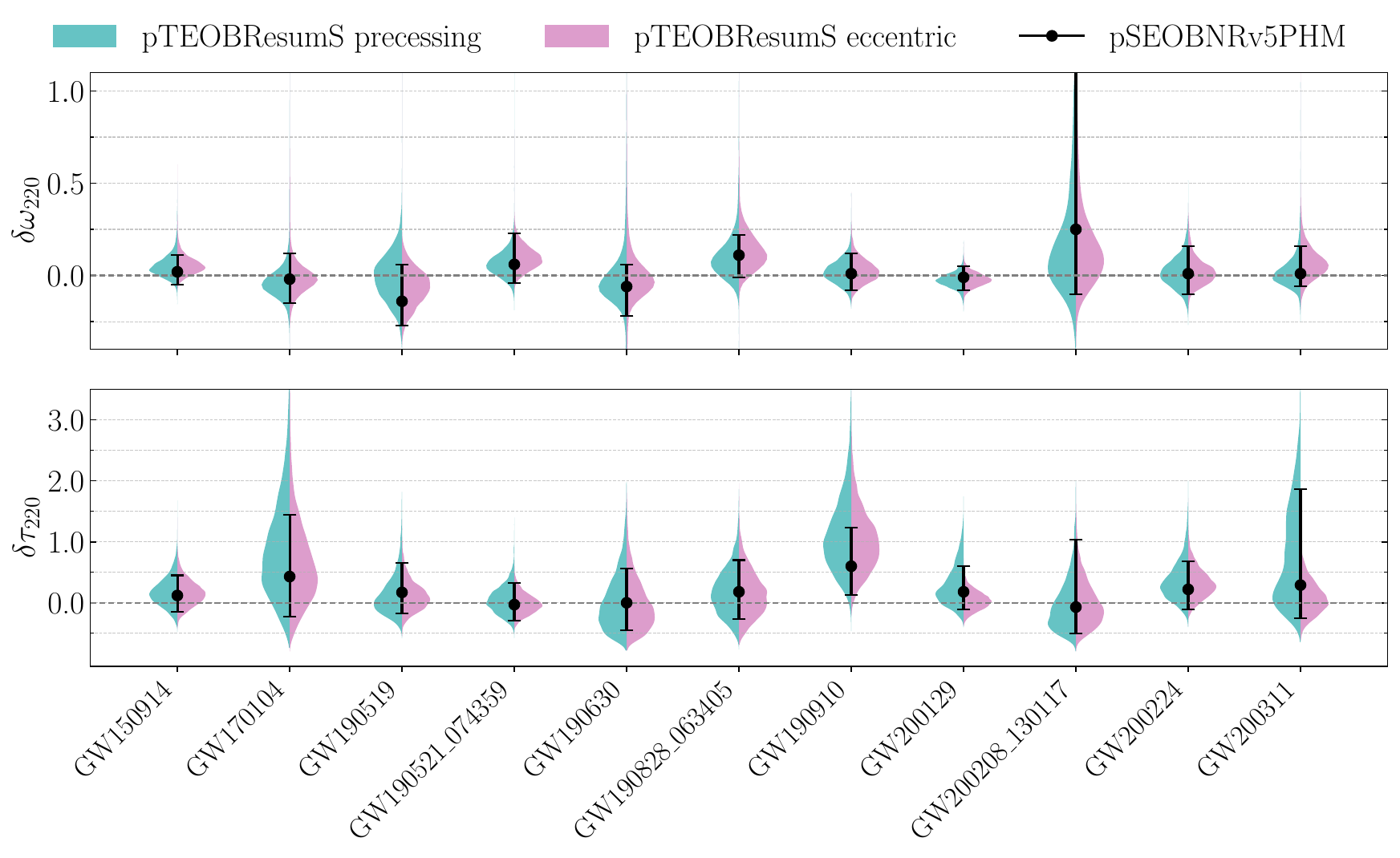}
	\caption{Posterior distributions of deviations in the \ac{qnm} mode parameters, $\delta \tau_{220}$ (top) and $\delta \omega_{220}$ (bottom), for the selected merger events. For each, half–violin plots show the marginalized posteriors obtained with the \TEOBB{} waveform model including only precession (left, green) or only eccentricity (right, pink). Black markers with error bars indicate median values and $90\%$ \acp{ci} inferred with \pseob, taken from~\cite{Pompili:2025cdc}. Horizontal dashed gray lines mark the \ac{gr} prediction of zero deviations.
    \label{fig:results_real_data}}
\end{figure*}

\begin{figure*}[t]
	\centering
    \includegraphics[width=0.47\textwidth]{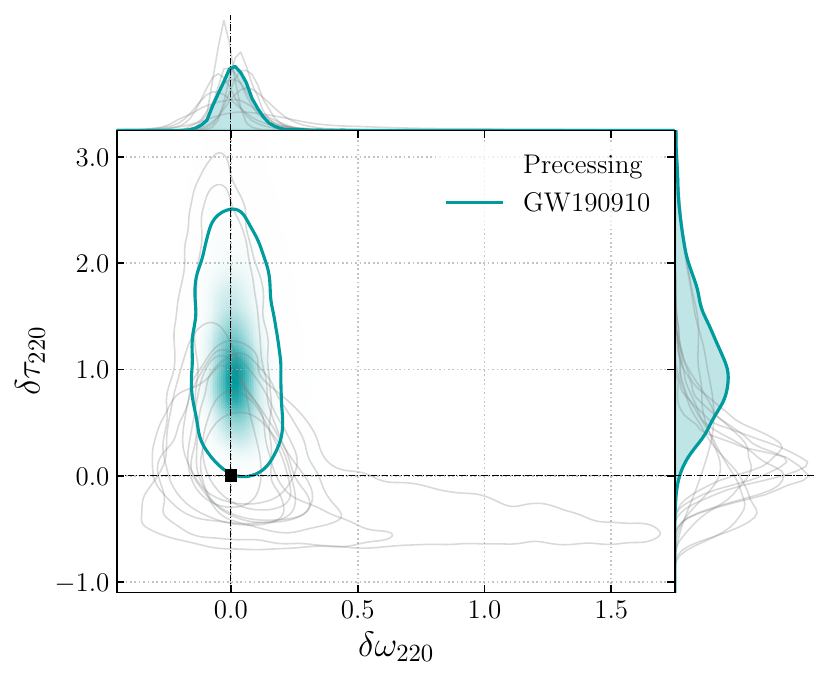}
    \includegraphics[width=0.47\textwidth]{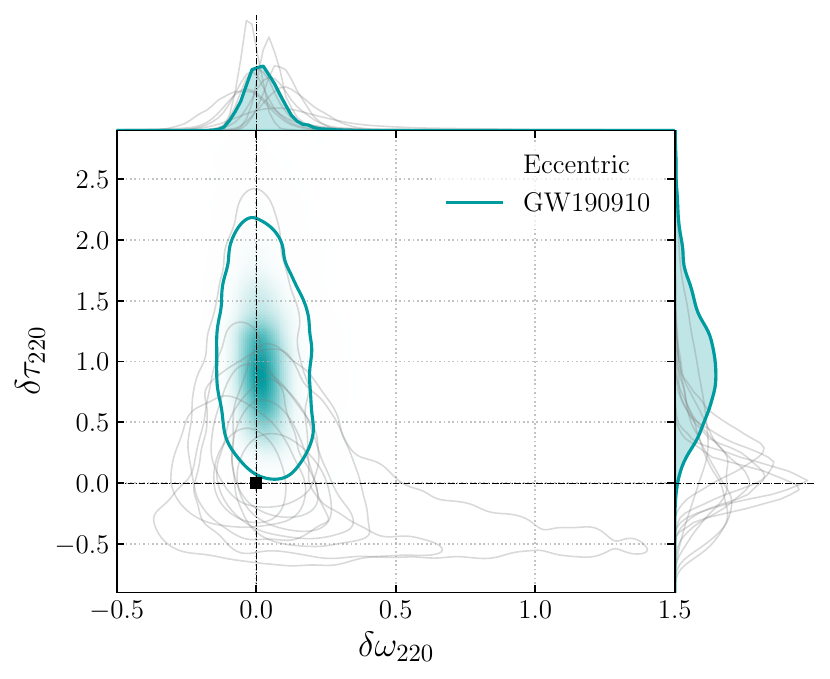}
	\caption{Posterior distributions for the \ac{qnm} frequency and damping time deviation parameters for all
    events analyzed in this work. \textit{Left:} results of \ac{pe} in the quasi-circular, precessing-spin case; \textit{right:}
    \ac{pe} results for the eccentric, spin-aligned case. 
    The black marker and dashed lines mark the \ac{gr} prediction; events for which the 90\%
    \ac{ci} of either parameter excludes 0 are highlighted in color. Contours in the two-dimensional plots correspond to the 90\% credible regions.
	\label{fig:corners}}
\end{figure*}

\begin{figure}[t]
	\centering
    \includegraphics[width=0.47\textwidth]{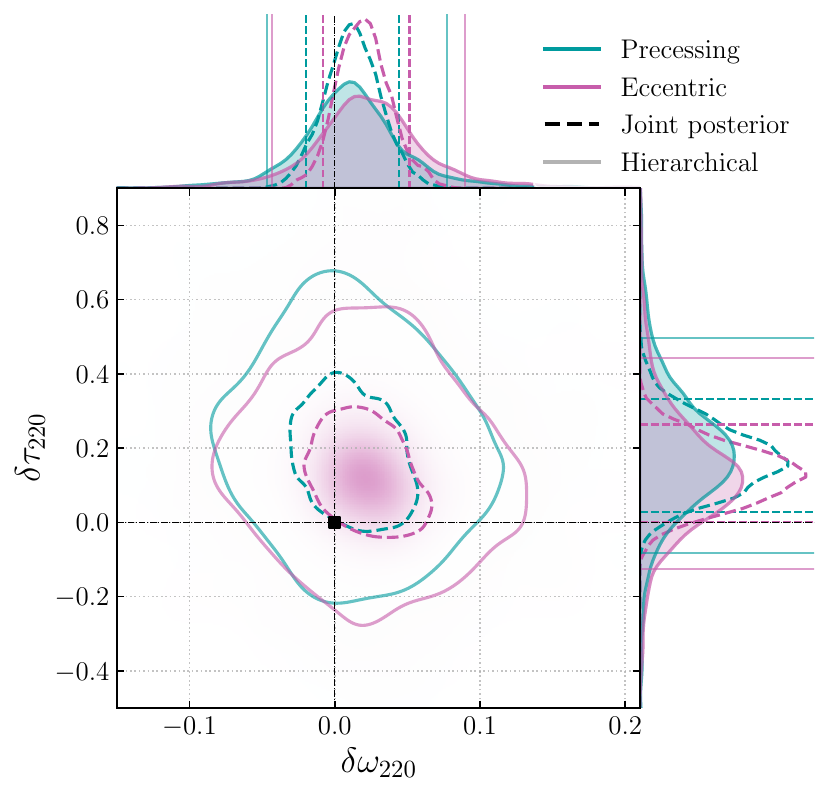}
	\caption{Combined posterior distributions for the \ac{qnm} frequency and damping time deviation parameters.
    We show results that combine either all of the
    eccentric (pink) or the precessing (green) analyses, multiplying the individual likelihoods (dashed)
    or via a hierarchical procedure (solid, lighter).
    The black marker and straight lines mark the \ac{gr} prediction; the straight lines in the
    1D plots and the contours in the 2D plot mark the 90\% credibility bounds.
    Joint posteriors exclude 0 at this credibility for the damping
    time deviation, which is biased towards positive values, though the hierarchical constraints are wider and \ac{gr}-consistent.
		\label{fig:combined_corner}}
\end{figure}

After validating our model, we apply \TEOBB\ to \Nevents\ \ac{bbh} signals from the \ac{gwtc}-3 catalog to search
for generic deviations from \ac{gr}. Following Ref.~\citep{LIGOScientific:2021sio, LIGOScientific:2025obp}
we select events with a \ac{far} $\leq 10^{-3}\,\mathrm{yr}^{-1}$ and a network \ac{snr}
greater than 8 in both the inspiral and post-inspiral phases, and test for deviations---under quasi-spherical and eccentric,
aligned-spin \ac{bbh} hypotheses---from the predicted fundamental \ac{qnm} frequency and damping time, $\delta\omega_{220}$ and $\delta\tau_{220}$.
For all events, we generate model waveforms including the (co-precessing) $(2,\pm 2)$, $(2,\pm 1)$, $(3,\pm 3)$, and $(4,\pm 4)$ modes.

\subsubsection*{Overview of results}

Fig.~\ref{fig:results_real_data} summarizes our findings.
The violin plots compare the \TEOBB\ posteriors in the precessing case (left, green) and the eccentric case (right, pink) for the selected events. For reference, the \pseob\ results are also shown as error bars, corresponding to the 90\% \acp{ci}.
Our findings overlap with the \pseob\ model's under both the precessing and eccentric hypotheses in almost every case.
Overall, the results of our two analyses are in good agreement with each other, with two notable exceptions in GW200129 and GW200311.
The case of GW200129 deserves additional discussion, which will be addressed below.
As for GW200311, Fig.~\ref{fig:results_real_data} highlights how the $\delta \tau_{220}$ posterior from the precessing analysis features much more support for positive values than the one from the eccentric run; simultaneously, the $\delta \omega_{220}$ eccentric distribution skews more towards a positive deviation, compared to the precessing one that is approximately symmetric about zero.
\citet{Pompili:2025cdc}~obtained similar results when comparing posteriors inferred with \pseob~and its nonprecessing counterpart \texttt{pSEOBNRv5HM}.
A more detailed analysis there revealed correlations between the deviation parameters and the luminosity distance, inclination, and effective precessing spin $\chi_{\rm p}$, with a secondary likelihood peak located at smaller $D_L$, higher $\iota$, and larger, positive $\delta \tau_{220}$.
Our complete \ac{pe} results in the precessing case closely match the same pattern, also finding the maximum likelihood point in the aforementioned secondary mode in the distribution, corroborating their conclusion that the combined effect of spin-precession and ringdown deviations allows the model to fit this signal well in a broader region of parameter space.

Across all cases, we find no events whose fundamental \ac{qnm} damping time and frequency are simultaneously incompatible with GR at 90\% credibility.
Small differences between precessing and eccentric reconstructions remain visible in the $\delta \tau_{220}$ and $\delta \omega_{220}$ distributions.
Fig.~\ref{fig:corners} shows the one- and two-dimensional posterior distributions for these parameters across all events, for both the precessing (left panel) and eccentric (right panel) cases, with significant deviations from zero highlighted.
Only in one instance, GW190910, do both analyses result in medians for one of the parameters, $\delta \tau_{220}$, that cleanly exclude zero at the 90\% credible level.

\subsubsection*{Combination of events}

Employing the two methods introduced in Sec.~\ref{sec:pe}, we combine the results of these analyses to derive
more meaningful constraints on potential deviations from \ac{gr}.
When using the hierarchical method, we adopt uniform priors for the means $\mu_{\delta \omega}, \mu_{\delta \tau}
\in [-0.8, 4]$ and the correlation $\rho_{\delta \omega, \delta \tau} \in [-1,1]$, and half-normal priors peaking
at 0 for the variances $\sigma_{\delta \omega}, \sigma_{\delta \tau} \geq 0$.
Fig.~\ref{fig:combined_corner} shows the combined posteriors for the deviation parameters, for both the eccentric (pink) and precessing (green) runs;
joint posteriors obtained multiplying individual likelihoods are drawn in dashed lines, while the population-marginalized
distributions from the hierarchical combination of the events are in solid lines with lighter colors.
Medians and 90$\%$ \acp{ci} for both methods are included in Tab.~\ref{tab:real_results};
the results for the hyperparameters of the underlying Gaussian distribution are in Tab.~\ref{tab:hierarchical}.
The two methods deliver close medians for both deviations, the main difference being the larger uncertainty in the hierarchically
combined results.

In all cases, the inferred bounds on the frequency deviation are compatible with zero at 90\% credibility; this is true
also for the mean value of the Gaussian in the hierarchical case, while a strict upper bound is placed on the variance.
Interestingly, the eccentric analyses yield a stronger preference for positive values of $\delta \omega_{220}$ than the precessing ones.
This could be partly caused by slight differences in the intrinsic parameters inferred under the two hypotheses, in combination with systematic effects due to different details of the waveform model in the two cases.
One such effect is the prediction of a higher remnant spin when including spin precession, as the norm of $\bm{\abhf}$ receives a contribution from the in-plane spin components~\cite{Pratten:2020ceb}; the remnant mass model is instead the same as the spin-aligned case.
This in turn similarly affects the baseline values of the \ac{qnm} frequencies.
Notably, because of the form of the spin priors we adopt, even when the data do not strongly favor a precessing interpretation, \ac{pe} results still show significant support for non-zero in-plane spins.
In some eccentric analyses, their effect could thus be mimicked by a positive shift in $\delta \omega_{220}$, consistently with the observed trend.

The joint posteriors for $\delta \tau_{220}$ are more decisively leaning towards positive values, especially in the precessing case; the multiplied posteriors and the hierarchical mean value exclude \ac{gr} at 90\% credibility.
The population-marginalized distribution for $\delta \tau_{220}$ still places 0 within its 90\% \ac{ci} though.
Such preference for longer damping times is not unheard of, as it already emerged in the results of the \texttt{pSEOBNR} tests~\cite{LIGOScientific:2021sio,Pompili:2025cdc}.
Like in those cases, we do not see this as sufficient evidence to back the claim of a violation of \ac{gr}, as a number of unaccounted for systematic effects could be at play. 
For instance, in~\cite{LIGOScientific:2020tif,Ghosh:2021mrv} it was found that improper noise modeling at the time of a
detection can lead to an overestimation of the \ac{qnm} damping time.
Waveform systematics also likely have a hand in this: Ref.~\cite{LIGOScientific:2025cmm}, while investigating
apparent \ac{gr} violations in $\tau_{220}$ for the GW230814 event, recovered similar biases in analyses of synthetic signals,
including from \ac{nr}, in zero noise with \pseob. In addition, the limited size of
the event sample means that we cannot exclude the possibility of this being a statistical fluctuation~\cite{Pacilio:2023uef};
more results from newer detections that meet the selection criteria are needed to fully evaluate the significance of these findings.

Our joint posteriors in the precessing case are in good agreement with those obtained by~\citet{Pompili:2025cdc}.
As for the hierarchically combined results, we find generally wider constraints on all of the hyperparameters,
which carry over to those for the deviations themselves.
\citet{Pompili:2025cdc} carry out the hierarchical combination separately for $\delta \omega_{220}$ and $\delta \tau_{220}$,
while here we consider a two-dimensional underlying Gaussian distribution for the pair.
The generalization to a bivariate population introduces additional freedom via the correlation between deviations;
this alone may not explain the broader constraints we obtain however, as analyses performed separately for each deviation yield consistent conclusions.
We note, however, that our results are structurally similar to those of~\cite{LIGOScientific:2026wpt}, albeit with wider posteriors owing to the smaller
sample of events we consider.

\subsubsection*{Evidence for eccentricity among the analyzed events}

\begin{figure}[t]
	\centering
	\includegraphics[width=0.47\textwidth]{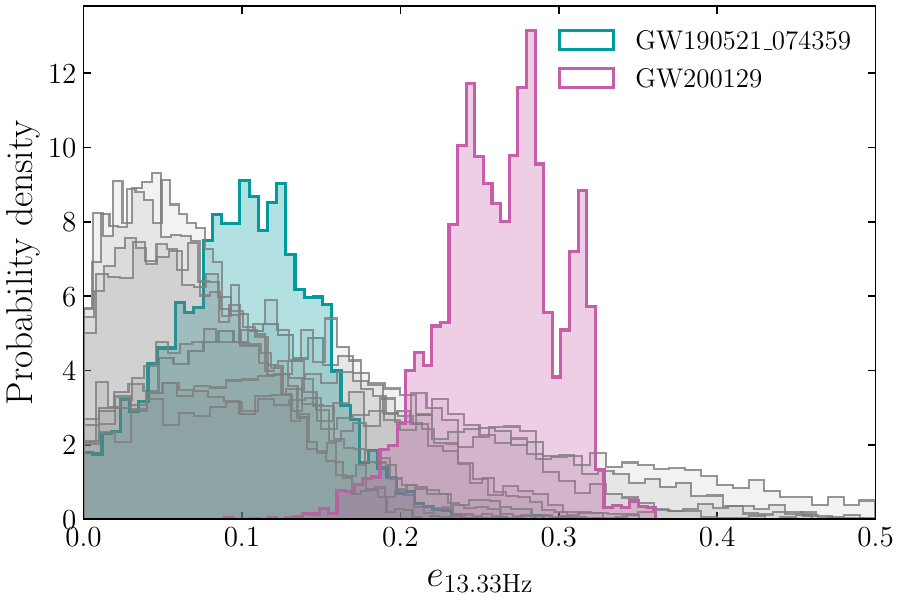}
	\caption{1D posterior distributions for the eccentricity at 13.33 Hz for the events of Tab.~\ref{tab:real_results}.
    We highlight two events with particularly interesting results. In green is GW190521\_074359, for which we find $e_{\rm 13.33 Hz} = 0.1^{+0.07}_{-0.08}$. 
    In pink, we show GW200129, discussed in more depth in the main text, for which the analysis finds clear evidence of
    eccentricity, with $e_{\rm 13.33 Hz} = 0.26^{+0.05}_{-0.07}$ (median and 90\% \ac{ci}), though this result may be affected by data quality
    issues.
    \label{fig:ecc_posts_real}}
\end{figure}

The majority of the analyzed events are consistent with quasi-circular binaries.
This is illustrated in Fig.~\ref{fig:ecc_posts_real}, which displays the posterior probability distributions for
the eccentricity $e_{\rm 13.33 Hz}$ measured at the waveform starting frequency of 13.33 Hz for our analyses.
The only events that show noticeable support for non-zero eccentricity are GW190521\_074359 and GW200129, highlighted in green and pink, respectively. 
In the case of GW190521\_074359, the posterior distribution peaks at $e_{\rm 13.33 Hz} \simeq 0.1$, with 90\% of samples above $0.04$.
The symmetric 90\% \ac{ci} about the median is $e_{\rm 13.33 Hz} = 0.1^{+0.07}_{-0.08}$.
The eccentric model yields a slightly higher maximum likelihood, $\ln \mathcal{L}_{\rm ecc}^{\rm max}/\mathcal{L}_{\rm prec}^{\rm max} \simeq 0.99$, but the log \ac{bf} between the two hypotheses is inconclusive, with $\log_{10}\bayes{\rm ecc}{\rm prec}\simeq 0.075$.
The Savage-Dickey ratio contrasting the beyond-\ac{gr} and \ac{gr} hypotheses is strongly in favor of \ac{gr} in both cases. Overall, our analysis does not reveal significant evidence in favor of an eccentric interpretation for this event; similar results were already found in Ref.~\cite{Gupte:2024jfe}.


\begin{figure}
    \centering
    \includegraphics[width=0.47\textwidth]{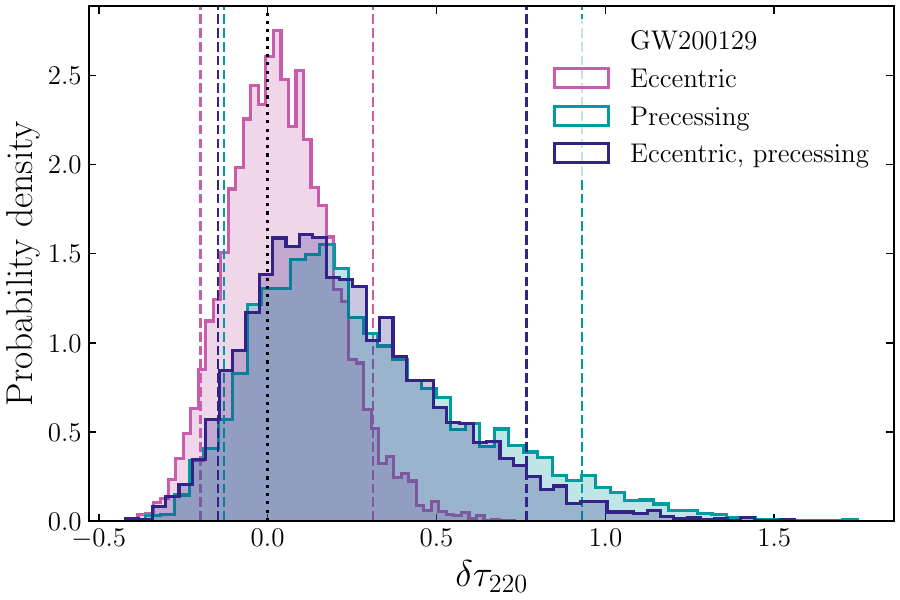}
    \caption{Comparison of the posterior probability distributions for the \ac{qnm} damping time deviation $\delta \tau_{220}$ for the GW200129 event
    under the eccentric (pink) and precessing (green) hypotheses. While both are compatible with the \ac{gr} prediction of zero (black dotted line),
    the eccentric analysis yields a tighter constraint, with a more symmetric distribution about zero.
    \label{fig:gw200129_tau}}
\end{figure}

GW200129 (pink in Fig.~\ref{fig:ecc_posts_real}) stands out in our analysis for its strong preference toward an eccentric interpretation. 
Previous studies~\cite{Gupte:2024jfe,Planas:2025jny} have similarly identified possible eccentricity, and despite differing parameter definitions,
our findings are consistent with a moderately eccentric ($e_{\rm 10\,Hz} \sim 0.27$), near--equal-mass binary with low effective spin. 
The event has also been linked to spin precession~\cite{Hannam:2021pit,CalderonBustillo:2024akj} and to tentative deviations from \ac{gr}~\cite{Maggio:2022hre}.
However, it coincided with excess noise from an electro-optic modulator, that was subsequently linearly subtracted in \ac{lvk} analyses~\cite{KAGRA:2021vkt,Davis:2022ird}\footnote{For the Livingston detector, we use the 
\texttt{L1:DCS-CALIB\_STRAIN\_CLEAN\_SUB60HZ\_C01\_P1800169\_v4} channel from the de-glitched frame file of~\cite{ligo_scientific_collaboration_and_virgo_2021_5546680}.}. 
Data-quality studies~\cite{Payne:2022spz,Macas:2023wiw,Gupte:2024jfe,Planas:2025jny} indicate that the apparent evidence for precession or eccentricity is driven primarily by the Livingston data 
and depends on the details of the glitch subtraction.

From our analysis, we infer $e_{\rm 13.33\,Hz} = 0.26^{+0.05}_{-0.07}$ at an orbit-averaged frequency of 13.33\,Hz; the mean anomaly posterior is uninformative.
Using \texttt{gw\_eccentricity}, this maps to $e_{\rm 10 Hz}^{\rm GW} = 0.34^{+0.06}_{-0.08}$.
Interestingly, our eccentricity posterior displays a multi-modal structure.
This feature is retained in subsequent reanalyses (see below, and App.~\ref{app:gw200129}), and has been observed before in \ac{pe} of \ac{gw} events with eccentric waveforms~\cite{Kacanja:2025kpr}.
Although a detailed assessment of glitch-mitigation effects lies beyond our scope, the data favor an aligned-spin eccentric binary over a precessing non-eccentric one. 
The Bayes factor, $\log_{10}{\cal B}^{\rm ecc}_{\rm prec} = 4.5$, and the maximum likelihood ratio, $\ln(\mathcal{L}_{\rm ecc}^{\rm max}/\mathcal{L}_{\rm prec}^{\rm max}) = 4.1$, 
both strongly support the eccentric hypothesis. 
In both scenarios, the posteriors for $\delta \omega_{220}$ and $\delta \tau_{220}$ remain consistent with zero, but the eccentric model yields tighter constraints and improved agreement with \ac{gr} (Fig.~\ref{fig:gw200129_tau}, Tab.~\ref{tab:real_results}); for only one other event (GW200311, discussed above) do we find such a marked difference between the two results.
A similar contrast was observed in~\citet{Pompili:2025cdc} between quasi-circular precessing and non-precessing analyses using \pseob~and \texttt{pSEOBNRv4HM}, suggesting that neglect of precession explains the reduced support for positive $\delta \tau_{220}$ in the eccentric posterior.
Notably, the authors there however also found the spin-precessing hypothesis to be strongly favored over the quasi-circular, spin-aligned one.

To delve deeper into this particularly interesting event, we re-analyze GW200129 while considering simultaneously eccentricity, spin-precession and \ac{qnm} deviations.
Fig.~\ref{fig:gw200129_tau} compares the $\delta \tau_{220}$ posterior distributions for the three analyses; more detailed results can be found in App.~\ref{app:gw200129}.
The damping time constraints obtained in the precessing analyses display the same tail toward positive $\delta \tau_{220}$, with the eccentric and precessing run favoring only slightly lower values.
Aside from that parameter though, the complete analysis returns lower support for precessing spins and in general much more closely matches the eccentricity-only run, further corroborating an eccentric interpretation for GW200129.

\section{Conclusions}
\label{sec:conclusions}
This work introduces \TEOBB, a new waveform model designed for \ac{gw}-based parametrized tests of \ac{gr} in the plunge, merger, and ringdown phases. 
The model builds on the eccentric, precessing \ac{eob} waveform model \dali, allowing it to identify beyond-\ac{gr} effects even in eccentric \ac{bbh} signals. In particular, \TEOBB\ enables a broad range of beyond-\ac{gr} parametrizations, including:
\begin{inparaenum}[(1)]
    \item \textit{Mode-by-mode QNM spectrum deviations}, through modifications to the fundamental frequency $\omega_{\ell m 0}$ or damping time $\tau_{\ell m 0}$;
    \item \textit{Merger--ringdown matching adjustments}, via changes in the mode amplitudes $A_{\ell m}^\mathrm{peak}$ or matching frequencies $\omega_{\ell m}^\mathrm{peak}$;
    \item \textit{Late-inspiral dynamical perturbations}, by varying high-order \ac{pn} calibration parameters $\delta a_{6}^c$ and $\delta c_{\mathrm{N^{3}LO}}$, which affect the \ac{eob} dynamics and impact all modes;
    \item \textit{Deviations in \ac{nr}-informed remnant property predictions}, used in the post-merger modeling of \ac{eob} waveforms.
\end{inparaenum}

We validated \TEOBB\ through a series of simulation studies using synthetic signals. These demonstrated that the model is self-consistent, correctly recovering signal parameters, except when the inference intentionally omitted part of the parameter set. The tests revealed correlations between measured deviations and estimates of the system's mass and eccentricity, and showed that neglecting eccentricity can bias inference.
We next applied \TEOBB\ to numerically simulated \ac{bbh}, boson star and \ac{dcs}-corrected merger waveforms and found that, in all instances, \TEOBB\ correctly identified them.

Finally, we analyzed \Nevents\ \ac{bbh} events from the GWTC-3 catalog~\citep{KAGRA:2021vkt, LIGOScientific:2021sio} under two competing hypotheses ---one including precession, the other eccentricity--- and sampled deviations in the fundamental \ac{qnm} complex frequency.
The two approaches yielded mutually consistent results, with those of the non-eccentric analyses agreeing with previous findings from the \pseob\ model~\citep{Pompili:2025cdc}.
No event provided compelling evidence for deviations from \ac{gr}, with Bayes factors favoring the null hypothesis even when the posteriors supported beyond-\ac{gr} effects.
Combining events within either scenario led to tighter constraints on the deviation parameters.
However, the joint posterior on the \ac{qnm} damping time, obtained multiplying the single posterior distributions under the assumption that the true value of $\delta \tau_{220}$ is common to all events, from both sets of analyses excluded the \ac{gr} value at 90\% credibility, consistent with earlier reports in the precessing case.
A more agnostic hierarchical combination of the posteriors, assuming the deviations to be drawn from an underlying
distribution characterized by its first two moments, yielded broader constraints compatible with 0 for both parameters, although the \ac{gr} prediction still lay outside the 90\% \ac{ci} for the mean value of $\delta \tau_{220}$ inferred from the precessing analyses.
These results reflect a preference found in many individual cases for larger damping times, and is likely owing to noise and/or waveform systematics that will be investigated further in future work.
Given the small sample size and possible systematics, we interpret our findings as indicative rather than conclusive, and anticipate stronger constraints from high-\ac{snr} events in upcoming catalogs such as GWTC-4~\citep{LIGOScientific:2025slb}.

Looking ahead, the most direct route to improving \TEOBB\ and meeting the increasing requirements for waveform accuracy is to improve \dali.
A crucial upgrade will be to incorporate additional \ac{nr} information into the plunge-merger–ringdown sector of \dali, particularly from eccentric~\cite{Carullo:2023kvj} and spin-precessing simulations.
These will reduce the risk of systematic biases in recovering deviation parameters arising from incomplete or inadequate waveform modeling.
Additionally, expanding the model’s calibration set to cover a broader region of the astrophysical parameter space, including high-mass-ratio and high-spin configurations, as well as more generic orbital morphologies, will further strengthen its robustness. 

\section*{Acknowledgments}

The authors are grateful to S.~Albanesi for useful discussions and comments on the manuscript.
The authors also thank L.~Pompili, T.~Islam, and S.~Mukherjee for their helpful comments.
DC thanks L.~G.~Jade for their constant support throughout this project.
RG and KC are also grateful to A.~Choi, K.~Eun-jae, A.~Nuna, and R.~Ami for inspiration during
the early stages of this work.
JL acknowledges support from the Italian Ministry of University and Research (MUR) via the PRIN 2022ZHYFA2, 
{\it GRavitational wavEform models for coalescing compAct binaries with eccenTricity} (GREAT).
KC acknowldeges support from NSF grants AST-2307147, PHY-2207638 and  PHY-2308886.
RG acknowledges support from NSF Grant PHY-2020275
(Network for Neutrinos, Nuclear Astrophysics, and Symmetries (N3AS)).
This material is based upon work supported by NSF's LIGO Laboratory which is a major facility
fully funded by the National Science Foundation.
The authors are grateful for the computational resources provided by the LIGO Laboratory and supported by
National Science Foundation Grants PHY-0757058 and PHY-0823459.
We are grateful for computational resources provided by the Leonard E Parker
Center for Gravitation, Cosmology and Astrophysics at the University of
Wisconsin-Milwaukee.

\paragraph*{Software} Analyses in this work made use of \texttt{NumPy}~\cite{Harris:2020xlr}, \texttt{SciPy}~\cite{Virtanen:2019joe}, 
\texttt{Bilby}~\cite{Ashton:2018jfp}, \texttt{Bilby\_pipe}~\cite{Romero-Shaw:2020owr}, \texttt{Dynesty}~\cite{Speagle:2019ivv},
\texttt{LALSuite}~\cite{lalsuite,swiglal}, \texttt{pyCBC}~\cite{alex_nitz_2024_10473621}, \texttt{gwPy}~\cite{Macleod:2021goi},
\texttt{gw\_eccentricity}~\cite{Shaikh:2023ypz}, and \texttt{pocoMC}~\cite{Karamanis:2022alw,Karamanis:2022ksp}.
Figures in this work were produced using \texttt{matplotlib}~\cite{Hunter:2007}, \texttt{palettable}~\cite{palettable},
\texttt{PESummary}~\cite{Hoy:2020vys}, and \texttt{corner}~\cite{corner}.

\paragraph*{Data availability} Data supporting this work is publicly available at~\cite{data_repo}.

\appendix

\section{Additional analysis of GW200129}
\label{app:gw200129}
\begin{table}[t]
	\centering 
    \caption{Medians and boundaries of the 90\% \acp{ci} for the eccentricity at 13.33 Hz, the effective precessing spin, and the \ac{qnm} deviations for our three analyses of GW200129.
    The last two lines list \acp{bf} comparing the eccentric runs with the quasi-circular, precessing one, and each of the three with the corresponding null hypothesis of zero deviation.
    \label{tab:gw200129_results}}
	\begin{tabularx}{0.8\linewidth}{*{1}{C} c c c}
		\toprule
		 & \textbf{Precessing} & \textbf{Eccentric} & \textbf{Ecc., prec.} \\ 
        \midrule
        \bmath{$e_{13.33 \rm{Hz}}$}   &  & $0.27^{+0.05}_{-0.06}$ & $0.26^{+0.05}_{-0.06}$ \\[0.5em]
        
        \bmath{$\chi_{\rm p}$}   & $0.52^{+0.33}_{-0.32}$ & & $0.36^{+0.32}_{-0.24}$ \\[0.5em]
        
        \bmath{$\delta \tau_{220}$}   & $0.23^{+0.71}_{-0.35}$ & $0.04^{+0.28}_{-0.23}$ & $0.20^{+0.57}_{-0.34}$\\[0.5em]
        
        \bmath{$\delta \omega_{220}$} & $-0.027^{+0.066}_{-0.067}$ & $-0.023^{+0.064}_{-0.060}$ & $-0.042^{+0.068}_{-0.055}$\\
        \midrule
        \bmath{$\log \mathcal{B}^X_{\rm Prec.}$} & & 4.48 & 4.56 \\
        \bmath{$\log \mathcal{B}_{\rm GR}$} & 1.96 & 2.21 & 1.94 \\
		\bottomrule
	\end{tabularx}
\end{table}

\begin{figure*}[t]
    \centering
    \includegraphics[width=0.47\textwidth]{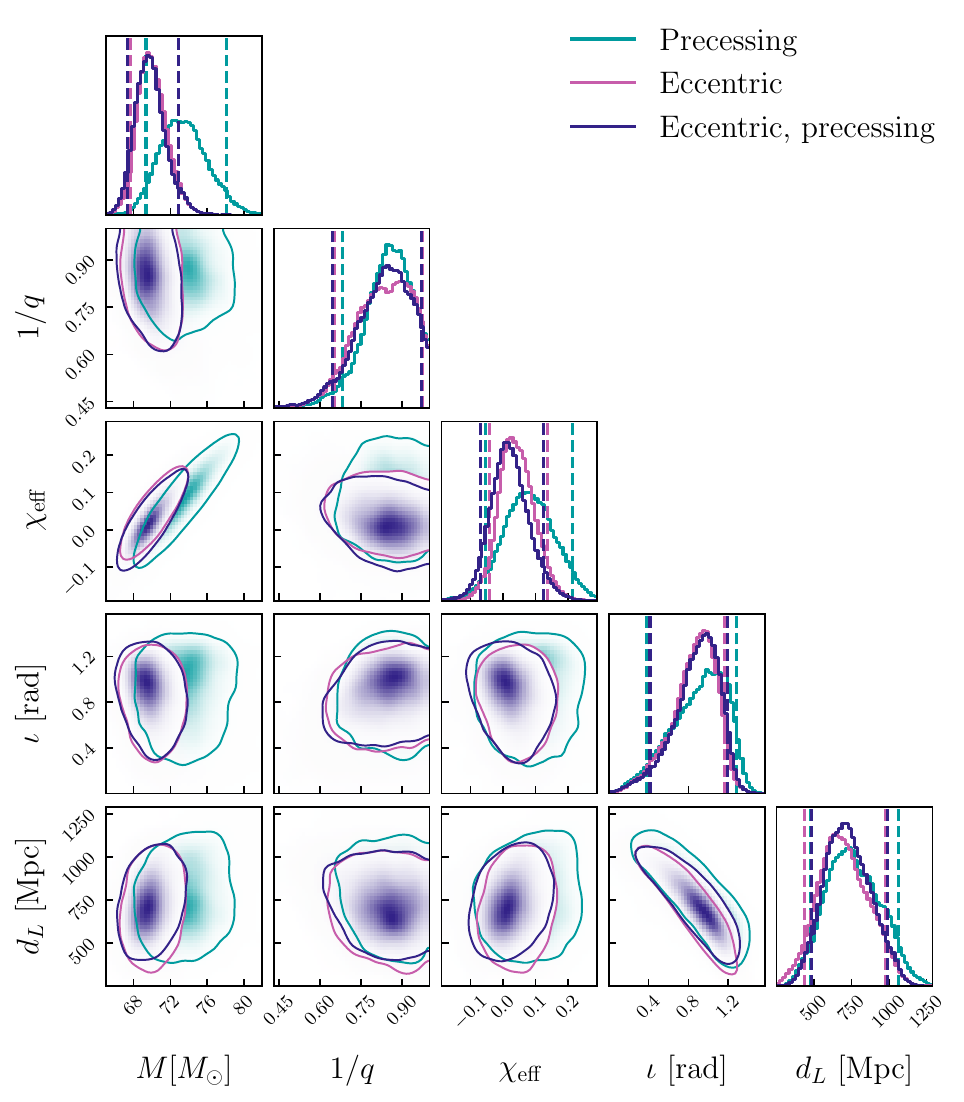}
    \includegraphics[width=0.395\textwidth]{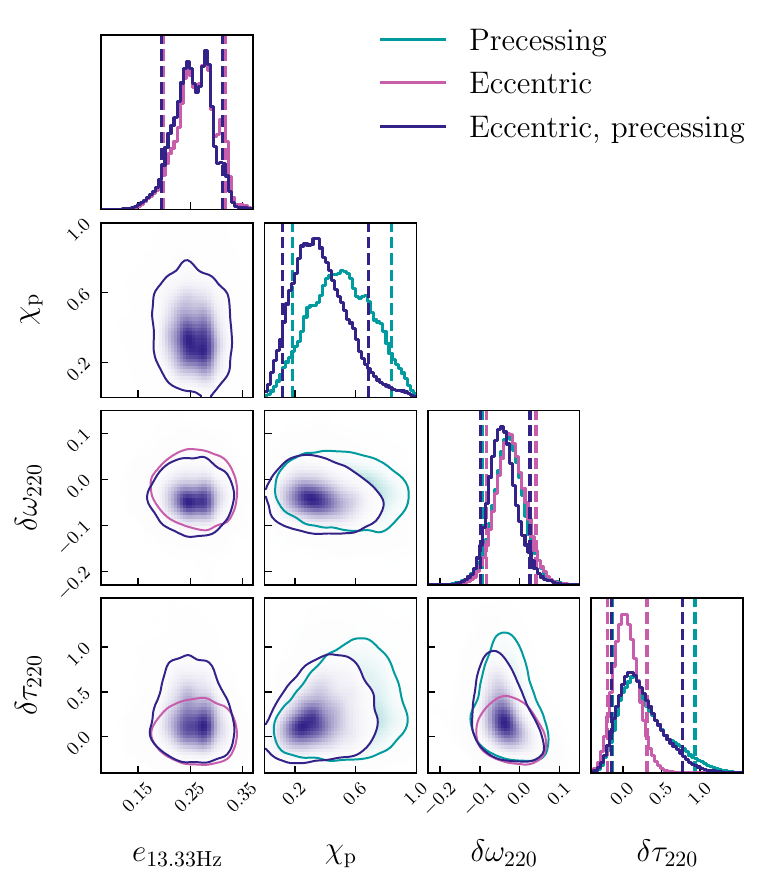}
    \caption{Corner plots displaying our results for the GW200129 event across three hypotheses: quasi-circular, precessing (green); eccentric, non-precessing (pink); and eccentric, precessing (purple).
    On the left, the main intrinsic and extrinsic parameters common to all three analyses; on the right, the eccentricity, $\chi_{\rm p}$, and deviation parameters.
    Vertical lines and solid boundaries mark 90\% credibility regions.
    The eccentric analyses return consistent results for all parameters but $\delta \tau_{220}$, while neglecting eccentricity leads to significantly different inference in some cases.
    \label{fig:gw200129_corners}}
\end{figure*}

\begin{figure}[t]
    \centering
    \includegraphics[width=0.47\textwidth]{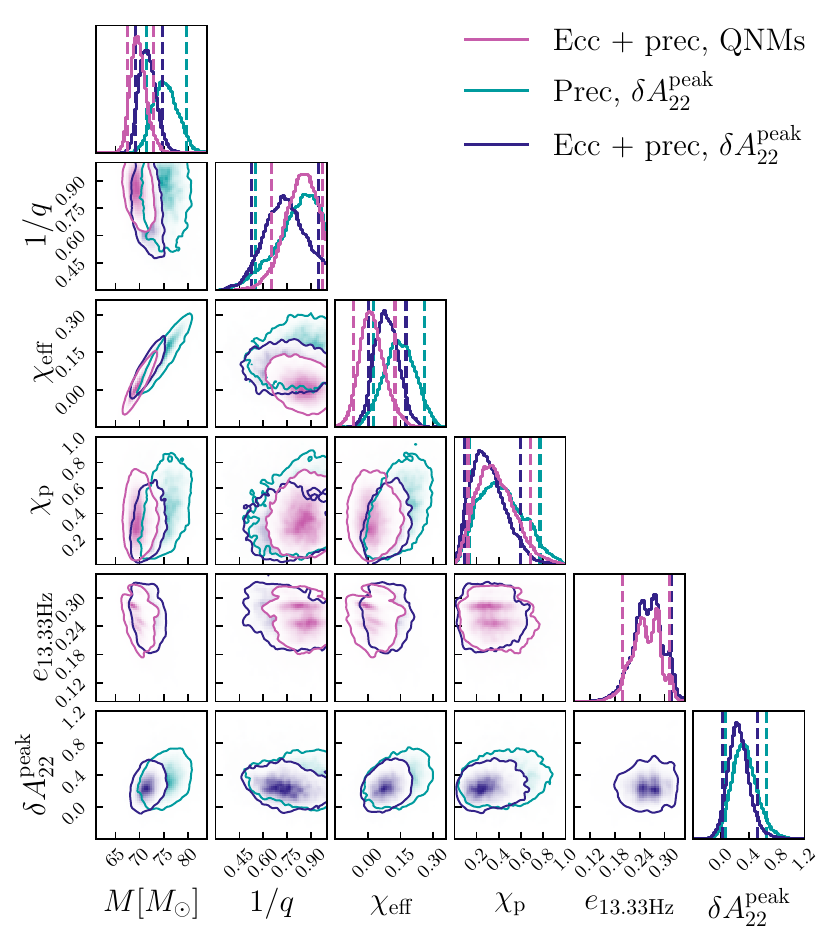}
    \caption{\ac{pe} results for GW200129 analyzed while sampling on the peak time amplitude deviation $\delta A_{22}^{\rm peak}$, in the \ac{qc}, precessing (green) and eccentric, precessing (purple) scenarios.
    We also include eccentric, precessing results found considering \ac{qnm} deviations for reference.
    Vertical dashed lines in the 1D marginalized posteriors mark 90\% \acp{ci}, as do the contours in the 2D distributions.
    \label{fig:gw200129_deltaA}}
\end{figure}

In this Appendix we share additional results regarding the GW200129 event.
Fig.~\ref{fig:gw200129_corners} displays 1D and 2D posterior distributions for the main \ac{bbh} parameters obtained in our three analyses (eccentric, non-precessing (pink); quasi-circular, precessing (green); and eccentric, precessing (purple)), with contours and vertical dashed lines marking 90\% \acp{ci}.
Table~\ref{tab:gw200129_results} lists a few key results, and \acp{bf} comparing our runs with each other and with the null, \ac{gr} hypothesis.
Results from the eccentric analyses closely align for almost all parameters, while those from the quasi-circular, precessing one show at times significant deviations.
Particolarly notable cases are the total mass and $\chi_{\rm eff}$, as well as $\chi_{\rm p}$, with the run incorporating both eccentricity and precession returning a preference for lower in-plane spins.
The addition of precession to the non-circular analysis only minimally shifts the posterior for $e_{13.33 {\rm Hz}}$.
All of these facts seem to signal, as mentioned in the main text, that eccentricity, rather than spin-precession, is the key physical feature necessary to accurately model this \ac{gw} event.
We remark though that all of these conclusions should be carefully evaluated taking into account the details of the glitch mitigation procedure applied to Livingston data.
See Sec.~\ref{sec:events} for a discussion of the deviation parameter posteriors.

In a parametrized analysis of this event, \citet{Maggio:2022hre} found evidence for a violation of \ac{gr} in the peak amplitude of the $(2,2)$ mode of the waveform.
Sampling on $\delta \omega$ and $\delta A$, equivalent to our $\delta \omega_{22}^{\rm peak}$ and $\delta A_{22}^{\rm peak}$, they found the former to be consistent with \ac{gr}, while the $\delta A$ posterior excluded 0 at 90\% credibility: $\delta A = 0.44^{+0.38}_{-0.28}$.
Notably, this result was found using the nonprecessing \texttt{pSEOBNRv4HM} model.
Supported by a series of injection-recovery studies, the authors attributed this apparent deviation to the combination of lack of modeling of spin-precession, and the LIGO Livingston glitch, whose effect on inference has been tied to evidence of precession for this event.
We attempt here to reproduce this result with \TEOBB.
We analyze the event under the quasi-circular, precessing and the eccentric, precessing hypotheses, each time sampling on $\delta A_{22}^{\rm peak}$ in addition to the standard \ac{bbh} parameters.
We adopt for $\delta A_{22}^{\rm peak}$ a uniform prior in $[-0.8, 2]$.
Fig.~\ref{fig:gw200129_deltaA} shares the results: the \TEOBB~analysis also recovers a posterior for the peak amplitude deviation that is inconsistent with \ac{gr} at 90\% credibility.
We find $\delta A_{22}^{\rm peak} = 0.34^{+0.32}_{-0.27}$ when including only spin-precession, and $\delta A_{22}^{\rm peak} = 0.25^{+0.27}_{-0.22}$ when also considering orbital eccentricity.
There is some overlap between our results and~\citet{Maggio:2022hre}, although we find slightly lower values and tighter constraints on the deviation.
While there are differences between our analyses' setup and theirs, it is significant that, even while accounting for both orbital eccentricity and spin-precession, both of which where neglected by the \texttt{pSEOBNRv4HM} analysis, we still recover an apparent deviation from \ac{gr}.
Residual modeling systematics could still be at the root of this -- we do not carry out a follow-up injection campaign to further delve into the matter -- but these results otherwise underscore the importance of properly taking data quality into account in the interpretation of GW200129.

\section{Detailed results for the \ac{dcs} injections}
\label{app:dcs}

\begin{figure}[t]
    \centering
    \includegraphics[width=0.47\textwidth]{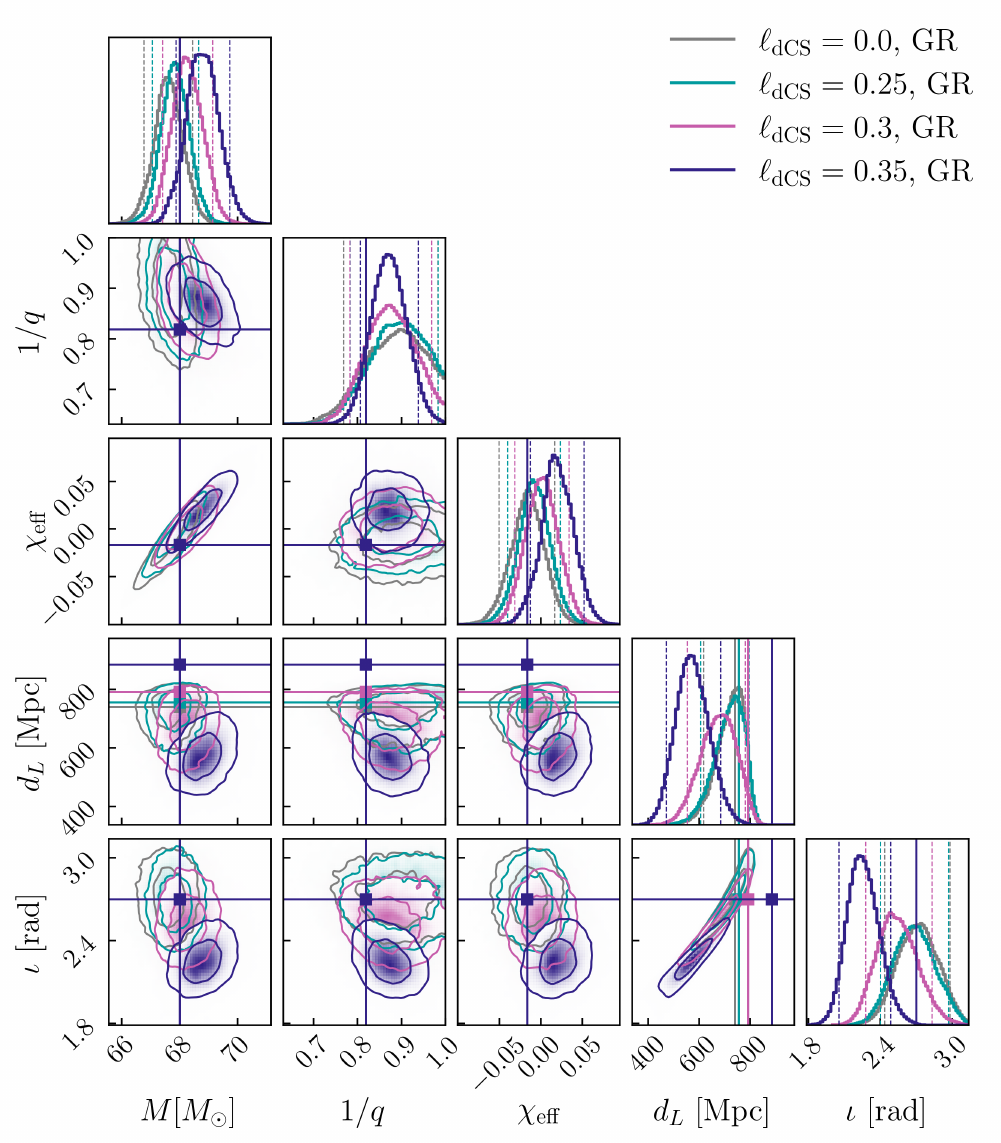}
    \caption{Analysis of a GW150914-like \ac{nr} simulation and three signals built on it to incorporate \ac{dcs} corrections to \ac{gr}. Here we sample only on the standard \ac{bbh} parameters.
    We show one- and two-dimensional posterior distributions, with expected values marked by squares and straight lines.
    A \ac{gr}-consistent description becomes a progressively worse fit for the data as $\ell_{\rm dCS}$ grows, with large biases appearing in the parameter estimates.
    \label{fig:dcs_gr_corner}}
\end{figure}

\begin{figure}[t]
    \centering
    \includegraphics[width=0.47\textwidth]{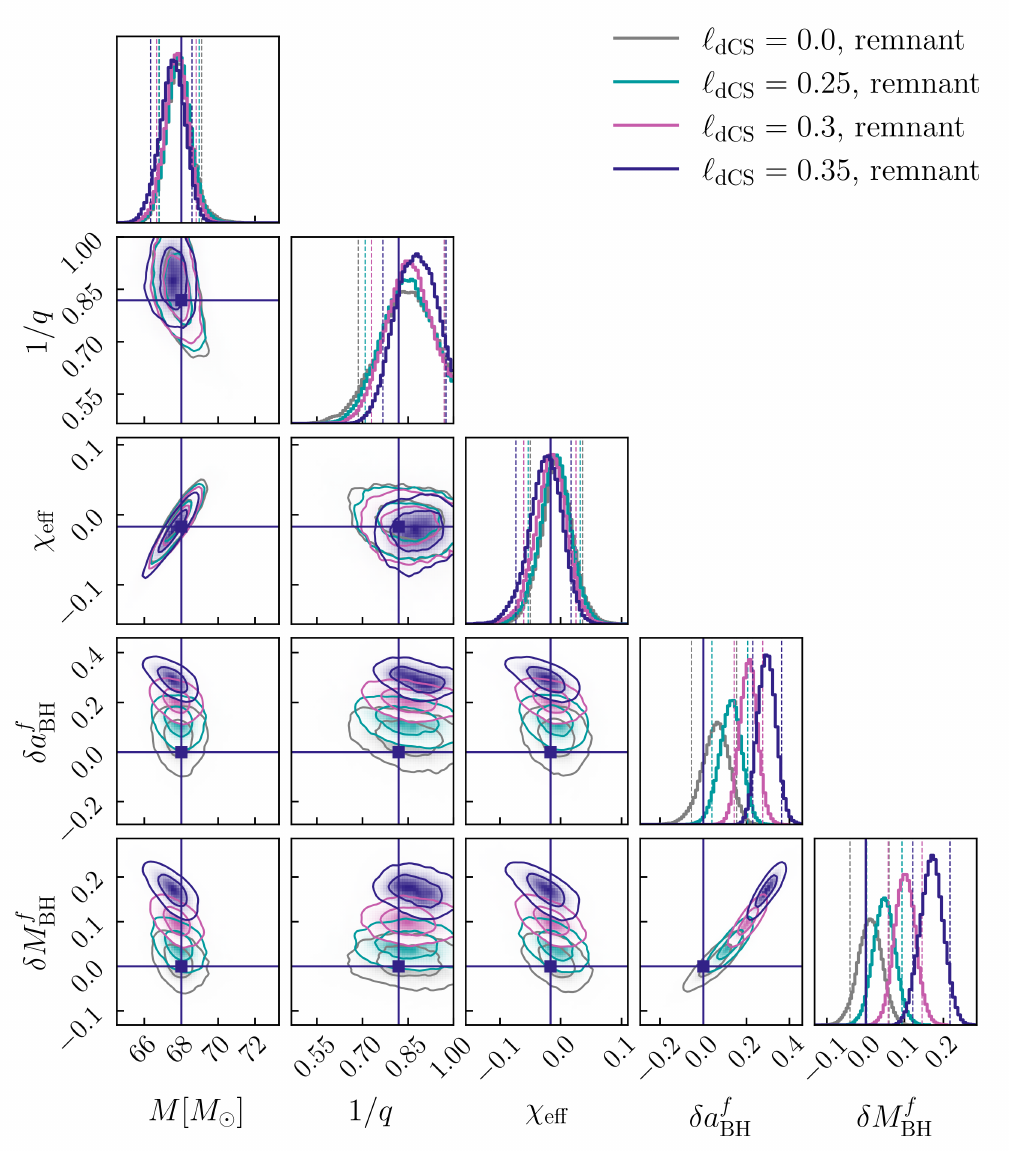}
    \caption{Analysis of a GW150914-like \ac{nr} simulation and three signals built on it to incorporate \ac{dcs} corrections to \ac{gr}, sampling on deviations from the remnant's mass and spin.
    We show one- and two-dimensional posterior distributions, with expected values marked by squares and straight lines.
    Most intrinsic parameters are accurately recovered for all runs here, while posteriors for the \ac{gr} deviations quickly move away from 0.
    \label{fig:dcs_rmn_corner}}
\end{figure}

\begin{figure}
    \centering
    \includegraphics[width=0.47\textwidth]{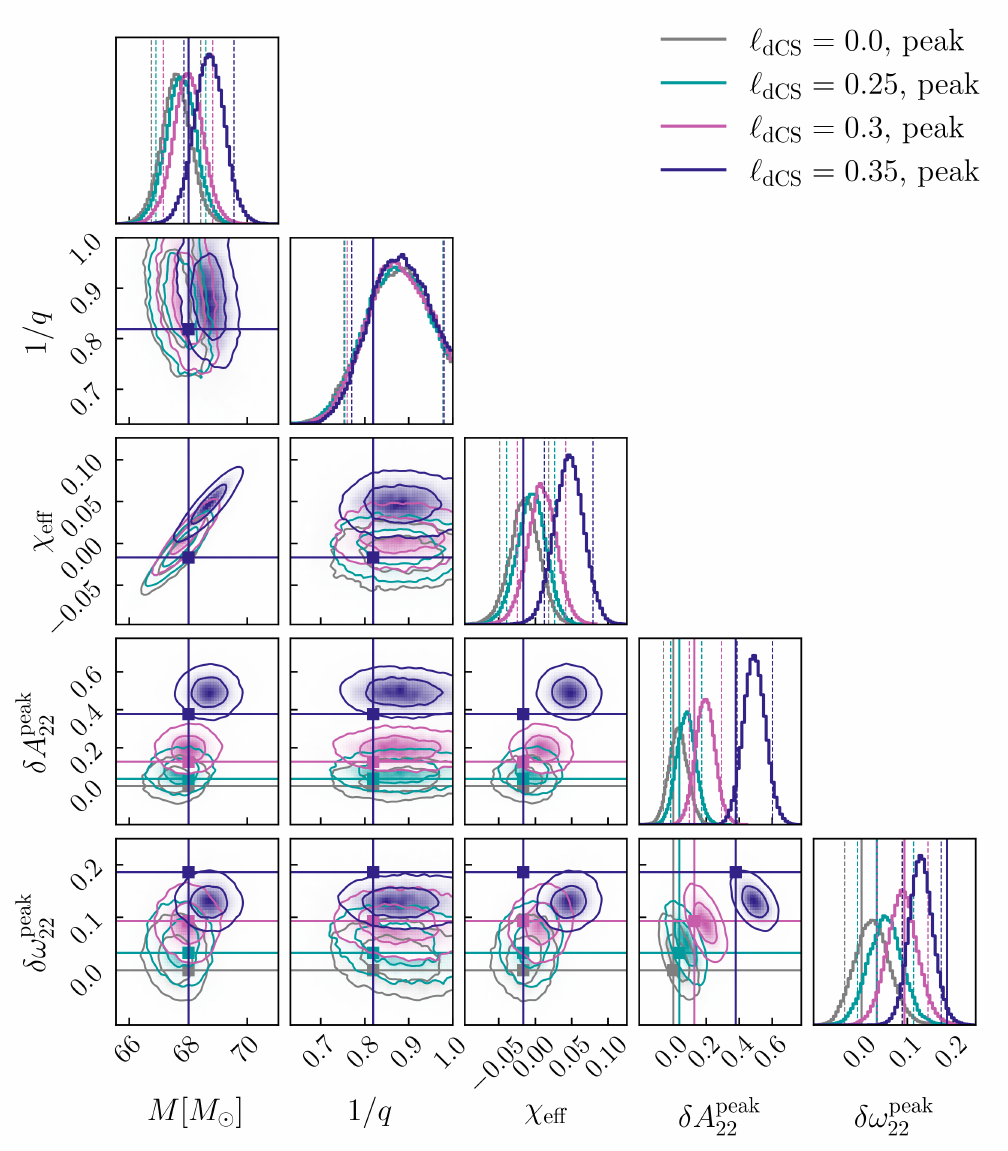}
    \caption{Analysis of a GW150914-like \ac{nr} simulation and three signals built on it to incorporate \ac{dcs} corrections to \ac{gr}, sampling on deviations from the $(2,2)$ mode's peak time amplitude and frequency.
    We show one- and two-dimensional posterior distributions, with expected values marked by squares and straight lines.
    We find some lingering bias in the estimate of $M$ and $\chi_{\rm eff}$ even when using the peak time deviations, indicating that they alone do not provide a perfect fit to the signal.
    \label{fig:dcs_mrg_corner}}
\end{figure}

We collect in this Appendix plots displaying detailed \ac{pe} results for the analyses of synthetic, \ac{dcs}-corrected signals discussed more widely in Sec.~\ref{sec:nrwf}.
Figs.~\ref{fig:dcs_gr_corner},~\ref{fig:dcs_rmn_corner}, and~\ref{fig:dcs_mrg_corner} show posterior distributions for the analyses sampling only on \ac{gr} parameters, on the remnant properties, and the peak quantities, respectively, each comparing results for the base \ac{nr} simulation and its three \ac{dcs} versions.

\end{document}